%% file: main.tex
\documentclass[a4paper,11pt]{article}

\usepackage{jheppub}

\usepackage[utf8]{inputenc}
\usepackage[american]{babel}

\usepackage{comment}

\usepackage{soul}
\usepackage{letterint}
\usepackage{brackety}
\usepackage{mathrsfs}

\usepackage{booktabs}
\usepackage{array}
\newcolumntype{C}{>{$}c<{$}} 
\newcolumntype{R}{>{$}r<{$}} 
\newcolumntype{L}{>{$}l<{$}} 

\usepackage{hyperref}
\hypersetup{%
    colorlinks,%
    breaklinks,%
    citecolor=blue,urlcolor=blue,linkcolor=blue,%
    hypertexnames=false,%
    pdftitle={The HVP at NNNLO},%
    pdfauthor={L. Lellouch, A. Lupo, M. Sjo, K. Szabo, P. Vanhove}%
}
\makeatletter
\renewcommand\HyPsd@CatcodeWarning[1]{}     
\makeatother

\usepackage{cleveref}
\crefname{equation}{eq.}{eqs.}
\crefname{section}{sec.}{secs.}
\Crefname{equation}{Eq.}{Eqs.}
\Crefname{section}{Sec.}{Secs.}
\newcommand{\rcite}[2][]{ref.~\cite[#1]{#2}}
\newcommand{\rrcite}[2][]{refs.~\cite[#1]{#2}}

\usepackage{placeins}   
\usepackage{afterpage}

\DeclareMathOperator*{\diag}{diag}

\usepackage{pgf,tikz}
\usetikzlibrary{%
    calc,
    positioning,
    patterns,
    decorations.pathreplacing,
    decorations.markings,
    decorations.pathmorphing,
    shapes.geometric,
    arrows.meta,
    external%
}

\usepackage{pgfplots}
\pgfplotsset{compat=1.18}
\usepgfplotslibrary{fillbetween}

\usepackage{mathtools} 
\usepackage{xfrac} 
\usepackage{tipa} 

\newcommand{\eqnbreak}{\notag\\&\quad}

\newcommand{\intx}{\int\hspace{-.75ex}} 
\renewcommand{\d}{\mathrm{d}}

\newcommand{\p}{\partial}
\newcommand{\Z}{\mathbb{Z}}
\newcommand{\F}{\mathcal{F}}
\newcommand{\U}{\mathcal{U}}
\newcommand{\D}{\mathcal{D}}

\renewcommand{\O}{\mathcal{O}}
\newcommand{\inv}{{-1}}
\newcommand{\Gram}{\mathbb{G}}
\newcommand{\lagr}{\mathcal{L}}

\newcommand{\form}{\texttt{FORM}}
\newcommand{\formversion}{\texttt{FORM~4.2}}
\newcommand{\chpt}{\ifmmode{\mathrm{ChPT}}\else{ChPT}\fi}
\newcommand{\LO}{\ifmmode{\mathrm{LO}}\else{LO}\fi}
\newcommand{\NLO}{\ifmmode{\mathrm{NLO}}\else{NLO}\fi}
\newcommand{\NNLO}{\ifmmode{\mathrm{NNLO}}\else{NNLO}\fi}
\newcommand{\NNNLO}{\ifmmode{\mathrm{N^3LO}}\else{N\textsuperscript{3}LO}\fi}

\newcommand{\NL}{{N_L}}
\newcommand{\NP}{{N_p}}
\newcommand{\NK}{{N_k}}

\DeclareMathOperator{\Li}{Li}

\DeclareMathOperator{\E}{\mathcal E}

\newcommand{\I}{\mathcal I}
    
\newcommand{\Mpi}{M_\pi}
\newcommand{\Fpi}{F_\pi}
\newcommand{\Mpiz}{M_0}
\newcommand{\Fpiz}{F_0}

\newcommand{\Lpi}{L_\pi}

\newcommand{\vnu}{{\vec\nu}}

\newcommand{\rat}{\mathrm{rat}}

\newcommand{\tadbub}[1]{T_{#1}}
\newcommand{\Jbub}[1]{J_\bub^{(#1)}}

\definecolor{bubblegum}{rgb}{0.99, 0.76, 0.8}

\def\mev{\mathrm{Me\kern-0.1em V}}
\def\gev{\mathrm{Ge\kern-0.1em V}}
\def\tev{\mathrm{Te\kern-0.1em V}}
\def\fm{\mathrm{fm}}

\newcommand{\formrefs}{\cite{Vermaseren:2000nd,Ruijl:2017dtg}}
\newcommand{\literedrefs}{\cite{Lee:2012cn,Lee:2013mka}}
\newcommand{\psdrefs}{\cite{Borowka:2017idc,Borowka:2018goh,Heinrich:2021dbf,Heinrich:2023til}}

\input{tikz-settings.tex}
\input{scriptdiagr.tex}

\newlength{\tikzineqyshift}
\NewDocumentCommand{\tikzineq}{ O{} O{\tikzineqyshift} m }{%
    {%
        \tikz[%
        anchor=base,%
        baseline={([yshift=#2]current bounding box.center)},%
        #1]{#3}%
    }}

\title{Hadronic vacuum polarization to three loops in chiral perturbation theory}

\author[a]{Laurent Lellouch,}
\author[a]{Alessandro Lupo,}
\author[a]{Mattias Sjö,}
\author[b,c]{Kálmán Szabo,}
\author[d]{and Pierre Vanhove}

\affiliation[a]{Aix Marseille Univ, Université de Toulon, CNRS, CPT, Marseille, France}
\affiliation[b]{Department of Physics, University of Wuppertal, D-42119 Wuppertal, Germany}
\affiliation[c]{Jülich Supercomputing Centre, Forschungszentrum Jülich, D-52428 Jülich, Germany}
\affiliation[d]{Institut de Physique Théorique, Université Paris-Saclay, CEA, CNRS,\\F-91191 Gif-sur-Yvette Cedex, France}

\abstract{
Hadronic vacuum polarization at low virtualities limits the precision of experimental tests of the standard model via important physical observables.
Here we compute that effect in two-flavor chiral perturbation theory to three loops.
Among the master integrals that describe the amplitude, six are elliptic functions of the momentum.
Of these, five are new to this work, although all can be related to the three-loop sunset integral.
The renormalizability of the amplitude hinges on relations between the master integrals that were not previously known and that are not consequences of the integration-by-parts reduction.
Our result is intended to serve as a starting point for phenomenological calculations, as well as the computation of finite-volume corrections in lattice QCD.
}

\begin{document}

\maketitle

\section{Introduction}

The propagation of photons in the vacuum is modified by quantum fluctuations of the quark and gluon fields. 
This phenomenon is known as hadronic vacuum polarization (HVP). 
At low virtualities, our current knowledge of HVP limits the precision of experimental tests of the standard model via important observables such as the muon anomalous magnetic moment~\cite{Muong-2:2025xyk} or the value of the electromagnetic coupling at the electroweak scale~\cite{FCC:2018byv}.
To align the standard-model prediction for the muon $g-2$ with the current experimental uncertainty~\cite{Muong-2:2025xyk}, the precision of the HVP contribution must improve by a factor of 4 relative to the world average of \cite{Aliberti:2025beg} or by a factor of 2 compared to the result in \rcite{Boccaletti:2024guq}.
Furthermore, stringent tests of the standard model via precision electroweak observables at future facilities such as CERN’s FCC-ee require reducing uncertainties by a factor of 3 on the running of $\alpha$ at low virtualities~\cite{FCC:2018byv}.
To various degrees, HVP also enters the computation of quantities such as the anomalous magnetic moments of the electron and the $\tau$, as well as the ground-state hyperfine splitting of muonium and muonic helium (see e.g. \rcite{Keshavarzi:2019abf}).

At low energies, the challenge of computing HVP stems from the fact that quark and gluon interactions are highly non-linear and not amenable to the usual perturbative methods of quantum field theory (QFT): the strong interaction, described by quantum chromodynamics (QCD), is nonperturbative in that regime. 
There are two main approaches to addressing this challenge: massively parallel numerical simulations in a discretized version of QCD known as lattice QCD, and a data-driven approach.
The latter combines the measurement of the $e^+e^-\to\text{hadrons}$ cross section as a function of center-of-mass energy with fundamental features of relativistic QFT, such as unitarity and analyticity properties of two-point Green's functions (see e.g.  \rcite{Aliberti:2025beg} for a review).

A third approach is Chiral Perturbation Theory (\chpt) \cite{Gasser:1983yg,Gasser:1984gg}. 
$\chpt$ is the low-energy effective field theory of QCD and, as such, is a key tool for studying hadronic interactions at low energies. 
While the predictive power of this approach is limited by the appearance of an increasing number of {\em a priori} unknown couplings at each new loop order (low-energy constants, LECs), it does constrain the functional form of strong-interaction observables at low energies. 
In particular, it provides important information on the very-low-energy QCD contributions to HVP, which are dominated by two-pion contributions. 
This information can then be combined with the predictions of the main approaches to improve their accuracy.
Here we present a calculation of those contributions to three loops in $\mathrm{SU}(2)$ \chpt~\cite{Gasser:1983yg}.

One motivation for this calculation is to obtain a precise, model-independent determination of the finite-volume corrections that must be applied to lattice results for quantities sensitive to long-distance HVP effects, such as the HVP contribution to the muon anomalous magnetic moment.
Other model-independent approaches have also been proposed.%
\footnote{
    These are the Hansen-Patella method~\cite{Hansen:2019rbh, Hansen:2020whp} and the Meyer-Lellouch-Lüscher approach~\cite{Luscher:1990ux, Lellouch:2000pv, Meyer:2011um}.}
It is understood that finite-volume effects arise from long-distance, low-energy hadronic fluctuations, making $\mathrm{SU}(2)$ \chpt\ the natural framework to describe them~\cite{Aubin:2020scy}. 
Intuitively, the lightest propagating states will be the most sensitive to the finite spacetime boundaries, since they have the slowest rate of exponential decay with the spatial size $L$ of the lattice.
In the isospin-$1$ channel, which dominates the HVP contribution to the muon $g-2$, these are two-pion states.

Finite-volume effects on the HVP contribution to the muon $g-2$ have been computed to one~\cite{Aubin:2015rzx} and two~\cite{Bijnens:2017esv,Aubin:2019usy,Borsanyi:2020mff} loops. 
Assuming a geometric progression of these effects, the three-loop contribution and its uncertainty can be estimated, with the latter taken to be the sum of the remainder of the series.
Using the results in \rcite{Borsanyi:2020mff}, one obtains $(17.9\pm 1.0)\times 10^{-10}$ for lattices of spatial (and temportal) extent $L=6.3\,\fm$ ($T=3L/2$) that are typical of those used in state-of-the-art computations.
This result is fully compatible with the direct lattice determination of that correction~\cite{Borsanyi:2020mff}, but with an uncertainty that is $2.5$ times smaller.
Moreover, three loops is the first order in ChPT at which this uncertainty is smaller—by a factor of $1.5$---than the one on the current world average of measurements of the muon $g-2$~\cite{Muong-2:2025xyk}, guaranteeing that this uncertainty is not a limiting factor in testing the Standard Model via this muon property.
Of course, such a determination will require knowledge of several LECs.
However, as discussed later, at three loops only six combinations of LECs contribute (three each from the NLO~\cite{Gasser:1983yg} and NNLO~\cite{Bijnens:1999sh} Lagrangians), all of which can be determined from quantities such as the pion vector form factor at two loops.%
\footnote{
    In \rcite{Aubin:2020scy} it is shown that a new counterterm involving muon fields is needed in the three-loop $\chpt$ computation of $a_\mu$. However, because it is a contact term, it will only contribute to the finite-volume corrections to $a_\mu$ at four loops. Moreover, no such counterterm is needed in the computation of the HVP function itself.}
    
In the meantime, a precise lattice-plus-data-driven determination of the HVP contribution to $a_\mu$ was obtained~\cite{Boccaletti:2024guq}. 
In that approach, finite-volume corrections were reduced by a factor of roughly 2 and their uncertainty by approximately 3. 
This means that two-loop $\chpt$ should be sufficient to reach the precision required by experiment. 
Nevertheless, a three-loop determination would provide an important cross-check of the corresponding uncertainty estimate. 
Moreover, a number of lattice collaborations are currently calculating the HVP contribution to $a_\mu$ entirely on the lattice~\cite{Djukanovic:2024cmq,RBC:2024fic,FermilabLatticeHPQCD:2024ppc} and for those the improvement brought by a three-loop calculation is relevant. 
In addition, such a calculation will reduce the uncertainty on lattice calculations of the low-energy running of $\alpha$.

As a first step in computing those FVEs, we calculate here HVP in infinite volume to three loops in \chpt. 
Beyond the determination of FVEs, the results of this calculation are interesting in and of themselves. 
They not only allow the study of the slope, but also of the curvature and turnaround of the HVP function for spacelike momenta. 
In the timelike region, its imaginary part not only describes the interference between the one-~\cite{Gasser:1983yg} and two-loop~\cite{Bijnens:1998fm} pion form-factor contributions to the $e^+e^-\to\pi^+\pi^-$ cross section, but also the four-pion contribution, which is absent at lower orders. 
Moreover, it is only the second three-loop \chpt\ calculation performed
and therefore pushes the envelope of such computations.%
\footnote{
    The only other one that we know of is the three-loop computation of the pion mass and decay constant by Bijnens \&~Hermansson-Truedsson \cite{Bijnens:2017wba}.}

The evaluation of the three-loop amplitude presented in this work
requires state-of-the-art integration techniques of Feynman integrals
and the use of new identities between master integrals, which are not derivable from the commonly used
integration-by-part algorithms,
and which were previously unknown. The one- and two-loop
contributions to HVP involve only logarithms and polylogarithms, whereas the three-loop amplitude leads to elliptic integrals.

The elliptic master integrals needed in this work correspond to two-point functions with a single mass and no massless internal propagators.
Using Tarasov dimension shifting relations~\cite{Tarasov:1996br}, the elliptic master integrals are reduced to integrals in two dimensions that are both ultraviolet and infrared finite.
Compared to previous three-loop computations in other theories, the fact that all internal lines are massive here leads to new elliptic master integrals that were
unknown in the literature. 
All the needed elliptic master integrals can be obtained by differentiation or integration of the three-loop sunset integral in two dimensions. 
This integral arises in several precision calculations of the standard model~\cite{Groote:1998wy,Groote:2004qq,Groote:2005ay},
quark and gluon self-energies~\cite{Broadhurst:1993mw},  
QED calculations~\cite{Lee:2019wwn,Lee:2020mvt,Forner:2024ojj} 
and in Higgs boson decay amplitudes~\cite{Wang:2024ilc}.
It has been the subject of many studies, including \rrcite{Bailey:2008ib,Bloch:2014qca,Primo:2017ipr,Broedel:2019kmn,Pogel:2022yat}.

Using the algorithms developed in~\rcite{Lairez:2022zkj,delaCruz:2024xit}, we give the differential equations satisfied by the new elliptic master integrals. 
The resolution of these differential equations provides fast, high-precision evaluations of the master integrals     in the complex energy plane and complete control of the theoretical uncertainties at this order in \chpt.
The renormalization of the ultraviolet divergences of the three-loop amplitude is obtained thanks to the derivation of new relations between the elliptic master integrals evaluated in two dimensions.
These relations are not a consequence of integration-by-part identities as they arise in fixed spacetime dimension. 
We provide two methods for deriving them: one based on the familiar Schouten identities~\cite{Tancredi:2015pta} and another, on the differential equations satisfied by the master integrals.

While we give an analytic derivation of these elliptic master integrals here, we defer details concerning their numerical implementation to forthcoming work~\cite{LLSV}.
Likewise, due to the complexity of the present calculation, we focus here on the computation of the three-loop HVP amplitude itself,
postponing phenomenological applications and the computation of finite-volume effects to later work.

We conclude this introduction with a summary of earlier ChPT HVP calculations.
It was calculated at one-loop order in Gasser~\&~Leutwyler's original \chpt\ papers~\cite{Gasser:1983yg,Gasser:1984gg}.
The two-loop calculation was performed by Golowich \&~Kambor~\cite{Golowich:1995kd} and then at greater generality by Bijnens et al.~\cite{Amoros:1999dp,Bijnens:2011xt}.
Its finite-volume counterpart has been computed in partially quenched \chpt~\cite{Bijnens:2017esv} and using the time-momentum representation~\cite{Aubin:2020scy,Aubin:2015rzx}.
Few calculations have been carried out to three-loop order in \chpt. As mentioned above, to the best of our knowledge, there is only this work and the pion mass and decay constant calculated by Bijnens \&~Hermansson-Truedsson \cite{Bijnens:2017wba}.
Reproduction of the results of \rrcite{Bijnens:2011xt,Bijnens:2017wba}, whose \form~\formrefs\ implementations have been made available to us, served as helpful verification throughout the present calculation.

The remainder of this paper is structured as follows.
\Cref{sec:chpt} describes the relevant theoretical background and lists the Feynman diagrams.
\Cref{sec:loops,sec:more-loops} are devoted to the treatment of the loop integrals, with lengthy expressions delegated to \cref{app:diffeq,app:dimshift,app:expr-4d,app:expr-2d}.
We present the results in \cref{sec:results}, with further results in \cref{app:unsub-results}.
A description of how the calculations were implemented can be found in \cref{app:implementation}, along with links to our codes.

\section{Theoretical background}\label{sec:chpt}

The basic quantity of interest is the vacuum polarization
\begin{equation}\label{eq:Pi-def}
    \Pi^{\mu\nu}(q) \coloneq \int \d^4 x\, e^{iqx} \tr[\big]{0\big|\,T\big\{j^\mu(x) j^\nu(0)\big\}\big|0}\,,
\end{equation}
where $j_\mu(x)$ is the electromagnetic current.
HVP consists of the hadronic contributions to $\Pi^{\mu\nu}(q)$, i.e., those where the right-hand side is computed from some model of hadrons, for which we use two-flavor \chpt\ as described in detail below.
From now on, we will assume that $\Pi^{\mu\nu}(q)$ contains only this hadronic part.

As a rank-$2$ tensor depending on a single four-vector $q^\mu$, Lorentz invariance dictates that $\Pi^{\mu\nu}(q)$ can be projected onto just two orthogonal structures,
\begin{equation}
    \Pi^{\mu\nu}(q) = (q^\mu q^\nu - q^2\eta^{\mu\nu}) \Pi_T(q^2) + q^\mu q^\nu \Pi_L(q^2)\,,
\end{equation}
where (in $d$ dimensions with $\eta^{\mu\nu}\eta_{\mu\nu}=d$)
\begin{equation}\label{eq:LT-def}
    \Pi_L(q^2) \coloneq \frac{1}{(q^2)^2} q_\mu q_\nu \Pi^{\mu\nu}(q)\,,\qquad
    \Pi_T(q^2) \coloneq \frac{1}{1-d}\bigg[\frac{1}{q^2} \eta_{\mu\nu}\Pi^{\mu\nu}(q) - \Pi_L(q^2)\bigg]
\end{equation}
are its longitudinal and transverse components, respectively.
By the Ward--Takahashi identity, $\Pi_L(q^2) = 0$.%
\footnote{
    Most individual Feynman diagrams contributing to $\Pi^{\mu\nu}(q)$ possess a longitudinal component, and the vanishing of $\Pi_L(q^2)$ only becomes manifest after summing all diagrams and performing the master integral reduction (see \cref{sec:loops}).
    Therefore, the Ward--Takahashi identity is a powerful sanity check on our calculations, ruling out most possible errors such as missing diagrams, erroneous symmetry factors, malformed Feynman rules, or misconfigured master integral reduction.}

\subsection{The \chpt\ Lagrangian}
We use two-flavor \chpt\ in the isospin limit, describing the low-energy dynamics of a $\mathrm{SU}(2)$ triplet of pion fields of mass $\Mpi$ coupled to an external, non-dynamical photon field $A_\mu$.
We need the Lagrangian to the first four orders in the chiral power counting,
\begin{equation}
    \lagr_\chpt = \lagr_\LO + \lagr_\NLO + \lagr_\NNLO + \lagr_\NNNLO + \cdots.
\end{equation}
The pions are contained in a matrix field $U$, which we parametrize as
\begin{equation}\label{eq:param}
    U = \frac{i\Phi}{\Fpiz\sqrt2} + \sqrt{1 - \frac{\tr{\Phi^2}}{4F_0^2}}\,,
\end{equation}
where $\Fpiz$ is the bare pion decay constant
and $\Phi = \sum_i \phi^i \sigma^i 
$ is the matrix of pion fields.
We normalize our Lie algebra as $\tr{\sigma^i\sigma^j} = \delta^{ij}$, where $\tr{\cdots}$ denotes an $\mathrm{SU}(2)$ trace.
The same physical results are obtained from any parametrization that is equivalent to \cref{eq:param} up to $\O(\Phi^2)$, such as $U=\exp(i\Phi/\Fpiz\sqrt2)$.

The leading Lagrangian is
\begin{equation}
    \lagr_\text{LO} = \frac{\Fpiz^2}{4} \tr[\big]{D_\mu U (D^\mu U)^\dag + \chi U^\dag + U \chi^\dag}\,,
\end{equation}
featuring the covariant derivative $D_\mu U \coloneq \p_\mu U - ir_\mu U + iU\ell_\mu$. 
In this calculation, the chiral gauge fields are simply $\ell_\mu=r_\mu=A_\mu Q$, where $Q=\frac e3\diag(+2,-1)$ is the quark charge matrix.
Likewise, $\chi$, which contains the quark mass matrix, is simply the bare pion mass $\Mpiz^2$.
However, we retain $\ell_\mu,r_\mu$ and $\chi$ to preserve the manifestly chiral-invariant appearance of the Lagrangian. 
We also keep $\ell_\mu$ and $r_\mu$ because the axial current, proportional to $r_\mu-\ell_\mu$, determines the pion decay constant which appears in our calculation (see \cref{sec:massdecay}). 

The next-to-leading Lagrangian is (following \rcite{Gasser:1983yg} but using somewhat modernized notation)%
\footnote{
    There is a subtlety here due to the distinct form of the original 2-flavor Lagrangian compared to the later 3- or $N$-flavor ones~\cite{Gasser:1984gg}.
    If done too naïvely, a reduction from $N$ to 2 flavors via the Cayley--Hamilton theorem leads to errors in the part of our result where $\lagr_4$ enters quadratically (diagrams 10--20 in~\cref{tab:NNNLO}) due to not fully taking into account the equations of motion remarked in \rcite{Bijnens:1999hw}.
    }
\begin{equation}
    \begin{aligned}
    \lagr_\text{NLO}
        &= \frac{l_1}{4}\tr[\big]{D_\mu U (D^\mu U)^\dag}^2
        + \frac{l_2}{4}\tr[\big]{D_\mu U (D_\nu U)^\dag}\tr[\big]{D^\mu U (D^\nu U)^\dag} 
        + \frac{l_3}{16}\tr[\big]{\chi U^\dag + U \chi^\dag}^2\\
        &+ \frac{l_4}{2}\tr[\big]{D_\mu\chi^\dag D^\mu U}
        + l_5\tr[\big]{U^\dag F_R^{\mu\nu} U F_{L\mu\nu} }
        + i\frac{l_6}{2}\big[\tr{F_R^{\mu\nu} U_\mu U^\dag_\nu} + \tr{F_L^{\mu\nu} U^\dag_\mu U_\nu}\big]
        \\
        &- 2h_2 \tr[\big]{F_R^{\mu\nu} F_{R\mu\nu} + F_L^{\mu\nu} F_{L\mu\nu}}+\cdots\,,
    \end{aligned}
\end{equation}
where the ellipses denote terms which do not appear in our calculation. 
The parameters $l_i$ and $h_i$ are the bare LECs and the field strengths are, for our purposes, $F_L^{\mu\nu}=F_R^{\mu\nu}=Q(\p^\mu A^\nu - \p^\nu A^\mu)$.%
\footnote{
    The Lagrangian is derived under the assumption that $Q$ (and therefore $F_{L,R}^{\mu\nu}$) is traceless, which is not the case here.
    However, $\tr{Q^2}$ and $\tr{Q}^2$ appear here only as $\tr{Q^2}-\tr{Q}^2$, making
    the number of degrees of freedom the same as if $\tr{Q}=0$ and requiring no additional terms in the Lagrangian.
    Seen another way, such additional terms would only act as contact terms in the present calculation, so their contribution would not be independent of that of $h_2$.
    The same remains true at higher orders.}
Note that the $h_2$ term is a \emph{contact term} and does not contain pion fields.

The lengthy higher-order Lagrangians $\lagr_\NNLO$ and $\lagr_\NNNLO$ are given in \rrcite{Bijnens:1999hw,Bijnens:2018lez}.
We have implemented the full \chpt\ lagrangian in \form~\formrefs, and the attached code (see \cref{app:implementation}) includes efficient on-the-fly derivation of any Feynman rule from \mbox{$\lagr_\text{LO}+\cdots+\lagr_\NNNLO$}.

\subsection{Renormalization}\label{sec:renorm}
The Lagrangians beyond $\lagr_\text{LO}$ provide counterterms.
At NLO, the renormalization of the LECs in $d=4-2\epsilon$ dimensions at a scale $\mu$ (conventionally, $\mu=770~\mev\approx M_\rho$) is
\begin{equation}\label{eq:renorm-NLO}
    l_i = \pi_{16}(c\mu)^{d-4}\Big[l^q_i(\mu) + \tfrac{\gamma_i}{d-4}\Big]\,,\qquad
    h_i = \pi_{16}(c\mu)^{d-4}\Big[h^q_i(\mu) + \tfrac{\delta_i}{d-4}\Big]\,,
\end{equation}
where the $l^q_i(\mu)$ are the renormalized LECs
and $\pi_{16}\coloneq\tfrac{1}{16\pi^2}$.%
\footnote{
    Note that in \rcite{Bijnens:1999sh}, renormalized quantities are given as functions of $d$, which would imply that they possess an $\epsilon$ expansion whose subleading terms would appear in diagrams containing both counterterms and loops.
    However, such subleading terms can always be absorbed into the leading part of higher-order LECs, and this is indeed done. 
    (We thank Johan Bijnens for clarifying this point.)
    }
We follow \rcite{Bijnens:2017wba} in using a convention that more uniformly distributes powers of $\pi_{16}$; the conversion to that of, e.g., \rrcite{Bijnens:1999sh,Bijnens:2014lea} is $l_i^r(\mu)=\pi_{16} l_i^q(\mu)$.
The coefficients $\gamma_i$ are~\cite{Gasser:1983yg}
\begin{equation}
    \gamma_1 = \tfrac13\,,\quad
    \gamma_2 = \tfrac23\,,\quad
    \gamma_3 = -\tfrac12\,,\quad
    \gamma_5 = -\tfrac16\,,\quad
    \gamma_6 = -\tfrac13\,,\quad
    \gamma_7 = 0\,,\qquad
    \delta_2 = -\tfrac1{12}\,.
\end{equation}
The scheme-dependent coefficient $c$ is conventionally%
\footnote{
    This is the $\overline{\mathrm{MS}}$ scheme but with $-1$ added in the exponent, which is the standard scheme for \chpt.}
\begin{equation}\label{eq:cmu}
    (c\mu)^{d-4} 
        = e^{-\epsilon(\gamma_E + \log\mu^2 - \log4\pi - 1)}
        = C(\epsilon,\mu^2) e^\epsilon\,,
\end{equation}
where $d=4-2\epsilon$ and $\gamma_E$ is the Euler--Mascheroni constant.
For later convenience, we have defined $C(\epsilon,z)\coloneq e^{-\epsilon(\gamma_E+\log z-\log4\pi)}$ (see \cref{sec:loops}).

Similarly, the LECs $c_i$ of $\lagr_\text{NNLO}$ are renormalized as
\begin{equation}\label{eq:renorm-NNLO}
    c_i = \frac{\pi_{16}^2 (c\mu)^{2(d-4)}}{\Fpiz^2}\Big[c_i^q(\mu) - \tfrac{\gamma_i^{(2)}}{(d-4)^2} - \tfrac{\gamma_i^{(1)} + \gamma_i^{(L)}(\mu)}{\pi_{16}(d-4)}\Big]\,,
\end{equation}
with coefficients listed in \rcite{Bijnens:1999sh}.
The $\gamma_i^{(2)}$ and the $\gamma_i^{(1)}/\pi_{16}$ are rational numbers like the $\gamma_i$, while the $\gamma_i^{(L)}/\pi_{16}$ are rational linear combinations of the $l_j^q$.
The factor of $1/\Fpiz^2$ ensures that the physical LECs are dimensionless.

There currently exists no \NNNLO\ counterpart to \rcite{Bijnens:1999sh}.
The renormalization of the \NNNLO\ LECs would look analogous to \cref{eq:renorm-NNLO}, including a $1/(d-4)^3$ term, with all coefficients unknown.
However, the \NNNLO\ LECs only appear in very simple terms in our amplitude and need only absorb a limited kind of local divergences.
Therefore, the coefficient combinations that appear can be inferred from whatever is left after the more involved NLO$+$NNLO renormalization.
We state these expressions in \cref{sec:renorm-N3LO}.

\newcommand{\arrayfudge}{\rule[-4.5ex]{0ex}{10.1ex}}
\begin{table}[tp!]
    \centering
    \input{diagram-table-lower.tex}
    \caption{%
        All \NLO\ (top) and \NNLO\ (bottom) diagrams, sorted by ascending number of loops, then roughly by complexity.
        Unmarked vertices are \LO, dots \NLO, and squares \NNLO.
        The rows and columns are numbered so that individual diagrams can be easily referenced, such as ``14'' for the double bubble.
        The mirror images of the non-symmetric diagrams (03 and 12) are not listed independently.
        Note that all 2-loop diagrams factor into 1-loop integrals; there are no sunset diagrams.
        }
    \label{tab:NLO-NNLO}
\end{table}
\begin{table}[p!]
    \centering
    {\renewcommand{\diagramscale}{1.15}
    \input{diagram-table.tex}}
    \caption{%
        All \NNNLO\ diagrams, presented similarly to~\cref{tab:NLO-NNLO}.
        Unmarked vertices are \LO, dots \NLO, squares \NNLO, and triangles \NNNLO.
        The rows and columns are numbered so that individual diagrams can be easily referenced.
        The mirror images of the non-symmetric diagrams (03, 13, 32--41, etc.) are not listed independently.
        Note that, unlike at \NNLO, there are six 3-loop diagrams (91--101) which do not factor into 1-loop integrals.
        }
    \label{tab:NNNLO}
\end{table}

\subsection{Diagrams}

For an N$^k$LO diagram, $k$ is the order in ChPT's expansion parameters, not the number of loops or the number of orders above lowest that contributes (thus, the leading HVP contribution is labeled as NLO).
The full set of N$^k$LO diagrams consists of the $k$-loop diagrams constructed using $\lagr_\LO$, plus lower-loop diagrams containing vertices from the higher-order Lagrangians, such that in each diagram, $k$ equals the number of loops plus the total power-counting order of the vertices (informally, the number of N's).
For $\Pi^{\mu\nu}(q^2)$, the diagrams at NLO and NNLO are given in~\cref{tab:NLO-NNLO}, and those at \NNNLO\ in~\cref{tab:NNNLO}.
In particular, note how the lack of vertices with an odd number of pion lines---a consequence of G-parity---strongly limits the available topologies and completely removes the two-loop sunset.

As is usual with \chpt\ calculations, the algebraic expressions corresponding to these diagrams are very lengthy, so we used \formversion~\formrefs\ throughout our calculations.
To handle the large number of diagrams in a less error-prone way, we used a newly written program that, given a simple description of the diagram topologies, generates the \form\ code that would have been written by hand in earlier work.
See \cref{app:implementation} for further information.

\afterpage{\clearpage} 

\subsection{Mass and decay constant}\label{sec:massdecay}
The Lagrangian contains the bare pion decay constant, $\Fpiz$, and all masses coming from vertices and propagators are nominally the bare pion mass, $\Mpiz$.
These are related to their renormalized counterparts $\Fpi$ and $\Mpi$ (which are scheme-independent physical observables) via
\begin{equation}\label{eq:massdecay}
    \Mpiz^2 \to \Mpi^2\bigg[1 + \sum_{i=1}^\infty \xi^i \delta M_i\bigg]\,,\qquad
    \Fpiz   \to \Fpi  \bigg[1 + \sum_{i=1}^\infty \xi^i \delta F_i\bigg]\,,\\
\end{equation}
where $\xi\coloneq \pi_{16}\Mpi^2/\Fpi^2$ and the expressions for $\delta M_i$ and $\delta F_i$ are given below.
An N$^k$LO amplitude is then obtained by summing, over $k'\leq k$, all contributing N$^{k'}$LO amplitudes in which mass and decay constant renormalization has been performed to N$^{k-k'}$LO, via the expressions above.

Given the lack of LO diagrams in our case, we only need mass and decay constant renormalization up to NNLO.
This was first determined by Bürgi~\cite{Burgi:1996qi}, although we use the form given along with the \NNNLO\ case in \rcite{Bijnens:2017wba}:
\begin{equation}
    \delta M_i = \sum_{j=0}^i b^M_{ij} \Lpi^i\,,\qquad
    \delta F_i = \sum_{j=0}^i b^F_{ij} \Lpi^i\,,
\end{equation}
where $\Lpi\coloneq \log(\Mpi^2/\mu^2)$ and
\begin{equation}
    \begin{aligned}
        &b^M_{10} = -2l_{3}^q\,,\quad
        b^M_{11} = -\tfrac{1}{2}\,,\qquad
        b^F_{10} = -l_{4}^q\,,\quad
        b^F_{11} = 1\,,\\
        &b^M_{20} = -\tfrac{163}{96} - 64c_{18}^q - 32c_{17}^q - 96c_{11}^q - 48c_{10}^q + 16c_{9}^q + 32c_{8}^q + 16c_{7}^q + 32c_{6}^q \\
        &\qquad- 4l_{3}^ql_{4}^q + 8(l_{3}^q)^2 - 2l_{2}^q - l_{1}^q\,,\\
        &b^M_{21} = \tfrac{13}{3} - l_{4}^q + 11l_{3}^q + 8l_{2}^q + 14l_{1}^q\,,\qquad
        b^M_{22} = -\tfrac{5}{8}\,,\\
        &b^F_{20} = \tfrac{13}{192} - 8c_{9}^q - 16c_{8}^q - 8c_{7}^q - (l_{4}^q)^2 + 2l_{3}^ql_{4}^q + l_{2}^q + \tfrac{1}{2}l_{1}^q\,,\\
        &b^F_{21} = -\tfrac{29}{12} + 3l_{4}^q - 4l_{2}^q - 7l_{1}^q\,,\qquad
        b^F_{22} = -\tfrac{1}{4}\,.
    \end{aligned}
\end{equation}

\section{Loop integral overview}\label{sec:loops}

In this section, we summarize the methods used to reduce and evaluate the loop integrals appearing in $\Pi^{\mu\nu}(q)$,
unifying the notation relative to the various sources from which they are taken.

\subsection{Definitions}\label{sec:loop-manip}
To write down a family of $\NL$-loop integrals with loop momenta $\{\ell_i\}_{i=1}^\NL$ and $\NP$ independent external momenta $\{p_i\}_{i=1}^\NP$,
one first constructs a set of $\NK$ momenta $\{k_j\}_{j=1}^\NK$ such that all propagators appearing in the integrals have the form
\begin{equation}\label{eq:Dj-def}
    \frac{1}{D_j} \coloneq \frac{1}{k_j^2 - m_j^2 + i\epsilon}\,,\qquad j\in\{1,\ldots,\NK\}\,,
\end{equation}
and such that any product $\ell_i\cdot\ell_j$ or $\ell_i\cdot p_j$ can be written as a linear combination of inverse propagators $D_j$ and the masses; this condition fixes $\NK=\NP\NL + \NL(\NL+1)/2$.%
\footnote{
    To see that the number of distinct propagators never exceeds this $\NK$,
    consider a connected graph with $N_V$ vertices, $N_E$ internal edges and $\NL$ loops,
    ignoring 2-point vertices since these do not affect the number of distinct propagators.
    Thus, each vertex has at least $3$ legs, each of which is either an external leg or half an edge, so that $\NP+2N_E \geq 3N_V$.
    Combining this with Euler's formula, \mbox{$N_V - N_E + \NL=1$}, yields $N_E \leq 3(\NL+1) + \NP$, which implies $N_E<\NK$ for all $\NL\geq1,\NP\geq 0$.}

In the present case, we only need to consider $m_j=\Mpiz$.
All of the integrals in $d$ dimensions (which is taken to be an integer minus $2\epsilon$) can be written as
\begin{equation}\label{eq:Inu-D}
    \big[\pi_{16} C(\epsilon, \Mpiz^2)\big]^\NL I^{(d)}_\vnu\big(\{p_i\cdot p_j\},\Mpiz\big) \coloneq 
        \intx\frac{\d^d \ell_1}{\pi^{d/2}}\cdots
        \intx\frac{\d^d \ell_\NL}{\pi^{d/2}}
        \prod_{j=1}^\NK \frac{1}{D_j^{\nu_j}}\,,
\end{equation}
where $\vec\nu\in\Z^\NK$ is a vector of integers that distinguishes members of the integral family.
This accommodates integrals with nontrivial numerators as long as they consist only of scalar products, since these can be accounted for with negative $\nu_j$.
The Wick rotation to Euclidean space takes the form $I_\vnu \to i^\NL (-1)^{\sum_j\nu_j} I^E_\vnu$.

Note in~\cref{eq:Inu-D} how we use the same normalization factor $C(\epsilon,\Mpiz^2)$ as in~\cref{eq:cmu}, albeit with $\Mpiz^2$ instead of $\mu^2$; this simplifies dimensional regularization and also ensures that $I^{(d)}_\vnu$ has integer mass dimension.
After the conversion $C(\epsilon,\mu^2)=C(\epsilon,\Mpiz^2) e^{\epsilon L}$, all diagrams at N$^k$LO will, when expressed in terms of $I_\vnu$ and renormalized LECs, contain $C(\epsilon,\Mpiz^2)^k$, which can be kept as-is throughout the calculations until the $\epsilon\to0$ limit can be safely taken with $C(0,\Mpiz^2)=1$.
The $\pi_{16} (4\pi)^\epsilon$ contained in the normalization of $I_\vnu$ compensates for how we normalize with $\pi^{d/2}$ in~\cref{eq:Inu-D} (which simplifies many formulae below) rather than $(2\pi)^d$ (which is how the integrals appear in the diagrams).

There exists a large number of techniques for manipulating loop integrals, and in \cref{sec:more-loops} we provide some details on those used in this work.
In summary, we first use integration-by-parts relations to express all integrals in terms of a small number of \emph{master integrals} listed below.
All subsequent calculation efforts can then be concentrated on these.
Tarasov's dimensional shift~\cite{Tarasov:1996br} is then used to relate master integrals in $d$ dimensions to those in $d-2$, simplifying renormalization since lower-dimensional integrals are typically less divergent.
We also derive novel Schouten relations which, in the vicinity of integer dimensions, provide additional relations between the master integrals.
Lastly, we derive differential equations satisfied by the integrals, allowing them to be computed without explicitly integrating any loop momenta.

There exist several excellent programs for performing the reduction to master integrals, but in this work we exclusively use \texttt{LiteRed 2}~\literedrefs,%
\footnote{
    We have also unsuccessfully tried to perform the reduction in other programs, but have been hampered by crashes or extremely long run times.
    Out of the box, \texttt{LiteRed 2} requires $\O(\text{1 hour})$ to determine the general reduction rules for our propagators, which needs to be done only once, and can then reduce all integrals in $\Pi^{\mu\nu}(q)$ in $\O(\text{1 minute})$.}
along with \texttt{pySecDec}~\psdrefs\ for numerically verifying the various integral manipulations.
    
\subsection{Master integrals of $\Pi^{\mu\nu}(q^2)$ at three-loop order}\label{sec:masters}

We use the following basis, which allows writing all three-loop integrals appearing in $\Pi^{\mu\nu}(q)$:
\begin{equation}\label{eq:basis}
    \{k_j\}_{j=1}^9 = \left\{
        \ell_1,\ell_2,\ell_3,\quad
        \ell_1-q,\ell_2-q,\ell_3-q,\quad
        \ell_1+\ell_3, \ell_2+\ell_3, \ell_1-\ell_2
        \right\}\,.
\end{equation}
For 2-loop integrals, we use the subset $\{k_1,k_2,k_4,k_5,k_9\}$, 
and for 1-loop integrals, $\{k_1,k_4\}$.
To simplify our expressions, we write the master integrals as functions of the dimensionless parameter $t \coloneq q^2/\Mpi^2$, suppress their dependence on $\Mpi$, and render all master integrals dimensionless at $d=2$ with an appropriate power of $\Mpi$.%
\footnote{
    Note that while the bare mass $\Mpiz$ is used in \cref{eq:Inu-D}, here we use the physical mass $\Mpi$, since the quantities in this section are the ones that will appear in the renormalized result.}

At one-loop order, the set of master integrals is simply the tadpole ($\tad$) and bubble ($\bub$):%
\begin{equation}
           I^{(d)}_{1,0}(q^2,\Mpi) \eqcolon I_\tad(d)\,,\qquad
    \Mpi^2 I^{(d)}_{1,1}(q^2,\Mpi) \eqcolon I_\bub(d;t)\,.
\end{equation}
Thanks to our choice of $C(\epsilon,\Mpi^2)$, the expression for the former is exceedingly simple:
\begin{equation}\label{eq:tadpole}
    I_\tad(4-2\epsilon) 
        = -ie^{\epsilon\gamma_E}\Gamma(\epsilon-1) 
        = \frac{i\exp\Big[\sum_{k=2}^\infty (-\epsilon)^k \frac{\zeta(k)}{k}\Big]}{\epsilon(1-\epsilon)}\,,
\end{equation}
using the relation between $\log\Gamma(z)$ and the Riemann zeta function $\zeta(z)=\sum_{n=1}^\infty n^{-z}$.
The latter is similar, and textbook manipulations put it in the form
\begin{equation}\label{eq:bubble}
    I_\bub(4-2\epsilon; t) 
        = (1-\epsilon)I_\tad(4-2\epsilon) 
        \int_0^1\frac{\d x}{\big[1 - x(1-x)t\big]^\epsilon}\,,
\end{equation}
where we write the Feynman parameter integral as%
\footnote{
    The notation is inspired by the standard bubble function $\bar J(\Mpi^2, q^2) = -\pi_{16}\Jbub1(t)$.}
\begin{equation}\label{eq:Jbub}
    \int_0^1\frac{\d x}{\big[1 - x(1-x)t\big]^\epsilon} = 1 + \sum_{n=1}^\infty (-\epsilon)^n \frac{\Jbub{n}(t)}{n!}\,,\qquad
    \Jbub{n}(t) \coloneq \int_0^1 \d x\, \log^n\big[1 - x(1-x)t\big]
\end{equation}
and express our results in terms of $\Jbub{n}(t)$, which are given explicitly in \cref{app:bubble}.

We have three two-loop master integrals, but, as noted in \cref{tab:NLO-NNLO}, they factorize into tadpoles and bubbles.
Adopting the tadpole-bubble product notation
\begin{equation}\label{eq:tadbub}
    \tadbub{ab}(d;t) \coloneq \big[I_\tad(d)\big]^a \big[I_\bub(d;t)\big]^b\,,
\end{equation}
they are
\begin{equation}\label{eq:tadbub-2loop}
    \begin{gathered}
               I^{(d)}_{1,1,0,0,0}(q^2,\Mpi) = \tadbub{20}(d)\,,\qquad
        \Mpi^2 I^{(d)}_{1,1,1,0,0}(q^2,\Mpi) = \tadbub{11}(d;t)\,,\\
        \Mpi^4 I^{(d)}_{1,1,1,1,0}(q^2,\Mpi) = \tadbub{02}(d;t)\,.
    \end{gathered}
\end{equation}

\begin{figure}
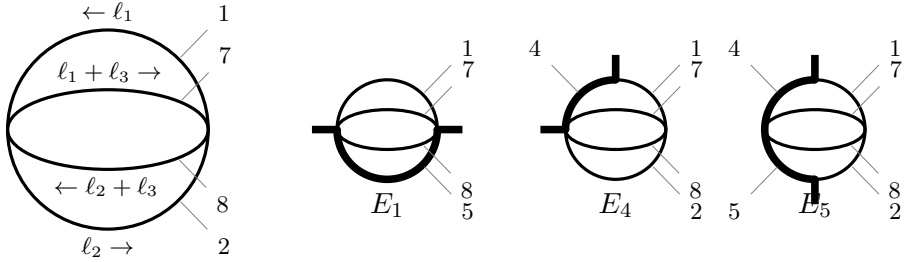

    \centering
    \hspace{-2cm}\rule[-2cm]{0cm}{2cm}
    \renewcommand{\diagramscale}{1}
    {\setlength{\bubblesize}{4\bubblesize}
    \begin{diagram}
        \renewcommand{\bubbleaspect}{1}
        \path[use as bounding box] (0,0) circle[radius=1.5\bubblesize];
        \draw (0,0)
            pic            {bubble={45,90,270,315}}
            pic[yscale=.4] {bubble={44,89,269,314}}
            ;
        \draw (node90.north)  + (0,+6pt) node {\footnotesize$\leftarrow\ell_1$};
        \draw (node270.south) + (0,-6pt) node {\footnotesize$\ell_2\rightarrow$};
        \draw (node89.north)  + (0,+5pt) node {\footnotesize$\ell_1+\ell_3\rightarrow$};
        \draw (node269.south) + (0,-5pt) node {\footnotesize$\leftarrow\ell_2+\ell_3$};
        \node[anchor=center, inner sep=0, pin=+45:{\footnotesize 1}] at (node45) {}; 
        \node[anchor=center, inner sep=0, pin=+45:{\footnotesize 7}] at (node44) {}; 
        \node[anchor=center, inner sep=0, pin=-45:{\footnotesize 8}] at (node314) {}; 
        \node[anchor=center, inner sep=0, pin=-45:{\footnotesize 2}] at (node315) {}; 
    \end{diagram}}
    \hspace{+1.5cm}
    {\setlength{\bubblesize}{2\bubblesize}\setlength{\photonlength}{1.5\bubblesize}\tikzset{q/.style={line width=1mm}}
    \begin{diagram}
        \renewcommand{\bubbleaspect}{1}
        \draw (0,0)
            pic            {bubble={45,315}}
            pic[yscale=.4] {bubble={44,314}}
            ;
        \tikzset{pion/.style={q}}
        \begin{scope}
            \clip (-\photonlength,0) rectangle (+\photonlength,-\photonlength);
            \draw (0,0)
                pic            {bubble}
                ;
        \end{scope}
        \fill[black] (-\bubblesize,0) circle[radius=.5mm];
        \fill[black] (+\bubblesize,0) circle[radius=.5mm];
        \draw[pion] (-\photonlength,0) -- (-\bubblesize,0) (+\bubblesize,0) -- (+\photonlength,0);
        \node[anchor=center, inner sep=0, pin=+45:{\footnotesize 1}] at (node45) {}; 
        \node[anchor=center, inner sep=0, pin=+45:{\footnotesize 7}] at (node44) {}; 
        \node[anchor=center, inner sep=0, pin=-45:{\footnotesize 8}] at (node314) {}; 
        \node[anchor=center, inner sep=0, pin=-45:{\footnotesize 5}] at (node315) {}; 
        \node at (0,-1.5) {$E_1$};
    \end{diagram}
    \hspace{2em}
    \begin{diagram}
        \renewcommand{\bubbleaspect}{1}
        \draw (0,0)
            pic            {bubble={45,135,315}}
            pic[yscale=.4] {bubble={44,314}}
            ;
        \tikzset{pion/.style={q}}
        \begin{scope}
            \clip (-\photonlength,0) rectangle (0,+\photonlength);
            \draw (0,0)
                pic            {bubble}
                ;
        \end{scope}
        \fill[black] (-\bubblesize,0) circle[radius=.5mm];
        \fill[black] (0,+\bubblesize) circle[radius=.5mm];
        \draw[pion] (-\photonlength,0) -- (-\bubblesize,0) (0,+\bubblesize) -- (0,+\photonlength);
        \node[anchor=center, inner sep=0, pin=+45:{\footnotesize 1}] at (node45) {}; 
        \node[anchor=center, inner sep=0, pin=+45:{\footnotesize 7}] at (node44) {}; 
        \node[anchor=center, inner sep=0, pin=-45:{\footnotesize 8}] at (node314) {}; 
        \node[anchor=center, inner sep=0, pin=-45:{\footnotesize 2}] at (node315) {}; 
        \node[anchor=center, inner sep=0, pin=135:{\footnotesize 4}] at (node135) {}; 
        \node at (0,-1.5) {$E_4$};
    \end{diagram}
    \hspace{1em}
    \begin{diagram}
        \renewcommand{\bubbleaspect}{1}
        \draw (0,0)
            pic            {bubble={45,135,225,315}}
            pic[yscale=.4] {bubble={44,314}}
            ;
        \tikzset{pion/.style={q}}
        \begin{scope}
            \clip (0,-\photonlength) rectangle (-\photonlength,+\photonlength);
            \draw (0,0)
                pic            {bubble}
                ;
        \end{scope}
        \fill[black] (0,-\bubblesize) circle[radius=.5mm];
        \fill[black] (0,+\bubblesize) circle[radius=.5mm];
        \draw[pion] (0,-\photonlength) -- (0,-\bubblesize) (0,+\bubblesize) -- (0,+\photonlength);
        \node[anchor=center, inner sep=0, pin=+45:{\footnotesize 1}] at (node45) {}; 
        \node[anchor=center, inner sep=0, pin=+45:{\footnotesize 7}] at (node44) {}; 
        \node[anchor=center, inner sep=0, pin=-45:{\footnotesize 8}] at (node314) {}; 
        \node[anchor=center, inner sep=0, pin=-45:{\footnotesize 2}] at (node315) {}; 
        \node[anchor=center, inner sep=0, pin=135:{\footnotesize 4}] at (node135) {}; 
        \node[anchor=center, inner sep=0, pin=225:{\footnotesize 5}] at (node225) {}; 
        \node at (0,-1.5) {$E_5$};
    \end{diagram}}
    \caption{From left to right: the basic topology and momentum routing common to all $E_i$ integrals, corresponding in that form to the vacuum integral $E_0$; the same with an external momentum $q$ routed through the thicker line, giving $E_1$; ditto for $E_4$; ditto for $E_5$.
    Each line is tagged with its index $i$: it carries momentum $k_i$ given in \cref{eq:basis} and occurs with power $\nu_i$ in $I_\vnu$.}
    \label{fig:elliptics}
\end{figure}

At three loops, we have 11 master integrals.
Of these, four are again factorizable into tadpole-bubble products:
\begin{equation}\label{eq:tadbub-3loop}
    \begin{alignedat}{2}
               I^{(d)}_{1,1,1,0,0,0,0,0,0}(q^2,\Mpi) &= \tadbub{30}(d)\,,\qquad&
        \Mpi^4 I^{(d)}_{1,1,1,1,1,0,0,0,0}(q^2,\Mpi) &= \tadbub{12}(d;t)\,,\\
        \Mpi^2 I^{(d)}_{1,1,1,1,0,0,0,0,0}(q^2,\Mpi) &= \tadbub{21}(d;t)\,,\qquad&
        \Mpi^6 I^{(d)}_{1,1,1,1,1,1,0,0,0}(q^2,\Mpi) &= \tadbub{03}(d;t)\,.
    \end{alignedat}
\end{equation}
The remaining ones are highly nontrivial, and we name them as follows:%
\footnote{
    For comparison, \rcite{Bijnens:2017wba} has two such non-factorizable three-loop master integrals (plus one at two loops, namely the sunset $I_{1,0,0,1,1}$).
    In our notation, they are $E_0$ and $I_{1,1,0,0,0,1,1,1,0}$, 
    the latter of which does not appear in our set.
    Both are known from earlier gauge theory calculations \cite{Melnikov:2000zc}.}
\begin{equation}\label{eq:E}
    \begin{alignedat}{2}
        \Mpi^2 I^{(d)}_{1,1,0,0,0,0,1,1,0}(q^2,\Mpi) &\eqcolon E_0(d)   \,,\qquad&
        \Mpi^4 I^{(d)}_{1,1,0,1,0,0,1,1,0}(q^2,\Mpi) &\eqcolon E_4(d;t) \,,\\
        \Mpi^2 I^{(d)}_{1,0,0,0,1,0,1,1,0}(q^2,\Mpi) &\eqcolon E_1(d;t) \,,\qquad&
        \Mpi^6 I^{(d)}_{1,1,0,1,1,0,1,1,0}(q^2,\Mpi) &\eqcolon E_5(d;t) \,,\\
        \Mpi^4 I^{(d)}_{2,0,0,0,1,0,1,1,0}(q^2,\Mpi) &\eqcolon E_2(d;t) \,,\qquad&
        \Mpi^8 I^{(d)}_{2,1,0,1,1,0,1,1,0}(q^2,\Mpi) &\eqcolon E_6(d;t) \,,\\
        \Mpi^6 I^{(d)}_{3,0,0,0,1,0,1,1,0}(q^2,\Mpi) &\eqcolon E_3(d;t) \,,
    \end{alignedat}
\end{equation}
where ``$E$'' refers to the fact that their evaluation involves elliptic integrals (see \cref{app:expr-2d}).
The vacuum integral $E_0(d)$ is simply $E_1(d;0)$, and $E_2$ and $E_3$ can be related to the first and second $t$-derivatives of $E_1$ (see \cref{app:diffeq}); likewise, $E_6$ is related to the derivative of $E_5$.
The general layout of the integrals is illustrated in \cref{fig:elliptics}.

All $E_i$ are divergent in four dimensions but finite in two, so we employ Tarasov's dimensional shift to extract their divergences.
Going forward, we use the following notation for the master integrals as series in $\epsilon$:
\begin{equation}\label{eq:eps-series}
    E_i(4-2\epsilon;t) = \sum_{r=-3}^{\infty} \epsilon^r E_i^{(r)}(4;t)\,,\qquad
    E_i(2-2\epsilon;t) = \sum_{r=0}^{\infty} \epsilon^r E_i^{(r)}(2;t)\,.
\end{equation}
Since the integrals are finite in two dimensions, we have $E_i(2;t)=E_i^{(0)}(2;t)$, whereas we refer to the dimensionless finite part in four dimensions as $\bar E_i(4;t)\coloneq \Mpi^{-6}E_i^{(0)}(4;t)$.

\section{Loop integral details}\label{sec:more-loops}

We apply various techniques to reduce the number of loop integrals and to compute them, some rather standard and others less so.
While the explicit expressions resulting from these techniques are lengthy enough that we defer them to the appendices, this section provides a thorough overview.
We use the notation of \cref{sec:loops} throughout, and abbreviate \cref{eq:Inu-D} as $I_\vnu=\int J_\vnu$.

The common theme of these techniques is the application of differential operators to the integrands (see, e.g., \rrcite{Smirnov:2012gma,Travaglini:2022uwo,Weinzierl:2022eaz} for expository texts).
Perhaps the most familiar example is integration by parts (IBP), namely the identity
\begin{equation}
    0 = \int \frac{\p}{\p\ell_i^\mu} \big(r^\mu J_\vnu\big)\,,
\end{equation}
where $\ell_i$ is any of the loop momenta and $r$ is any momentum, loop or external.
Evaluating the derivative gives a linear combination of integrands $J_{\vnu'}$ for various $\vnu'$, so each choice of $(\vnu,\ell_i,r)$ gives a linear combination of integrals $I_{\vnu'}$ that is known to vanish.
The resulting system of equations can be solved systematically using the Laporta algorithm~\cite{Laporta:2000dsw}, which is implemented in programs such as \texttt{LiteRed}~\literedrefs\ with various heuristic improvements. 
Thus, any Feynman integral $I_\vnu$, $\vnu \in \mathbb Z^{\NK}$ can be expressed in terms of a \emph{finite} basis of master integrals \cite{Smirnov:2010hn,Lee:2013hzt,Bitoun:2017nre}.

\subsection{Differential equations}\label{sec:diffeq}
A useful property of derivatives with respect to masses and momenta is that they can only change the power to which an existing propagator appears in the integrand, but never make a new propagator appear.
This produces a hierarchy among the master integrals based on the propagators they contain, which in our three-loop case looks schematically like
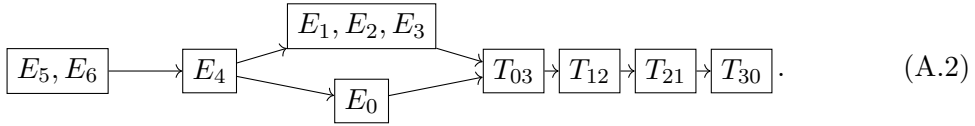
\begin{equation}\label{eq:hierarchy}
    \tikzineq{
        \tikzset{node distance=2cm}
        \node[draw,rectangle] (E5)               {$E_5,E_6$};
        \node[draw,rectangle] (E4) [right of=E5] {$E_4$};
        \node            (EX) [right of=E4] {};
        \node[draw,rectangle] (E1) at ($(EX) + (0,+.5)$) {$E_1,E_2,E_3$};
        \node[draw,rectangle] (E0) at ($(EX) + (0,-.5)$) {$E_0$};
        \node[draw,rectangle] (T0) [right of=EX] {$T_{03}$};
        \tikzset{node distance=1cm}
        \node[draw,rectangle] (T1) [right of=T0] {$T_{12}$};
        \node[draw,rectangle] (T2) [right of=T1] {$T_{21}$};
        \node[draw,rectangle] (T3) [right of=T2] {$T_{30}$};
        \draw[->] (E5) -- (E4);
        \draw[->] (E4) -- (E1);
        \draw[->] (E4) -- (E0);
        \draw[->] (E1) -- (T0);
        \draw[->] (E0) -- (T0);
        \draw[->] (T0) -- (T1);
        \draw[->] (T1) -- (T2);
        \draw[->] (T2) -- (T3);
        }\,.
\end{equation}
Here, 
\tikz[baseline=-0.75ex]{
    \node[draw,rectangle] (A) {$A$}; 
    \node[draw,rectangle] (B) [right of=A] {$B$};
    \draw[->] (A) -- (B)}
means that the set of propagators in $B$ is a subset of those of $A$, and we say that $B$ is lower in the hierarchy.
Applying a differential operator to a master integral gives, after reduction, a linear combination of master integrals at the same level or lower in the hierarchy as the original integral, never higher.

A salient case is $\p/\p t$, or in general the derivative with respect to any kinematical invariant.
Say that $\{\mathscr M_i\}_{i=1}^n$ are the master integrals living on some level in the hierarchy.
Then, schematically,
\begin{equation}\label{eq:deriv}
    \frac{\p}{\p t} \mathscr M_i(d;t) = \sum_{j=1}^n \rho_{ij}(d;t) \mathscr M_j(d;t) + \mathscr S_i(d;t)\,,
\end{equation}
where $\rho_{ij}$ are rational functions and $\mathscr S_i$ contains only integrals lower in the hierarchy.
Taking further derivatives produces the same general structure as \cref{eq:deriv}, yielding a system of relations that allow the other $\mathscr M_j$, $j\neq i$ to be traded for derivatives of $\mathscr M_i$.
Solving the system to eliminate the other $\mathscr M_j$ gives the inhomogeneous differential equation
\begin{equation}\label{eq:diffeq}
    \sum_{j=0}^{n} \pi_{ij}(d;t)\bigg[\frac{\p}{\p t}\bigg]^j \mathscr M_i(d;t) = \mathscr S'_i(d;t)\,,
\end{equation}
where $\pi_{ij}(d;t)$ are polynomials, and the inhomogeneous part $\mathscr S'_i$ consists of integrals lower in the hierarchy.
Solving such a differential equation (with the simpler integral obtained at $t=0$ as initial condition) is a viable alternative to direct integration [see, e.g., \rcite{Bloch:2014qca,Broedel:2019kmn,Pogel:2022yat}, which yield $E_1(2;t)$].
We list these equations (with $n=3$ for $E_1,E_2,E_3$, $n=2$ for $E_5,E_6$, and $n=1$ for $E_4$) in \cref{app:diffeq}.

The method outlined here is a simple way of obtaining differential equations satisfied by the master integrals in general dimensions~\cite{Primo:2017ipr}. In fixed dimensions one needs to resort to the Picard-Fuchs algorithm~\cite{Lairez:2022zkj,delaCruz:2024xit}. 
This algorithm also provides a direct way to derive the differential equations satisfied by a given Feynman integral, without requiring an explicit master integral reduction.

\subsection{Tarasov's dimensional shift}\label{sec:dimshift}
A more straightforward differential operator is $\p/\p m^2_i$, if we treat all propagators as having distinct masses; the equal-mass limit can be taken after it has been applied.
It acts as
\begin{equation}\label{eq:massderiv}
    \frac{\p}{\p m^2_i} \frac{1}{D_j^{\nu}} = \frac{\nu \delta_{ij}}{D_j^{\nu+1}}
    \quad\Rightarrow\quad
    \frac{\p}{\p m^2_i} I_\vnu = \nu_i I_{\vnu+\hat\imath},
\end{equation}
where the $i$th component of $\hat\imath$ is $1$ and the others $0$.

Following a trick due to Tarasov~\cite{Tarasov:1996br}, this operator is utilized to formally change the dimension of an integral in steps of $2$.
The formula is best introduced by first bringing the integrals to parametric form 
by using the identity
\begin{equation}
    \Gamma(\nu) \coloneq \int_0^\infty\d\alpha\, \alpha^{\nu - 1} e^{-\alpha} = D^\nu \int_0^\infty\d\alpha\, \alpha^{\nu-1} e^{-\alpha D}\,,
\end{equation}
where in the second equality we shifted $\alpha\to\alpha D$.%
\footnote{
    Note that this assumes $D>0$, which is only guaranteed in Euclidean space.
    However, since Wick rotation corresponds to a simple multiplicative factor for the integrals, the same identities hold also for the Minkowski-space integrals.}
This gives
\begin{equation}
    I_\vnu^{(d)} =
        \letterint{S}_\vnu
        \intx\frac{\d^d \ell_1}{\pi^{d/2}}\cdots
        \intx\frac{\d^d \ell_\NL}{\pi^{d/2}}
        \exp\bigg[-\sum_{j=1}^\NK \alpha_j D_j\bigg]\,,
\end{equation}
where for brevity, we have defined the ``Schwinger integral''
\begin{equation}
    \letterint{S}_\vnu \coloneq
        \int_0^\infty\frac{\d\alpha_1\,\alpha_1^{\nu_1-1}}{\Gamma(\nu_1)}
        \cdots
        \int_0^\infty\frac{\d\alpha_\NK\,\alpha_\NK^{\nu_\NK-1}}{\Gamma(\nu_\NK)}
        \,.
\end{equation}
Let us write [recall \cref{eq:Dj-def}]
\begin{equation}\label{eq:quadratic-form}
    \sum_{j=1}^\NK \alpha_j D_j = \sum_{i,j=1}^\NL A_{ij}[\ell_i \cdot \ell_j] + 2\sum_{i=1}^\NL B_i\cdot\ell_i + C\,,
\end{equation}
where $B$ and $C$ depend on the momenta $p_j$, and $C$ also on the masses in the form $-\sum_j \alpha_j m_j^2$, whereas $A$ is a constant matrix.
By shifting $\ell_{i=1}^\NL \to \ell_i - \sum_{j=1}^\NL A_{ij} B_j$ to complete the square in the exponential, one obtains a Gaussian integral that can be easily performed, leaving
\begin{equation}\label{eq:Inu-FU}
    I_\vnu^{(d)} =
        \frac{1}{\pi^{\NL d/2}} \letterint{S}_\vnu \big[\det(\pi A^\inv)\big]^{d/2}\exp\bigg[\sum_{i,j=1}^\NL A^\inv_{ij} B_i B_j - C\bigg]
        = \letterint{S}_\vnu
            \frac{e^{-\F/\U + i\epsilon}}{\U^{d/2}}\,,
\end{equation}
where $\U\coloneq\det(A)$ and $\F\coloneq\U\big(\sum_{ij} A^\inv_{ij}B_iB_j-C\big)$ are the \emph{Symanzik polynomials}; they are polynomials in $\{\alpha_j\}$ and are homogeneous of degree $\NL $ and $\NL +1$, respectively.\footnote{%
    We use the sign convention of \texttt{LiteRed}~\literedrefs\ for the second Symanzik polynomial, which differs by a sign from the definition used in~\rcite{Bloch:2014qca}.}

Now, note that $m_i^2$ only appears as the term $\alpha_i m_i^2$ inside the exponent $\F/\U$, so the effect of $\p/\p m_i^2$ on the right-hand side of \cref{eq:Inu-FU} is to bring down a factor of $\alpha_i$ inside the integrand.
Next, note that $d$ only appears as the exponent in $\U^{d/2}$, and recall that $\U$ is just a sum of products of $\alpha_i$'s; explicitly, let us write it as $\U \eqcolon U(\alpha_1,\alpha_2,\ldots,\alpha_\NK)$.
Then, maintaining that same functional form, it follows that
\begin{equation}
    \D\coloneq U\Big(\tfrac{\p}{\p m_1}, \tfrac{\p}{\p m_2}, \ldots, \tfrac{\p}{\p m_\NK}\Big)
\end{equation}
is a differential operator such that
\begin{equation}\label{eq:dimshift}
 \D I_\vnu^{(d)}
        =
        \letterint{S}
        \frac{U(\alpha_1,\alpha_2,\ldots,\alpha_\NK) e^{-\F/\U + i\epsilon}}{\U^{d/2}}
        =
        \letterint{S}
        \frac{e^{-\F/\U + i\epsilon}}{\U^{d/2-1}}
        =
        I_\vnu^{(d-2)} \,.
\end{equation}
On the other hand, $\D I_\vnu^{(d)}$ is a linear combination of the $d$-dimensional master integrals.
This produces a linear system that can be inverted to give the $d$-dimensional master integrals in terms of the less divergent $(d-2)$-dimensional ones.
We present the relations for our case in \cref{app:dimshift}.

A particularly elegant way of deriving $\D$ is the well-known (see \rrcite{Nakanishi:1971,Weinzierl:2022eaz}) graph-based determination of $\U$.
Consider the graph $G$ representing a given integral (as in \cref{fig:elliptics}) and consider all ways of removing $\NL$ edges from the graph in such a way that the remaining edges form a connected tree graph that joins all vertices, which is called a \emph{spanning tree}.
Then
\begin{equation}\label{eq:U}
    \U =
        \sum_{\substack{\text{spanning}\\\text{trees }T}}
        \quad
        \prod_{j\in G\setminus T} \alpha_j\,,
\end{equation}
where $G\setminus T$ is the set of $\NL$ edges that were removed to form the spanning tree $T$.
As above, $\D$ is obtained by replacing $\alpha_j\to\p/\p m_j^2$ in this formula.
\Cref{app:dimshift} details the application of this method to our master integrals.

\subsection{Schouten relations}\label{sec:schouten}

An important thing to note is that the set of master integrals is only minimal and irreducible with respect to IBP relations.
When working in fixed dimension or with fixed kinematics, additional relations may emerge that are not taken into account during master integral reduction.
A simple example is how $I_\tad$ (or $E_0$) is trivially related to $I_\bub$ (or $E_1$) by fixing $t=0$, even though it is formally an independent master integral.
However, the present amplitude involves less obvious relations, as can be inferred from renormalizability.
Taking the sum of diagrams 91--101 in \cref{tab:NNNLO} at face value, its divergent part contains the integrals $E_n(2;t)$,%
\footnote{
    The reason for this can be seen in \cref{app:dimshift}, specifically \cref{eq:dimshift-E5,eq:dimshift-E6}, where the coefficients relating $E_{5,6}(4-2\epsilon;t)$ to $E_n(2-2\epsilon;t)$ diverge, introducing the finite integrals $E_n(2;t)$ also into the divergent part.
    This is not the case for $E_0$ through $E_4$.}
but since these are elliptic functions of $t$, no combination of rational functions and polylogarithms (which is what diagrams 00--90 provide) can cancel them.
Thus, there must be some additional relation between the elliptic master integrals that makes this problematic divergence vanish.

Such subtle and novel relations indeed hold for $E_5$ and $E_6$, and can be viewed as consequences of Schouten%
\footnote{
    Pronounced \smash{[\textipa{sxA\textsubarch{U}t@n}]}.
    We make this point because difficult and oft-mangled names is a matter close to the heart of the author whose surname is pronounced \smash{[\textipa{\r\textturnmrleg\o:}]}.}
identities; by analogy, we refer to these master integral relations as \emph{Schouten relations}.
Schouten identities hold only at fixed integer dimension, and while the relations can be extended to the $\epsilon$ expansion around that dimension and translated to other dimensions through dimension shifting, they are not detected by the IBP relations, which operate in general $d$.%
\footnote{
    We thank Roman Lee and Lorenzo Tancredi for discussions regarding this point.}
The role of Schouten identities in reducing the number of independent master integrals was first noticed by Remiddi \&~Tancredi~\cite{Remiddi:2013joa} in the context of two-loop sunset integrals with three distinct masses, and is discussed further in~\rcite{Tancredi:2015pta}.
We have checked all relations numerically with \texttt{pySecDec}~\psdrefs.

In the remainder of this section, we will discuss Schouten relations from three angles, ranging from the more profoundly connected to its underlying cause, to the more practically applicable to the case at hand.

A Schouten identity is the statement that the Gram determinant of $n$ vectors vanishes in $d<n$ dimensions, since no more than $d$ vectors may be linearly independent.
In two dimensions, for three arbitrary vectors $u$, $v$ and $w$, this takes the form
\begin{equation}
    \Gram(u,v,w) \coloneq
    \begin{pmatrix}
        u^2        &   u\cdot v   &   u\cdot w    \\
        v\cdot u   &   v^2        &   v\cdot w    \\
        w\cdot u   &   w\cdot v   &   w^2         \\
    \end{pmatrix}
    \quad\Rightarrow\quad |\Gram(u,v,w)| = 0 \qquad \text{at $d=2$.}
\end{equation}
With $u,v,w$ drawn from the loop and external momenta, $|\Gram(u,v,w)|$ is a cubic polynomial in the $D_j$ of~\cref{eq:Dj-def} which vanishes at $d=2$.
Inserting that as the numerator of a suitably chosen integral and reducing the result yields a $(d=2)$-specific linear relation among the master integrals---a Schouten relation.

However, the above argument is only valid if the integral of interest is finite---the requirement that $d=2$ precludes dimensional regularization---and in most cases where a finite integral is obtained, the resulting linear combination is zero in any dimension, yielding no new information.
However, a finite yet nontrivial result in our basis of propagators is obtained from  $I^{(d)}_{1,1,0,1,1,0,2,1,0}$ and $|\Gram(\ell_1,\ell_2,\ell_3)|$: the integral
\begin{equation}
    G(d;t)\coloneq\int \frac{\d^d \ell_1 \d^d \ell_2 \d^d \ell_3}{\pi^{3d/2}} 
        \frac{|\Gram(\ell_1,\ell_2,\ell_3)|}{D_1D_2\, D_4D_5\, D_7^2 D_8}
\end{equation}
is finite at $d=2$ and reduces (using \texttt{LiteRed}~\literedrefs) to
\begin{multline}\label{eq:schouten-0}
    G(2;t) = (t+8)(t-4) E_1(2;t)
    - (t^3-20t^2-128(t-1)) E_2(2;t)
    - 2t(t-16)(t+4) E_3(2;t)
    \\- 12t(t+4) E_4(2;t)
    + 96(t-1)t E_5(2;t)
    - 48t^2(t-4) E_6(2;t)
    \\+ \frac{24t}{(t-4)^2}\,\big[2+\Jbub1(t)\big]\big[t^2 + 8t - 16 + 4t\Jbub1(t)\big]\,.
\end{multline}
On the other hand, $G(2;t)=0$ due to the presence of $|\Gram(\ell_1,\ell_2,\ell_3)|$.
Using this removes all elliptic functions from the $1/\epsilon^2$ pole of the 3-loop amplitude, which allows it to be canceled by the counterterms.
However, the $1/\epsilon$ pole contains $E_i^{(1)}(2;t)$ [recall \cref{eq:eps-series}] and is therefore difficult to tame with similar integer-dimension methods.

A more systematic way of finding Schouten identities at any order in $\epsilon$ comes from the differential equation for $E_5$.
In two dimensions, the second-order differential operator factorizes into two first-order ones; the explicit expressions are given in \cref{eq:diffeq-E5}, but the schematic form is
\begin{equation}\label{eq:LR}
    \Delta_L \Delta_R E_5(2;t) = \mathscr S_5\,,
\end{equation}
where $\Delta_{L,R}$ are linear differential operators and $\mathscr S_5$ consists of integrals lower in the hierarchy of \cref{eq:hierarchy}.
In fact, the same holds perturbatively in $\epsilon$:
\begin{equation}\label{eq:LR-r}
    \Delta_L \Delta_R E_5^{(r)}(2;t) = \mathscr S_5^{(r)}\,,
\end{equation}
where $r$ denotes the order in $\epsilon$ as given in \cref{eq:eps-series} and $\mathscr S_5^{(r)}$ consists of integrals lower in the hierarchy and of $E_5^{(r')}(2;t)$ and $E_6^{(r')}(2;t)$ with $r'<r$.

Now, given $\Sigma^{(r)}$ such that $\Delta_L\Sigma^{(r)} = \mathscr S_5^{(r)}$, we clearly get
\begin{equation}\label{eq:Sigma}
    \Delta_R E_5^{(r)}(2;t) = \Sigma^{(r)}\,,
\end{equation}
which is a linear differential equation; in fact, the $r=0$ version is the very same equation as that obtained through Griffiths-Dwork reduction~\cite{Lairez:2022zkj,delaCruz:2024xit}.
It can be contrasted with the plain derivative [explicitly, \cref{eq:E6toE5}]
\begin{equation}\label{eq:Sigma'}
    \frac{\p}{\p t} E_5^{(r)}(2;t) = \rho(t) E_6^{(r)}(2;t) + \tilde\Sigma^{(r)}\,,
\end{equation}
where $\rho$ is a rational function and $\tilde\Sigma^{(r)}$, like $\Sigma^{(r)}$, is lower in the hierarchy or lower in the $\epsilon$ expansion.
Canceling $\frac{\p}{\p t} E_5^{(r)}(2;t)$ between \cref{eq:Sigma,eq:Sigma'} gives a linear relation that for $r=0$ is equivalent to \cref{eq:schouten-0}.
The lengthier $r=1$ version is, after eliminating $E_6(2;t)$ with \cref{eq:schouten-0},
\begin{multline}\label{eq:schouten-1}
    0 =   \frac{2}{9t}E_0(2)
        + \frac{(t-4)(5t+64)}{144t}E_1(2;t)
        + \frac{(t-4)(t+8)}{144t}E_1^{(1)}(2;t)\\
        + \frac{t^3-20t^2+128(2t-4)}{144t} E_2(2;t)
        - \frac{t^3-20t^2+128(t-1)}{144t} E_2^{(1)}(2;t)\\
        + \frac{(t+4)(t-16)}{36} \Big(E_3(2;t) - \tfrac12E_3^{(1)}(2;t)\Big)
        + \frac{t-28}{12}E_4(2;t)
        - \frac{t+4}{12}E_4^{(1)}(2;t)\\
        - \frac{3t^2 - 36t + 16}{18}E_5(2;t)
        + \frac{2(t-1)}{3}E_5^{(1)}(2;t)
        - \frac{3}{t(t-4)}E_6^{(1)}(2;t)
        - \frac{2t}{(t-4)}\\
        - \frac{(t^2+16t-16)\Jbub2(t)
              (t^2+22t-72)\big[\Jbub1(t)\big]^2
            + 4t\Jbub1(t)\big[3t-4+2\Jbub2(t)\big]}{6(t-4)^2}  
\end{multline}
and eliminates the $1/\epsilon$ pole in the same way that \cref{eq:schouten-0} eliminated the $1/\epsilon^2$ one.
The procedure can in principle be carried out to all orders in $\epsilon$, but the challenge is the highly nontrivial step from $\mathscr S^{(r)}$ to $\Sigma^{(r)}$.
Nonetheless, this approach is interesting enough to warrant further study, since it can be applied to any master integral whose differential equation factorizes.

Ultimately, the most straightforward way of obtaining Schouten relations is to compare different ways of performing the $\epsilon$ expansion.
This obscures the origin of the relations and only works for the divergent part, but is sufficient for our purposes.
Thus, we examine the expansions
\begin{equation}\label{eq:expand-E5E6}
    \begin{aligned}
    E_5(4-2\epsilon;t) 
        &=\frac{E_5^{(-3)}(4;t)}{\epsilon^3}
        + \frac{E_5^{(-2)}(4;t)}{\epsilon^2}
        + \frac{E_5^{(-1)}(4;t)}{\epsilon^1}
        + \O(\epsilon^0)\,,\\
    E_6(4-2\epsilon;t) 
        &=\frac{E_6^{(-2)}(4;t)}{\epsilon^2}
        + \frac{E_6^{(-1)}(4;t)}{\epsilon^1}
        + \O(\epsilon^0)\,,\\
    \end{aligned}
\end{equation}
where the values of $E_{5,6}^{(-r)}(4;t)$ can be found by setting $d=4-2\epsilon$ in \cref{eq:dimshift-E5,eq:dimshift-E6}; this way, all but the $1/\epsilon^3$ divergences contain elliptic master integrals.
However, they can also be found by plugging \cref{eq:expand-E5E6} into the differential equations for $E_5$ in \cref{eq:diffeq-E5} and $E_6$ in \cref{eq:E6toE5}, where much of the complexity diverts to the $\O(\epsilon^0)$ terms.
Inserting the $\epsilon$ expansion for the lower-hierarchy integrals $T_{21}$, $T_{12}$, $E_0$, $E_1$, $E_2$ and $E_3$ near four dimensions given in \cref{app:tadbub-4d,app:E0-4d,app:E1E2E3-4d,app:E4-4d}, we deduce that the coefficients of the poles of $E_5$ and $E_6$ can be written entirely in terms of $\Jbub{n}(t)$, as stated in \cref{eq:E5-4d,eq:E6-finite}.
This is equivalent to \cref{eq:schouten-0,eq:schouten-1}.

The Schouten relations make it natural to express the amplitude in terms of $\bar E_5(4;t)$ and $\bar E_6(4;t)$, since in two dimensions they eliminate $E_6(2;t)$ and $E_6^{(1)}(2;t)$ but leave more problematic functions like $E_6^{(2)}(2;t)$.
Using the four-dimensional integrals becomes even more natural in the entirely four-dimensional approach of \cref{eq:expand-E5E6}.
On the other hand, the integrals that are not the subject of Schouten relations---$E_0$ through $E_4$---are more naturally used in their finite two-dimensional form.
This mixed two- and four-dimensional representation has the further benefit that $E_4$ completely drops out of the expressions, as shown in \cref{app:dimshift-E4}.

\section{Results}\label{sec:results}

In this section, we present our main result, namely $\Pi_T(q^2)$ [i.e., the transverse component of $\Pi^{\mu\nu}(q^2)$ as defined in \cref{eq:Pi-def,eq:LT-def}], computed to \NNNLO\ as outlined in previous sections.
Besides being a function of the pion mass $\Mpi\approx135~\mathrm{MeV}$, the pion decay constant $\Fpi\approx92.2~\mathrm{MeV}$~\cite{Bijnens:2014lea} and the photon momentum $q^\mu$ through $t\coloneq q^2/\Mpi^2$, it also contains the renormalized low-energy constants $l_i^q$ and $c_i^q$, as well as $r_i^q$, which are defined below in \cref{sec:renorm-N3LO}.
These are all functions of the renormalization scale $\mu$ in the customary \chpt\ renormalization scheme, while $\Pi_T$ itself is kept scale-independent through the appearance of $\Lpi\coloneq\log\frac{\Mpi^2}{\mu^2}$, which in turn emerges from the mass and decay constant renormalization, and from reconciling the factors of $C(\epsilon,\mu^2)$ in the LECs with those of $C(\epsilon,\Mpi^2)$ in the loop integrals (see \cref{sec:loop-manip}).
The constant $\zeta(3)=1.202\ldots$ also appears.

The $t$-dependence of the result is expressed using the functions $\Jbub{n}(t)$ defined in \cref{eq:Jbub} and evaluated in \cref{app:bubble}; $E_1(2;t)$, $E_2(2;t)$ and $E_3(2;t)$ defined in \cref{eq:E}, shifted to $2$ dimensions using the method described in \cref{sec:dimshift} and evaluated in \cref{app:expr-2d}; and $\bar E_5(4;t)$ and $\bar E_6(4;t)$ defined in \cref{eq:eps-series} and evaluated in \cref{app:expr-4d}.
The reason for using a mix of $2$- and $4$-dimensional integrals was explained at the end of \cref{sec:schouten}.

We have subjected the result to the following checks:
\begin{itemize}
    \item The longitudinal part $\Pi_L(q^2)$ vanishes, as required by the Ward--Takahashi identity;
    \item The renormalization removes all non-local divergences, and the local ones are of a form that allows them to be canceled by the \NNNLO\ counterterms (see \cref{sec:renorm-N3LO});
    \item The result is invariant under reparametrization of the Nambu--Goldstone manifold, e.g., replacing \cref{eq:param} by $U=e^{i\Phi/F_0\sqrt2}$, or more generally following \rcite[app.~B]{Bijnens:2022zsq};
    \item There are no singularities at $t=0$ despite the presence of factors of $\frac1t$;%
    \footnote{
        That such factors appear can be viewed as a consequence of our choice of master integrals (and $\Jbub{n}$).
        While a choice that absorbs them would certainly be possible---compare the use of $\overline B_{21}$ rather than the scalar bubble integral in \rcite{Golowich:1995kd}---we have found it simpler to retain the ``natural'' master integrals, at the cost of having to take the $t\to0$ limit more carefully.}
    \item The above checks also hold when the calculation is repeated at $3$ or $N$ mass-degenerate flavors, to ensure that no errors have been made whose effects just happen to vanish in the $2$-flavor case;
    \item Restricting the calculation to NLO or NNLO reproduces earlier results \cite{Gasser:1983yg,Gasser:1984gg,Golowich:1995kd,Amoros:1999dp,Bijnens:2011xt}; see \cref{eq:PiT-NLO,eq:PiT-NNLO}.
\end{itemize}

\subsection{Expressions for the HVP}
As discussed above, the longitudinal component, $\Pi_L(q^2)$, vanishes identically, so we only consider the transverse component $\Pi_T(q^2)$, which we renormalize on shell, as done in most physical applications. Thus, we present results for $\bar\Pi_T(q^2) \coloneq \Pi_T(q^2) - \Pi_T(0)$, 
deferring discussion of the unsubtracted quantity $\Pi_T(q^2)$ to \cref{app:unsub-results}.
We expand the result in powers of $\xi$, 
\begin{equation}\label{eq:main}
    \frac{1}{\pi_{16}}\bar\Pi_T(q^2) = \bar\Pi_T^\NLO(t) + \xi\bar\Pi_T^\NNLO(t) + \xi^2\bar\Pi_T^\NNNLO(t) + \O(\xi^3)\,,
\end{equation}
where, as in \cref{sec:massdecay}, $\xi\coloneq\pi_{16}\Mpi^2/\Fpi^2\approx0.014$ and $\pi_{16}\coloneq\frac{1}{16\pi^2}$.
Note that on the right-hand side, we deliberately write the $\bar\Pi$'s as dimensionless functions of $t$.

The first two terms in \cref{eq:main} are known from earlier work~\cite{Gasser:1983yg,Gasser:1984gg,Golowich:1995kd,Amoros:1999dp,Bijnens:2011xt}, but we restate them here in our notation:%
    \footnote{
        Refs.~\cite{Golowich:1995kd,Amoros:1999dp,Bijnens:2011xt} express the non-polynomial $t$-dependence in terms of the functions $\overline B_{21}$ and $\overline B_{22}$, which derive from the finite part of the rank-$2$ tensor bubble integral.
        Here, ``finite part'' is defined by writing the divergence in terms of $\lambda_0 = \big[1 + \tfrac1\epsilon\big]C(\epsilon,\Mpi^2) + \O(\epsilon)$.
        The relation to our conventions is thus
        \begin{equation*}
            \overline B_{21}(\Mpi^2,\Mpi^2,q^2) = \pi_{16}\Big(\tfrac{1 - t}{3t}\Jbub1(t) + \tfrac13 \Lpi - \tfrac{5}{18}\Big)\,,\qquad
            \overline B_{22}(\Mpi^2,\Mpi^2,q^2) = \pi_{16}\Big(\tfrac{t - 4}{12}\Jbub1(t) + \tfrac{t-6}{12} \Lpi + \tfrac{t}{36}\Big)\,.
        \end{equation*}}
\begin{align}
    \label{eq:PiT-NLO}
    \bar\Pi_T^\NLO(t) &= 
        2\frac{4-t}{3t}\Jbub1(t) + \frac{4}{9}\,,\\
    \label{eq:PiT-NNLO}
    \bar\Pi_T^\NNLO(t) &= 
        t\bigg[\frac{4-t}{3t}\Jbub1(t) 
            - \frac{1+3\Lpi}{9}\bigg]^2
        - 4t\bigg[
            \frac{4-t}{3t}\Jbub1(t)
            - \frac{1+3\Lpi}{9}
            \bigg] l_6^q
        - 8t c_{56}^q\,.
\end{align}
The novel third term is much lengthier, so for the sake of presentation, we partition it thematically,%
\footnote{
    While there is significant mixing due to master integral reduction and renormalization, $\bar\Pi_c$ can be said to mainly stem from diagrams 01--04 in \cref{tab:NNNLO}, $\bar\Pi_l$ from 10--53, $\bar\Pi_J$ from 54--90, and $\bar\Pi_E$ from 91--101.}
\begin{equation}
    \label{eq:PiT-N3LO}
    \bar\Pi_T^\NNNLO(t) = \bar\Pi_E(t) + \bar\Pi_J(t) + \bar\Pi_\zeta(t) + \bar\Pi_l(t) + \bar\Pi_c(t) + t r_1^q + t^2 r_2^q\,,
\end{equation}
where the \NNNLO\ LECs $r_i^q$ come from the contact diagram 00 in \cref{tab:NNNLO} and are discussed further in \cref{sec:renorm-N3LO}.
The part containing the elliptic master integrals $E_i$ (plus additional terms to ensure that it vanishes as $t\to0$) is
\begin{align}
    \bar\Pi_E(t)
        &= \bigg[\frac{107 t^3}{46656} - \frac{233 t^2}{3888} - \frac{47 t}{243} + \frac{17327}{1458} - \frac{20813}{486 t} + \frac{1952}{81 t^2}\bigg] E_1(2;t) \notag\\
        &\,- \bigg[\frac{143 t^4}{46656} - \frac{257 t^3}{1944} + \frac{2209 t^2}{1944} + \frac{32287 t}{2916} - \frac{32126}{243} + \frac{72214}{243 t} - \frac{7808}{81 t^2}\bigg] E_2(2;t)\notag\\
        &\,- \bigg[\frac{143 t^4}{23328} - \frac{166 t^3}{729} + \frac{316 t^2}{243} + \frac{29413 t}{1458} - \frac{99722}{729} + \frac{5216}{27 t}\bigg] E_3(2;t) \notag\\
        &\,- \bigg[\frac{t}{3} - \frac{13}{9} - \frac{2}{3t}\bigg] \bar E_5(4;t)
        + \bigg[\frac{t^3}{54} - \frac{19 t^2}{27} + \frac{143 t}{27} - \frac{100}{9}\bigg] \bar E_6(4;t)\notag\\
        &+ \bigg[\frac{1216}{27}+\frac{2 \zeta(3)}{9}-\frac{\pi^2}{18}\bigg]\frac{1}{t}
        - \bigg[\frac{26885}{972}-\frac{2131 \zeta(3)}{108}+\frac{91 \pi^2}{108}\bigg]\,,
    \label{eq:PiE}
\end{align}
the part containing the bubble functions $\Jbub{n}$ but no LECs (likewise with an additional term to ensure that it vanishes as $t\to0$) is
\begin{align}
    \bar\Pi_J(t)
        &= -\bigg[\frac{t^2}{648} + \frac{t}{324} - \frac{1}{108} - \frac{1}{9t}\bigg]\Jbub3(t)
        - \bigg[\frac{t^2}{54} - \frac{8t}{27} + \frac{4}{3} - \frac{2}{3t}\bigg]\Jbub2(t)\,\Jbub1(t)\notag\\
        &- \bigg[\frac{t^2}{54} - \frac{2t}{9} + \frac{8}{9} - \frac{32}{27t}\bigg]\big[\Jbub1(t)\big]^3
        - \bigg[\frac{5 t^2}{216} - \frac{103 t}{108} + \frac{151}{36} + \frac{1}{3 t}\bigg]\Jbub2(t)\notag\\
        &- \bigg[\frac{(1+\Lpi)t^2}{27} - \frac{(63+4\Lpi)t}{54} + \frac{275-64\Lpi}{54} + \frac{10 + 96\Lpi}{27t}\bigg]\big[\Jbub1(t)]^2\notag\\
        &+ \bigg[\frac{(97 -  9\pi^2)t^2}{3888} - \frac{(2285 + 9\pi^2)t}{1944} + \frac{407}{81}+\frac{\pi^2}{72} - \frac{4-\pi^2}{6t}\bigg] \Jbub1(t) \notag\\
        &- \bigg[\frac{t^2 \Lpi}{27} + \frac{(23 + 8\Lpi)t}{27} - \frac{92 + 48\Lpi}{27}\bigg]\Lpi\Jbub1(t)
        - \frac{4-\pi^2}{36}\,,
    \label{eq:PiJ}
\end{align}
the remainder containing neither $E_i$, $\Jbub{n}$ nor LECs is
\begin{align}
    \bar\Pi_\zeta(t)
        &= - \frac{t \zeta(3)}{9}
        - \frac{t^2 + 18t}{81} \Lpi^3
        - \frac{20t}{27} \Lpi^2 
        - \bigg[\frac{17t^2}{432} - \frac{845 t}{648}\bigg]\Lpi\notag\\
        &- \bigg[\frac{t^2}{216} - \frac{5 t}{27}\bigg]\pi^2
        + \frac{20245 t^2}{139968} - \frac{55511 t}{23328}\,,
    \label{eq:Piz}
\end{align}
the \NLO\ LEC part is
\begin{align}
    \bar\Pi_l(t)
        &= t\bigg[
                \frac{4-t}{3t}\Jbub1(t)
                - \frac{1+3\Lpi}{9}\bigg]^2
            \Big(
                \big[l_2^q - 2l_1^q - 2l_6^q\big]t + 2l_4^q
                \Big)
        + 4t\Lpi\big(l_6^q\big)^2
        \notag\\
        &- 2t\bigg[
                \frac{4-t}{3t}\Jbub1(t)
                - \frac{1+3\Lpi}{9}\bigg]
            \Big(
                2\Lpi
                \big[2l_1^q - l_2^q - 2l_6^q\big]
                + l_6^q\big[4l_4^q - t l_6^q\big]
                \Big)
        \,,
    \label{eq:Pil}
\end{align}
and, lastly, the \NNLO\ LEC part is
\begin{align}
    \bar\Pi_c(t)
        &= 4t\bigg[\frac{4-t}{3t}\Jbub1(t) - \frac{1+3\Lpi}{9}\bigg] \big(r_{V1}^q + t r^q_{V2}\big)
        + 8t\Lpi\big(r_{V2}^q - 4 c^q_{50}\big) + 16t(\Lpi - l_4^q)c_{56}^q\,,
    \label{eq:Pic}
\end{align}
where $r_{V1}^q\coloneq -4(2c_{53}^q+c_{35}^q+4c_{6}^q)$ and $r_{V2}^q\coloneq4(c_{53}^q-c_{51}^q)$ are the combinations of \NNLO\ LECs appearing in the $\O(t)$ and $\O(t^2)$ terms, respectively, of the small-$t$ expansion of the pion vector form-factor $F^\pi_V(t)$~\cite{Bijnens:1998fm}; these are discussed further in \cref{sec:LECs}.

\Cref{eq:PiE,eq:PiJ,eq:Piz,eq:Pil,eq:Pic} comprise, via \cref{eq:PiT-N3LO}, the main result of this work.

\subsection{\NNNLO\ renormalization}\label{sec:renorm-N3LO}
Unlike at lower order~\cite{Gasser:1983yg,Bijnens:1999sh}, the explicit renormalization of the \NNNLO\ LECs has not been worked out in the literature.
However, these LECs only enter through the contact diagram 00 in \cref{tab:NNNLO}, which evaluates to a quadratic polynomial in $t$:
\begin{equation}\label{eq:diagr-00}
    \begin{diagram}[baseline={(0,-.5ex)}]
        \draw[photon] (-\photonlength,0) -- (0,0)
            pic             {NNNLO}
            ;
        \draw[notoph] (+\photonlength,0) -- (0,0);
    \end{diagram}
    = r_0 + t r_1 + t^2 r_2\,.
\end{equation}
With $\tilde c_n$ the LEC of term $n$ in the 2-flavor \NNNLO\ Lagrangian~\cite{Bijnens:2018lez}, the coefficients $r_i$ are
\begin{equation}
    \begin{gathered}
        -r_0 = 64\big(\tilde c_{382}+\tilde c_{383}\big) + 4\tilde c_{457} + 8\big(\tilde c_{458}+\tilde c_{474}\big) + 16\tilde c_{475}\,,\\
        r_1 = 16\big(\tilde c_{332} - \tilde c_{333}\big)\,,\qquad
        r_2 = -8\tilde c_{459}\,,
    \end{gathered}
\end{equation}
which we symbolically renormalize via [compare \cref{eq:renorm-NLO,eq:renorm-NNLO}]
\begin{equation}
    r_i = \pi_{16}^3 (c\mu)^{3(d-4)} \bigg[ r_i^q(\mu) + \sum_{j=1}^3\frac{\gamma_{ij}}{(d-4)^j}\bigg]\,.
\end{equation}
The leading divergence of the sum of diagrams 01--101 in \cref{tab:NNNLO} is $\frac{8t^2}{81\epsilon^3}$, so we must have $\gamma_{03}=\gamma_{13}=0$ and $\gamma_{23} = - \tfrac{8}{81}$ in order to cancel this divergence.%
\footnote{
    Since $r_2$ only contains one LEC, the contact term $\tilde c_{459}\big(\tr{D^2 F_L^{\mu\nu} D^2 F_{L\mu\nu}} + \tr{D^2 F_R^{\mu\nu} D^2 F_{R\mu\nu}}\big)$ is the first (and, at the time of writing, only) term in the \NNNLO\ Lagrangian \cite{Bijnens:2018lez} to be explicitly renormalized.}
Likewise, from the subleading divergences we deduce
\begin{equation}
    \begin{aligned}
        \gamma_{01} &= - \tfrac{208}{27} - \tfrac{16}{3} l_2^q - \tfrac{8}{3} l_1^q - 32 l_3^q l_6^q + 64 l_3^q l_5^q - 64 c_{50}^q + 256 c_{47}^q - 128 c_{46}^q + 64 c_{44}^q \eqnbreak\quad - 64 c_{35}^q + 224 c_{34}^q + 16 c_{33}^q - 24 c_{32}^q - 24 c_{31}^q + 32 c_{30}^q - 96 c_{29}^q - 384 c_6^q  \,,\\
        \gamma_{02} &= - \tfrac{184}{9} - 16 l_6^q + 32 l_5^q - 16 l_2^q + 32 l_1^q\,,\\
        \gamma_{11} &= \tfrac{1013}{972} + 8 (l_6^q)^2 + \tfrac{256}{3} c_{53}^q - 64 c_{51}^q - 64 c_{50}^q + \tfrac{32}{3} c_{35}^q + \tfrac{128}{3} c_6^q \,,\\
        \gamma_{12} &= \tfrac{128}{81} + \tfrac{16}{3} l_2^q - \tfrac{32}{3} l_1^q\,,\\
        \gamma_{21} &= - \tfrac{169}{5832} - \tfrac{4}{3} (l_6^q)^2 - \tfrac{32}{3} c_{53}^q + \tfrac{32}{3} c_{51}^q \,,\\
        \gamma_{22} &= - \tfrac{8}{243} + \tfrac{8}{9} l_6^q - \tfrac{4}{9} l_2^q + \tfrac{8}{9} l_1^q\,.
    \end{aligned}   
\end{equation}
This makes our result finite.

\subsection{Dependence on the LECs}\label{sec:LECs}

The values of the \NLO\ LECs appearing in our result are~\cite{Bijnens:2014lea}%
\footnote{
    As can be seen in \cref{app:unsub-results}, also $l_5^q=-0.730\pm0.018$ appears in $\Pi_T(q^2)$.
    Note that \rrcite[etc.]{Bijnens:1998fm,Bijnens:2014lea} use the ``$r$'' normalization convention: $l_i^r=\pi_{16}l_i^q$ (recall \cref{sec:renorm}).}
\begin{equation}
    l_1^q=-0.65\pm0.10\,,\quad
    l_2^q=0.273\pm0.033\,,\quad
    l_4^q=0.92\pm0.20\,,\quad
    l_6^q=-1.96\pm0.07\,.
\end{equation}
However, as can be seen in \cref{eq:Pil}, $l_1^q$ and $l_2^q$ only appear in the combinations
\begin{equation}
    (l_2^q-2l_1^q-2l_6^q) = 5.9\pm0.4\,,\qquad
    (2l_1^q-l_2^q-2l_6^q) = 2.8\pm0.4\,.
\end{equation}

While all NLO LECs are known to some precision, the \NNLO\ LECs are generally not.
Fortunately, 
$r_{V2}^q$ [recall \cref{eq:Pic}] is among the few exceptions: $r_{V2}^q = 4.0\pm1.2$~\cite{Bijnens:2014lea}.%
\footnote{
    Again, note the normalization convention: $r_{Vi}^r=\pi_{16}^2 r_{Vi}^q$.
    Since the publication of the LEC values used here~\cite{Bijnens:2014lea}, there has been significant progress in the measurement of the observables from which they are determined (see, e.g.,~\rcite{Leplumey:2025kvv}), so some improvements should be possible.}
Also, $r_{V1}^q$ can be determined from measurements of the pion charge radius~\cite{NA7:1986vav,Ananthanarayan:2013dpa}, although it is difficult to disentangle from the NLO contributions; see \rcite{Colangelo:2021moe} for a recent treatment of this problem.
A naïve extraction assuming all other LECs fixed gives $r_{V1}^q = -5\pm8$, and vector meson dominance gives $r_{V1}^q\approx -6.2$~\cite{Bijnens:1998fm}.

Two more linear combinations of \NNLO\ LECs appear in \cref{eq:Pic}, but they are of lesser importance, since they lack $\Jbub1(t)$ and therefore do not contribute to the imaginary part, and since they combine with the entirely undetermined \NNNLO\ LEC $r_1^q$.
Moreover, in the computation of FVEs only terms associated with loops survive.
In \cref{eq:Pic}, these are $-32t\Lpi c_{50}^q$ (from diagram 01 in \cref{tab:NNNLO}) and $16t\Lpi c_{56}^q$ (via a tadpole loop in the decay constant renormalization), so for FVEs only the combination $2c_{50}^q-c_{56}^q$ is needed.
These LECs appear in different linear combinations at two loops in the vector-vector and axial-axial correlators (with $c_{50},c_{56}$ appearing as $C_{87},C_{93}$ at three flavors~\cite{Amoros:1999dp} and as $K_{109},K_{115}$ at $n$ flavors~\cite{Bijnens:2011xt}), so their values should be possible to determine if needed.

\section{Conclusions and outlook}\label{sec:conclusions}

We have computed the three-loop HVP amplitude $\Pi^{\mu\nu}(q)$ in two-flavor ChPT.
We have presented the result in terms of three (poly)logarithmic and five elliptic functions of the photon's virtuality, with four of the elliptic functions new to this work.
While the polynomial part of the amplitude depends on LECs whose values are not known, the (poly)logarithmic part depends on only five parameters, all more or less known from data, and the elliptic part has no free parameters.
Our result provides the next-to-next-to-leading contribution to the subthreshold part of the amplitude and its two-pion cut, and the leading contribution to the four-pion cut, which comes entirely from the elliptic part.

The process of evaluating the amplitude has required unexpectedly deep forays into the field of loop integrals.
Satisfying the Ward--Takahashi identity required state-of-the-art master integral reduction, and renormalization required a step beyond that with the novel Schouten relations.
While relatively accessible methods allowed all our elliptic integrals to be related to a single one, $E_1$, which is well-studied and amenable to powerful mathematical approaches, the sheer complexity of putting this to use has prompted us to split off the numerical implementation into a follow-up work~\cite{LLSV} to keep the present scope manageable.
As it stands, the amplitude can be evaluated using \texttt{pySecDec}~\psdrefs, although the precision is very limited for many virtualities, particularly above threshold.

The HVP amplitude, in particular when completed in \rcite{LLSV}, serves as a starting point for the wide range of phenomenological applications surveyed in the introduction.
Not least among these are FVEs, and even if the most direct approach then is to recompute the amplitude in a finite volume, this completed infinite-volume calculation can serve as a blueprint for that undertaking.
This work also leaves us in a better position to consider if other three-loop ChPT calculations may prove both useful and tractable.

\subsection*{Acknowledgements}
We would like to thank Antonin Portelli for invaluable discussions about the finite-volume HVP at three loops, toward which we hope the present calculation is a significant step.
We would also like to thank Roman Lee and Lorenzo Trancredi for very useful discussions regarding the integrals, and Johan Bijnens and Marc Knecht regarding \chpt.
MS would particularly like to thank Johan Bijnens for the use of his \form\ codes, which were an invaluable reference in the development of the \form\ code used here.

The work of PV was funded by the Agence Nationale de la Recherche (ANR) under the grant Observables (ANR-24-CE31-7996), and by the Munich Institute for Astro-, Particle and BioPhysics (MIAPbP) which is funded by the Deutsche Forschungsgemeinschaft (DFG, German Research Foundation) under Germany's Excellence Strategy -- EXC-2094 -- 390783311. 
The work of LL, AL and MS was funded in part by the French government under the France 2030 investment plan, as part of the ``Initiative d'Excellence d'Aix-Marseille Université -- A\kern-2pt*MIDEX'' under grant AMX-22-RE-AB-052, and by the ANR under grant ANR-22-CE31-0011.

\appendix
\allowdisplaybreaks

\section{Differential equations for the master integrals}\label{app:diffeq}

In this section, we present the differential equations satisfied by the master integrals (excluding the constants $I_\tad$ and $E_0$).
The procedure for obtaining them is outlined in~\cref{sec:diffeq}, and we will group them according to the hierarchy of~\cref{eq:hierarchy}.

\subsection{Differential equations for tadpoles and bubbles}\label{app:diffeq-tadbub}

The bubble integral satisfies the first-order differential equation
\begin{equation}\label{eq:diffeq-bub}
    \bigg[t(t-4)\frac{\p}{\p t} - \frac{4+(d-4)t}{2}\bigg] I_\bub(d;t) = (2-d) I_\tad(d)\,,
\end{equation}
so it follows directly from \cref{eq:tadbub} that
\begin{equation}\label{eq:diffeq-tadbub}
    \bigg[t(t-4)\frac{\p}{\p t} - b\frac{4+(d-4)t}{2}\bigg] T_{ab}(d;t) = b(2-d) T_{a(b-1)}(d;t)\,.
\end{equation}
At the level of $\Jbub{n}(t)$, \cref{eq:diffeq-bub} translates to
\begin{equation}\label{eq:diffeq-Jbub}
    \bigg[t(t-4)\frac{\p}{\p t} - 2\bigg]\Jbub{n}(t) = nt \Jbub{n-1}(t)\,,\qquad 
    \Jbub{0}(t) = 1\,.
\end{equation}

\subsection{Differential equations for $E_1, E_2$ and $E_3$}\label{app:diffeq-E1E2E3}

The master $E_1(d;t)$ satisfies the third-order differential equation [compare~\cref{eq:diffeq}]
\begin{multline}\label{eq:diffeq-E1}
    \bigg[
          \pi_{13} (d;t)\frac{\p^3}{\p t^3} 
        + \pi_{12} (d;t)\frac{\p^2}{\p t^2} 
        + \pi_{11} (d;t)\frac{\p}{\p t} 
        + \pi_{10}(d;t)\bigg] E_1(d;t) = -12 (d-2)^3 T_{30}(d)\,,
\end{multline}
with polynomial coefficients
\begin{equation}
    \begin{aligned}
        \pi_{13}(d;t) &= 4 (t-16) (t-4) t^2\,,\\
        \pi_{12}(d;t) &= -12 t \big[(d-4) t^2-10 (d-5) t-64\big]\,,\\
        \pi_{11}(d;t) &= (d-4) (11 d-36) t^2 - 4 (7 d^2-88 d+216)t - 64 (d-4) d\,,\\
        \pi_{10}(d;t) &= -(d-3) (3 d-8) \big[(d-4) t+2 d+4\big]\,.
    \end{aligned}
\end{equation}
The masters $E_2(d;t)$ and $E_3(d;t)$ are obtained from the derivatives of $E_1(d;t)$ using
\begin{align}
    \label{eq:E2toE1}
    E_2(d;t) &= \bigg[-\frac{t}{4}\frac{\p}{\p t} + \frac{3d-8}{8}\bigg] E_1(d;t)\,,\\
    \label{eq:E3toE1}
    E_3(d;t) &= \bigg[\frac{t}{2}\frac{\p^2}{\p t^2} + \frac{4d+(4-d)t}{16}\frac{\p}{\p t}  + \frac{(d-4)(3d-8)}{32} \bigg] E_1(d;t)\,.
\end{align}
Below, we often use this to eliminate $E_2$ and $E_3$ in favor of derivatives of $E_1$.

\subsection{Differential equation for $E_4$}\label{app:diffeq-E4}

The master $E_4(d;t)$ satisfies the first-order differential equation
\begin{equation}\label{eq:diffeq-E4}
    \bigg[2(4-t)t\frac{\p}{\p t} +(4+(d-4)t) \bigg] E_4(d;t) 
        = (d-2)E_0(d)
        + \bigg[\frac{d-4}{2} - 3 t\frac{\p}{\p t} \bigg] E_1(d;t)\,.
\end{equation}
In fact, we do not use this equation here, but include it for completeness.

\subsection{Differential equations for $E_5$ and $E_6$}\label{app:diffeq-E5E6}

The master $E_5(d;t)$ satisfies the second-order differential equation
\begin{multline}\label{eq:diffeq-E5}
    \bigg[
          \pi_{52}(d;t)\frac{\p^2}{\p t^2} 
        + \pi_{51}(d;t)\frac{\p}{\p t} 
        + \pi_{50}(d;t)\bigg] E_5(d;t) 
    = \bigg[
          \tilde\pi_{12}(d;t)\frac{\p^2}{\p t^2} 
        + \tilde\pi_{11}(d;t)\frac{\p}{\p t} 
        + \tilde\pi_{10}(d;t)\bigg] E_1(d;t)\\
    -24(d-3)(3d-8) E_4(d;t)
    +6 (3 d-8) (d-2) E_0(d)\\
    -6(d-2)^2 (t-2) T_{21}(d;t) + 12(d-3)(d-2) t T_{12}(d;t),
\end{multline}
with polynomial coefficients
\begin{equation}
    \begin{alignedat}{2}
        \pi_{52} &= 12 (t-4)^2 t^2\,,\qquad&
        \tilde\pi_{12} &= 2 t (t+2) (t+8)\,,\\
        \pi_{51} &= -24 (t-4)t\big[(d-5) t + 11-2 d\big]\,,\qquad&
        \tilde\pi_{11} &= -(t+8) \big[(5 d-16) t-2 d\big]\,,\\
        \pi_{50} &= 3(t-4)\big[3 (d-6) (d-4) t+8 (2 d-7)\big]\,,\qquad&
        \tilde\pi_{10} &= (3 d-8) \big[(d-3) t-2 (2 d-3)\big]\,.\\
    \end{alignedat}
\end{equation}
It should be noted that the differential operator on the left-hand side of~\cref{eq:diffeq-E5} factorizes in two dimensions, with $\epsilon$ corrections that are at most linear differential operators:
\begin{multline}\label{eq:factor-E5}
    \bigg[
          \pi_{52}(2-2\epsilon;t)\frac{\p^2}{\p t^2} 
        + \pi_{51}(2-2\epsilon;t)\frac{\p}{\p t} 
        + \pi_{50}(2-2\epsilon;t)\bigg]
    = 12(t-4)\bigg[t\frac{\p}{\p t} + 1\bigg] \circ \bigg[t(t-4)\frac{\p}{\p t} - 3(t-2)\bigg] \\
    + \bigg[48(t-4)(2t - 4)\frac{\p}{\p t} + 4(15t - 8)\bigg]\epsilon
    + 12t \epsilon^2\,. 
\end{multline}
The implications of this are discussed in \cref{sec:schouten}.

The master $E_6(d;t)$ is obtained from the derivative of $E_5(d;t)$ using
\begin{multline}\label{eq:E6toE5}
  E_6(d;t)
    = -\bigg[
        \frac{\p}{\p t} 
        + \frac{1}{t}\bigg] E_5(d;t)
    + \bigg[
        \frac{t-16}{48(d-3)}\frac{\p^2}{\p t^2} 
        - \frac{(5 d -16)t+16 d}{96 (d-3) t} \frac{\p}{\p t} 
        + \frac{3 d-8}{96 t}\bigg] E_1(d;t) \\
    + \frac{3 d-8}{8t} E_4(d;t) 
    -\frac{(d-2)^2}{16 (d-3) t} T_{21}(d;t)\,.
\end{multline}

\subsection{Initial conditions}\label{app:initial}
When working with the above differential equations, it is useful to know the values of the master integrals at $t=0$.
Assuming the integrals to be sufficiently regular, this can be determined by setting $q=0$, which following \cref{eq:basis} replaces $\{D_4,D_5,D_6\}\to\{D_1,D_2,D_3\}$, and then applying the master integral reduction.
Since we have only two $t$-independent master integrals, the result must be of the form
\begin{equation}\label{eq:t0}
    E_i(d;0) = \mu_i(d) E_0(d) + \tilde\mu_i(d) T_{30}(d)\,,
\end{equation}
where\footnote{
    Despite the appearance of $1/(d-4)$ multiplying the already cubically divergent integrals $E_0$ and $T_{30}$, the quartic (and for $E_3$ and $E_6$, also cubic) divergences cancel, as can be seen by combining~\cref{eq:t0} with the expressions in~\cref{app:expr-4d}.}
\begin{equation}
    \begin{alignedat}{2}
        \mu_1(d) &= 1\,,\qquad 
        \mu_2(d)=\mu_4(d) = \frac{3d-8}{8}\,,   &\qquad
        \tilde\mu_1(d)  &= \tilde\mu_2(d) = \tilde\mu_4(d) = 0\,,\\
        \mu_3(d)        &= \frac{(d-7) (3 d-10) (3 d-8)}{128 (d-4)}\,,    &\qquad
        \tilde\mu_3(d)  &= \frac{3 (d-2)^3}{64 (d-4)} \,,\\
        \mu_5(d)        &= \frac{(d-3) (3d-10 ) (3d-8)}{64 (d-4)}\,,    &\qquad
        \tilde\mu_5(d)  &= -\frac{(d-2)^3}{32 (d-4)} \,,\\
        \mu_6(d)        &= \frac{3 (d-6) (d-3) (3 d-10) (3 d-8)}{1024 (d-4)}\,, &\qquad
        \tilde\mu_6(d)  &= \frac{(d-2)^3(5d-14)}{512 (d-4)} \,.
    \end{alignedat}
\end{equation}
Along with derivative relations such as
\cref{eq:E2toE1,eq:E3toE1,eq:E6toE5}, this gives a complete set of
initial conditions for the differential equations.

Expanding these relations near four dimensions and using the expansion
of $E_0(4-2\epsilon)$ given in~\cref{eq:E0-4d} and $T_{30}(4-2\epsilon)$ given
in~\cref{eq:tadbub-3loop-4d},  we have
\begin{subequations}\label{eq:initial}
    \begin{align}
        \Mpi^{-6} E_1(4-2\epsilon;0)  
            &= \frac{2}{\epsilon^3}
            + \frac{23}{3\epsilon^2}
            + \frac{\pi^2 + 35}{2\epsilon}
            + \frac{275 + 23\pi^2 - 24\zeta(3)}{12}
            + \O(\epsilon)\,, \label{eq:initial:E1}\\
        \Mpi^{-6} E_2(4-2\epsilon;0)  
            &= \frac{1}{\epsilon^{3}} 
            + \frac{7}{3 \epsilon^{2}}
            + \frac{12+\pi^2}{4\epsilon}
            - \frac{20 - 7\pi^2 + 12\zeta(3)}{12}
            + \O(\epsilon)\,, \label{eq:initial:E2}\\
        \Mpi^{-6} E_3(4-2\epsilon;0)
            &= -\frac{5}{6 \epsilon^{2}}
            - \frac{5}{2 \epsilon}
            - \frac{140 + 5\pi^2 - 84\zeta(3)}{24}
            + \O(\epsilon) \,, \label{eq:initial:E3}\\
        \Mpi^{-6} E_4(4-2\epsilon;0)       
            &= E_2(4-2\epsilon;0) \,,\label{eq:initial:E4}\\
        \Mpi^{-6} E_5(4-2\epsilon;0)       
            &= \frac{1}{3 \epsilon^{3}}
            + \frac{1}{3 \epsilon^{2}}
            + \frac{4+\pi^{2}}{12\epsilon}
            + \frac{4 + \pi^2 - 32\zeta(3)}{12}
            + \O(\epsilon)\,,\label{eq:initial:E5}\\
        \Mpi^{-6} E_6(4-2\epsilon;0)      
            &= -\frac{1}{4 \epsilon^{2}}
            - \frac{1}{4 \epsilon}
            - \frac{4 + \pi^2 - 14\zeta(3)}{16}
            + \O(\epsilon)\,.\label{eq:initial:E6}
    \end{align}
\end{subequations}
We may also perform the expansion around $t=0$ of the finite master integrals to higher order by finding power-series solutions to the differential equations:
\begin{subequations}\label{eq:elliptics-t0}
    \begin{align}
        E_1(2;t) &= -7\zeta(3)+\frac{6-7\zeta(3)}{16}t+\frac{54-49\zeta(3)}{1024}
     t^2 + \O(t^3),\\
        E_2(2;t) &= \frac{7 \zeta(3)}{4}+\frac{7 \zeta(3)-6}{32} t +\frac{147 \zeta(3)-162}{4096} t^{2}+ \O(t^3),\\
        E_3(2;t) &= \frac{6-35 \zeta(3)}{32}+\frac{102-105 \zeta(3)}{512} t+\frac{458-399 \zeta(3)}{8192} t^{2} + \O(t^3),\\
        \bar E_5(4;t) &= \frac{4+\pi^2-32\zeta(3)}{12} + \frac{34+8\pi^2-105\zeta(3)}{192}t + \O(t^2),\\
        \bar E_6(4;t) &= \frac{14\zeta(3)-\pi^2-4}{16} + \frac{21\zeta(3) - \pi^2 - 15}{96} t +\O(t^2)\,.
    \end{align}
\end{subequations}

\section{Dimension-shifting relations}\label{app:dimshift}

In this appendix, we provide explicit expressions for the dimension-shifting relations derived using the methods described in \cref{sec:dimshift}.
Since the tadpole and bubble integrals do not simplify significantly in two dimensions, we keep them four-dimensional and only apply the dimension shift to the $E_i$.
We will express the relations using the notation
\begin{equation}\label{eq:dimshift-coeffs}
    E_i(d;t) 
        = \sum_{a,b} (d-2)^a \tau_i^{ab}(d;t) T_{ab}(d;t) 
        + \Mpi^6\sum_j \varepsilon_i^j(d;t) E_j(d-2;t)\,,
\end{equation}
with the coefficients $\tau_i^{ab}(d;t)$ and $\varepsilon_i^j(d;t)$ given below, again grouped by the hierarchy \cref{eq:hierarchy}.
Note that \cref{eq:dimshift} gives $E_j(d-2;t)$ in terms of $E_i(d;t)$, so \cref{eq:dimshift-coeffs} requires the inverse of that relation.
We keep $d$ general here; the $\epsilon$ expansion around $d=4$ can be found in \cref{app:expr-4d}.

\setlength{\tikzineqyshift}{-0.25ex}
\newcommand{\drawbball}[1]{%
    \renewcommand{\bubbleaspect}{1}%
    \draw (0,0)%
        pic[#1]           {bubble}%
        pic[#1,yscale=.4] {bubble}%
        ;%
    \fill[black] (-\bubblesize,0) circle[radius=.6pt];%
    \fill[black] (+\bubblesize,0) circle[radius=.6pt];%
    }
\newcommand{\drawfatcat}[1]{%
    \drawbball{#1}%
    \draw[#1,pion] (-\photonlength,0) -- (-\bubblesize,0);%
    \draw[#1,pion] (0,+\photonlength) -- (0,+\bubblesize);%
    \fill[black] (0,+\bubblesize) circle[radius=.6pt];%
    }
\newcommand{\drawcatseye}[1]{%
    \drawbball{#1}%
    \draw[#1,pion] (0,-\photonlength) -- (0,-\bubblesize);%
    \draw[#1,pion] (0,+\photonlength) -- (0,+\bubblesize);%
    \fill[black] (0,-\bubblesize) circle[radius=.6pt];%
    \fill[black] (0,+\bubblesize) circle[radius=.6pt];%
    }

\subsection{Dimension shifting for the master $E_0$}\label{app:dimshift-E0}

This is a simple case that is good for familiarizing oneself with the method described in \cref{sec:dimshift}.
The spanning trees (black, with removed lines grayed out) are
\begin{equation}
    \tikzineq{
        \drawbball{lightgray}
        \begin{scope}[overlay]
            \clip (-\photonlength,0) rectangle (+\photonlength,+\photonlength);
            \draw (0,0) pic {bubble};
        \end{scope}
        }
    \,,\qquad
    \tikzineq{
        \drawbball{lightgray}
        \begin{scope}[overlay]
            \clip (-\photonlength,0) rectangle (+\photonlength,+\photonlength);
            \draw (0,0) pic[yscale=.4] {bubble};
        \end{scope}
        }
    \,,\qquad
    \tikzineq{
        \drawbball{lightgray}
        \begin{scope}[overlay]
            \clip (-\photonlength,0) rectangle (+\photonlength,-\photonlength);
            \draw (0,0) pic[yscale=.4] {bubble};
        \end{scope}
        }
    \,,\qquad
    \tikzineq{
        \drawbball{lightgray}
        \begin{scope}[overlay]
            \clip (-\photonlength,0) rectangle (+\photonlength,-\photonlength);
            \draw (0,0) pic {bubble};
        \end{scope}
        }
    \,.    
\end{equation}
Reading the line indices off \cref{fig:elliptics}, \cref{eq:U} gives
\begin{equation}
    \U = \alpha_1\alpha_2\alpha_7\alpha_8\Big(\tfrac1{\alpha_1}+\tfrac1{\alpha_2}+\tfrac1{\alpha_7}+\tfrac1{\alpha_8}\Big)\,,
\end{equation}
and the operator $\D$ follows from $\alpha_j\to\p/\p m_j^2$.
Running this through \cref{eq:dimshift} gives \cref{eq:dimshift-coeffs} with coefficients
\begin{equation}\label{eq:dimshift-E0}
    \kappa_0(d) \tau_0^{30}(d;t) = \frac{2 (11 d-38)}{3}\,,\qquad
    \kappa_0(d) \varepsilon_0^0(d;t) = \frac{128 (d-4)}{3}\,,
\end{equation}
where we have pulled out a common denominator $\kappa_0(d) \coloneq (d-3) (3 d-10) (3 d-8)$.
Here and below, we omit coefficients $\tau_i^{ab}$ and $\varepsilon_i^j$ that vanish identically.
Note that since $\varepsilon^0_0(d;t)$ vanishes at $d=4$, $E_0(2)$ does not appear in $E_0(4-2\epsilon)$ until the first order in $\epsilon$.

\subsection{Dimension shifting for the masters $E_1$, $E_2$ and $E_3$} \label{app:dimshift-E1E2E3}

Here, $\U$ and $\D$ are the same as for $E_0$, except that index $2$ is replaced by $5$; compare \cref{fig:elliptics}.
The resulting coefficients in \cref{eq:dimshift-coeffs} for $E_1$ are
\begin{subequations}\label{eq:dimshift-E1}
    \begin{align}
        \kappa_1(d)\tau_1^{30}(d;t)    
            &= \frac{(d-4)^2 }{48} t^2
                + \frac{11 d^2-85 d+164}{6} t
                + \frac{2(38 d^2-268 d+473)}{3} \,,\\
        \kappa_1(d) \varepsilon_1^1(d;t)
            &= \frac{5 d^2-47 d+110}{24} t^3
                + \frac{2(15 d^2-128 d+275)}{3} t^2
                - \frac{2(276 d^2-2342 d+5027)}{3} t\eqnbreak
                - \frac{32(20 d^2-178 d+385)}{3} 
                + \frac{512(3 d^2-26 d+56)}{3}\frac{1}{t}\,,\\
        \kappa_1(d) \varepsilon_1^2(d;t)
            &= \frac{d-6}{24} t^4
                + \frac{d-10}{6} t^3
                - \frac{2(97 d-487)}{3} t^2\eqnbreak
                + \frac{16(73 d-354)}{3} t
                + \frac{128(29 d-119)}{3} 
                - \frac{4096 (d-4)}{3} \frac{1}{t-4}\,,\\
        \kappa_1(d) \varepsilon_1^3(d;t)
            &= -\frac{1}{6} t^4
               -\frac{10}{3} t^3
               + 112 t^2
               -\frac{640}{3} t
               -\frac{2048}{3}\,,
    \end{align}%
\end{subequations}
with $\kappa_1(d) \coloneq (d-3) (2 d-7) (2 d-5) (3 d-10) (3 d-8)$.
Those for $E_2$ are
\begin{subequations}\label{eq:dimshift-E2}
    \begin{align}
        \kappa_2(d)\tau_2^{30}(d;t)    
            &= \frac{9 d^2-70 d+136}{48} t
                + \frac{54 d^2-383 d+680}{12}\,,\\
        \kappa_2(d) \varepsilon_2^1(d;t)
            &= \frac{(33 d^2-295 d+662)}{24} t^2
                + \frac{(-108 d^2+921 d-1996)}{6}\eqnbreak
                -8 (9 d^2-80 d+175)
                + \frac{128(3 d^2-26 d+56)}{3} \frac{1}{t}\,,\\
        \kappa_2(d) \varepsilon_2^2(d;t)
            &= \frac{5 (d-6)}{24} t^3
                + \frac{254-49 d}{6} t^2
                + \frac{2 (33 d-179)}{3} t\eqnbreak
                + \frac{32 (31 d-131)}{3}
                -\frac{1024 (d-4)}{3}\frac{1}{t}\,,\\
        \kappa_2(d) \varepsilon_2^3(d;t)
            &= t^3
                + 14 t^2
                -\frac{512}{3}\,,
    \end{align}%
\end{subequations}
with $\kappa_2(d) \coloneq (d-3) (2 d-7) (3 d-10) (3 d-8)$.
Lastly, those for $E_3$ are
\begin{subequations}\label{eq:dimshift-E3}
    \begin{align}
        \kappa_3(d) \tau_3^{30}(d;t)  
            &= \frac{(d-4)^2}{32} t
                + \frac{69 d^3-753 d^2+2732 d-3296}{48 (2 d-7)}\,,\\
        \kappa_3(d) \varepsilon_3^1(d;t)
            &= \frac{6 d^2-55 d+126}{24} t^2
                + \frac{-111 d^3+1454 d^2-6270 d+8896}{24 (2 d-7)} t \eqnbreak
                + \frac{-264 d^3+3077 d^2-11907 d+15288}{6 (2 d-7)} \eqnbreak
                + \frac{8 (21 d^3-254 d^2+1016 d-1344)}{3 (2 d-7)}\frac{1}{t}\,,\\
        \kappa_3(d) \varepsilon_3^2(d;t)
            &= \frac{d-6}{24} t^3
                + \frac{-73 d^2+646 d-1368}{24 (2 d-7)} t^2
                + \frac{21 d^2-264 d+676}{6 (2 d-7)} t \eqnbreak
                + \frac{10 (47 d^2-355 d+664)}{3 (2 d-7)}
                - \frac{64 (7 d^2-52 d+96)}{3 (2 d-7)} \frac{1}{t}\,,\\
        \kappa_3(d) \varepsilon_3^3(d;t)
            &= -\frac{1}{6} t^3
                + \frac{33 d-116}{6 (2 d-7)} t^2
                +  \frac{2 (3 d-8)}{3 (2 d-7)} t
                - \frac{32 (7 d-24)}{3 (2 d-7)}\,,
    \end{align}%
\end{subequations}
with $\kappa_3(d) \coloneq (d-3) (3 d-10) (3 d-8)$.

\subsection{Dimension shifting for the master $E_4$} \label{app:dimshift-E4} 
Here, the spanning trees are
\begin{equation}
    \setlength{\photonlength}{1.5\bubblesize}
    \tikzineq{
        \drawfatcat{lightgray}
        \begin{scope}[overlay]
            \clip (-\photonlength,0) rectangle (+\photonlength,+\photonlength);
            \draw (0,0) pic {bubble};
        \end{scope}
        }
    \,,\qquad
    \tikzineq{
        \drawfatcat{lightgray}
        \begin{scope}[overlay]
            \clip (-\photonlength,0) rectangle (0,+\photonlength);
            \draw (0,0) pic {bubble};
        \end{scope}
        \begin{scope}[overlay]
            \clip (-\photonlength,0) rectangle (+\photonlength,+\photonlength);
            \draw (0,0) pic[yscale=.4] {bubble};
        \end{scope}
        }
    \,,\quad
    \tikzineq{
        \drawfatcat{lightgray}
        \begin{scope}[overlay]
            \clip (-\photonlength,0) rectangle (0,+\photonlength);
            \draw (0,0) pic {bubble};
        \end{scope}
        \begin{scope}[overlay]
            \clip (-\photonlength,0) rectangle (+\photonlength,-\photonlength);
            \draw (0,0) pic[yscale=.4] {bubble};
        \end{scope}
        }
    \,,\quad
    \tikzineq{
        \drawfatcat{lightgray}
        \begin{scope}[overlay]
            \clip (-\photonlength,0) rectangle (0,+\photonlength);
            \draw (0,0) pic {bubble};
        \end{scope}
        \begin{scope}[overlay]
            \clip (-\photonlength,0) rectangle (+\photonlength,-\photonlength);
            \draw (0,0) pic {bubble};
        \end{scope}
        }
    \,,\qquad
    \tikzineq{
        \drawfatcat{lightgray}
        \begin{scope}[overlay]
            \clip (+\photonlength,0) rectangle (0,+\photonlength);
            \draw (0,0) pic {bubble};
        \end{scope}
        \begin{scope}[overlay]
            \clip (-\photonlength,0) rectangle (+\photonlength,+\photonlength);
            \draw (0,0) pic[yscale=.4] {bubble};
        \end{scope}
        }
    \,,\quad
    \tikzineq{
        \drawfatcat{lightgray}
        \begin{scope}[overlay]
            \clip (+\photonlength,0) rectangle (0,+\photonlength);
            \draw (0,0) pic {bubble};
        \end{scope}
        \begin{scope}[overlay]
            \clip (-\photonlength,0) rectangle (+\photonlength,-\photonlength);
            \draw (0,0) pic[yscale=.4] {bubble};
        \end{scope}
        }
    \,,\quad
    \tikzineq{
        \drawfatcat{lightgray}
        \begin{scope}[overlay]
            \clip (+\photonlength,0) rectangle (0,+\photonlength);
            \draw (0,0) pic {bubble};
        \end{scope}
        \begin{scope}[overlay]
            \clip (-\photonlength,0) rectangle (+\photonlength,-\photonlength);
            \draw (0,0) pic {bubble};
        \end{scope}
        }
    \,,
\end{equation}
giving
\begin{equation}
    \U = \alpha_2 \alpha_7 \alpha_8 \Big[ 
        1
        + (\alpha_1 + \alpha_4)
        \Big(
            \tfrac{1}{\alpha_2} 
            + \tfrac{1}{\alpha_7} 
            + \tfrac{1}{\alpha_8}
        \Big)\Big]\,.
\end{equation}
The coefficients are
\begin{subequations}
    \begin{align}
        \kappa_4(d)\tau_4^{30}(d;t)
            &= \frac{(d-4)^2}{48} t
                - \frac{16 d^2-115 d+207}{6}\,,\\
        \kappa_4(d) \tau_4^{21}(d;t)
            &= \frac{(d-3) (2 d-7) (11 d-38)}{2} \,,\\
        \kappa_4(d) \varepsilon_4^0(d;t)
            &= \frac{32 (d-4) (2 d-7)}{3}\,,\\
        \kappa_4(d) \varepsilon_4^1(d;t)
            &= \frac{5 d^2-47 d+110}{24} t^2
                + \frac{-24 d^2+263 d-678}{6} t\eqnbreak
                + \frac{8(37 d^2-320 d+693)}{3}
                - \frac{128(3 d^2-26 d+56)}{3} \frac{1}{t}\,,\\
        \kappa_4(d) \varepsilon_4^2(d;t)
            &=  \frac{d-6}{24} t^3 
                + \frac{74-13 d}{6} t^2
                + \frac{2(61 d-315)}{3} t\eqnbreak
                - 32 (9 d-41)
                + \frac{1024 (d-4)}{3} \frac{1}{t}\,,\\
        \kappa_4(d) \varepsilon_4^3(d;t)
            &= -\frac{1}{6} t^3
                + 6 t^2
                - 64 t
                + \frac{512}{3}\,,\\
        \kappa'_4(d) \varepsilon_4^4(d;t)
            &=  -\frac{32(d-4)}{3} t
                +\frac{128 (d-4)}{3}\,,
    \end{align}%
\end{subequations}
with $\kappa_4(d)\coloneq (d-3)^2 (2 d-7) (3 d-10) (3 d-8)$ and $\kappa'_4(d) \coloneq (d-3)^2(9d^2-54d+80)$.

An interesting thing to note here is that $\varepsilon_4^4(4;t) = 0$, so $E_4(2;t)$ is not needed in order to evaluate $\bar E_4(4;t)$.
This does not necessarily mean that $E_4$ is eliminated by the dimension shift, though, since it still appears in $E_5$ and $E_6$ (see below) as well as in $E_i^{(r)}(4;t)$ for $r>0$.
However, this is why it does not appear in \cref{eq:PiE}.

\subsection{Dimension shifting for the masters $E_5$ and $E_6$} \label{app:dimshift-E5E6}

Here, the spanning trees are
\begin{equation}
    \setlength{\photonlength}{1.5\bubblesize}
    \tikzineq{
        \drawcatseye{lightgray}
        \begin{scope}[overlay]
            \clip (0,0) -- (+\photonlength,0) -- (+\photonlength,-\photonlength) -- (-\photonlength,-\photonlength) -- (-\photonlength,+\photonlength) -- (0,+\photonlength) -- cycle;
            \draw (0,0) pic {bubble};
        \end{scope}
        }
    \,,\qquad
    \tikzineq{
        \drawcatseye{lightgray}
        \begin{scope}[overlay]
            \clip (0,0) rectangle (+\photonlength,-\photonlength) (0,0) rectangle (-\photonlength,+\photonlength);
            \draw (0,0) pic {bubble};
        \end{scope}
        \begin{scope}[overlay]
            \clip (-\photonlength,0) rectangle (+\photonlength,+\photonlength);
            \draw (0,0) pic[yscale=.4] {bubble};
        \end{scope}
        }
    \,,\qquad
    \tikzineq{
        \drawcatseye{lightgray}
        \begin{scope}[overlay]
            \clip (0,+\photonlength) rectangle (+\photonlength,-\photonlength);
            \draw (0,0) pic {bubble};
        \end{scope}
        \begin{scope}[overlay]
            \clip (-\photonlength,0) rectangle (+\photonlength,+\photonlength);
            \draw (0,0) pic[yscale=.4] {bubble};
        \end{scope}
        }
    \,,
\end{equation}
plus horizontal and vertical reflections thereof.
Thus,
\begin{equation}
    \U = \alpha_7\alpha_8\big[\alpha_1+\alpha_2+\alpha_4+\alpha_5\big]
        + (\alpha_7+\alpha_8)\big[\alpha_1\alpha_5 + \alpha_2\alpha_4 + \alpha_1\alpha_4 + \alpha_2\alpha_5\big]\,.
\end{equation}
The coefficients are
\begin{subequations}\label{eq:dimshift-E5}
    \begin{align}
        \kappa_5(d;t) \tau_5^{30}(d;t)
            &=  -\frac{(d-4)^3}{96 (d-5) (d-3)^2} t^4
                + \frac{(d-4)^2}{24 (d-3)^2} t^3 \eqnbreak
                + \frac{9 d^5-191 d^4+1613 d^3-6755 d^2+14002 d-11480}{12 (d-5) (d-3)^2 (2 d-7) (3 d-10)} t^2\eqnbreak
                + \frac{-72 d^4+1075 d^3-5981 d^2+14700 d-13468}{3 (d-3)^2 (6 d^2-41 d+70)} t\eqnbreak
                +\frac{4 (d-4)^2 (39 d^2-261 d+434)}{3 (d-3)^2 (6 d^2-41 d+70)}\,,\\
        \kappa_5(d;t) \tau_5^{21}(d;t)
            &= \frac{(d-4)^3}{48 (d-5) (d-3)} t^4
                - \frac{(d-4)^2}{12 (d-3)} t^3
                + \frac{-7 d^3+100 d^2-475 d+748}{6 (d^2-8 d+15)} t^2\eqnbreak
                + \frac{4 (15 d^2-125 d+258)}{3 (d-3)} t
                - \frac{8 (5 d^2-41 d+84)}{d-3}\,,\\
        \kappa_5(d;t) \tau_5^{12}(d;t)
            &= \frac{8 (d^2-9 d+20)}{3} t^2
                - \frac{4 (23 d^2-195 d+410)}{3} t\eqnbreak
                + \frac{16 (11 d^2-93 d+196)}{3}\,,\\
        \kappa'_5(d) \varepsilon_5^0(d;t)
            &= \frac{2 (d-4)^3}{3} t
                + \frac{8 (d-4)^2}{9}
                - \frac{16 (d-4) (27 d^2-217 d+434)}{9 (3 d-10)} \frac{1}{t}\,,\\
        \kappa'_5(d) \varepsilon_5^1(d;t)
            &= \frac{8(d^2-9 d+20)}{3}  t^2
                - \frac{4(23 d^2-195 d+410)}{3}  t\eqnbreak
                + \frac{16 (11 d^2-93 d+196)}{3}\,,\\
        \kappa'_5(d) \varepsilon_5^2(d;t)
            &= -\frac{(d-6) (d-4)^2}{144 (d-5)} t^4   
                + \frac{10 d^3-133 d^2+586 d-856}{36 (d-5)} t^3 \eqnbreak
                + \frac{-9 d^5+327 d^4-3821 d^3+20111 d^2-49728 d+47180}{18 (d-5) (2 d-7) (3 d-10)} t^2 \eqnbreak
                + \frac{2 (177 d^4-2798 d^3+16759 d^2-45116 d+46060)}{9 (6 d^2-41 d+70)} t \eqnbreak
                + \frac{8 (111 d^4-1339 d^3+5186 d^2-5760 d-2688)}{9 (6 d^2-41 d+70)} \eqnbreak
                - \frac{128 (54 d^4-810 d^3+4537 d^2-11245 d+10402)}{9 (6 d^2-41 d+70)}\frac{1}{t}\,,\\
        \kappa'_5(d) \varepsilon_5^3(d;t)
            &= \frac{(d-4)^2}{36 (d-5)} t^4
                + \frac{-4 d^2+33 d-68}{9 (d-5)} t^3 \eqnbreak
                + \frac{2 (9 d^4-201 d^3+1499 d^2-4669 d+5250)}{9 (d-5) (2 d-7) (3 d-10)} t^2 \eqnbreak
                - \frac{8 (33 d^3-422 d^2+1809 d-2590)}{9 (6 d^2-41 d+70)} t
                - \frac{128 (3 d^3-10 d^2-53 d+182)}{9 (6 d^2-41 d+70)}\,,\\
        \frac{\kappa'_5(d)}{3d-14} \varepsilon_5^4(d;t)
            &= \frac{(d-4)^2 }{12} t^3
                - \frac{(2 d^2-17 d+36)}{3} t^2
                + \frac{2 (2 d^2-17 d+35)}{3} t\eqnbreak
                +\frac{8(4 d^2-31 d+59)}{3} \,,\\
        \kappa'_5(d) \varepsilon_5^5(d;t)
            &= -\frac{(d-6) (d-4)^2}{3} t^3
                + 2 (d^3-15 d^2+74 d-120) t^2 \eqnbreak
                -\frac{8(9 d^3-121 d^2+536 d-780)}{3} t
                + \frac{32(4 d^3-52 d^2+223 d-315)}{3}\,,\\
        \kappa'_5(d) \varepsilon_5^6(d;t)
            &= -\frac{2(d-4)^2}{3} t^4
                + \frac{8(2 d^2-17 d+36)}{3} t^3
                -\frac{16 (4 d^2-35 d+75)}{3} t^2\eqnbreak
                + \frac{64(2 d^2-17 d+35)}{3} t\,,                
    \end{align}%
\end{subequations}
with $\kappa_5(d;t)\coloneq (d-4)(3d-14)(3d-10)(t-4)^2$ and $\kappa'_5(d)\coloneq(d-4)^2(d-3)^2(3d-14)(3d-10)$.
For $E_6$, they are
\begin{subequations}\label{eq:dimshift-E6}
    \begin{align}
        \kappa_6(d;t) \tau_6^{30}(d;t)
            &= \frac{(d-4)^2 (3 d-10)}{192 (d-3)} t^3
                + \frac{-3 d^3+37 d^2-150 d+200}{96 (d-3)} t^2\eqnbreak
                + \frac{-21 d^3+274 d^2-1185 d+1690}{24 (d-3)} t
                + \frac{24 d^3-311 d^2+1329 d-1870}{6 (d-3)}\,,\\
        \kappa_6(d;t) \tau_6^{21}(d;t)
            &=  -\frac{(d-4)^2 (3 d-10)}{96} t^3
                + \frac{(3 d^3-37 d^2+150 d-200)}{48} t^2\eqnbreak
                + \frac{1}{12} (45 d^3-594 d^2+2585 d-3690) t
                - 18 d^3+233 d^2-993 d+1390\,,\\
        \kappa_6(d;t) \tau_6^{12}(d;t)
            &= -\frac{4 (d-5)^2 (d-3) (3 d-10)}{3} t
                + \frac{4 (d-5) (d-3) (3 d-10) (5 d-23)}{3}\,,\\
        \kappa'_6(d) \varepsilon_6^0(d;t)
            &= -\frac{(d-4)^2}{3} 
                + \frac{2 (d-4) (9 d^2-87 d+206)}{9 (3 d-10)}\frac{1}{t}\,,\\
        \kappa'_6(d) \varepsilon_6^1(d;t)
            &= \frac{6 d^3-79 d^2+346 d-504}{288 (d-5)} t^2
                + \frac{6 d^3-73 d^2+291 d-378}{144 (d-5)} t\eqnbreak
                + \frac{63 d^3-714 d^2+2670 d-3292}{36 (3 d-10)}
                + \frac{4 (36 d^3-465 d^2+1997 d-2850)}{9 (3 d-10)} \frac{1}{t}\,,\\
        \kappa'_6(d) \varepsilon_6^2(d;t)
            &= \frac{d^2-10 d+24}{288 (d-5)} t^3
                + \frac{-19 d^2+175 d-398}{144 (d-5)} t^2 
                + \frac{-11 d^2+93 d-186}{36 (d-5)} t\eqnbreak
                - \frac{2 (3 d^2+14 d-88)}{9 (3 d-10)}
                -\frac{16 (27 d^2-225 d+466)}{9 (3 d-10)} \frac{1}{t}\,,\\
        \kappa'_6(d) \varepsilon_6^3(d;t)
            &= \frac{4-d}{72 (d-5)} t^3
                + \frac{7 d-29}{36 (d-5)} t^2
                + \frac{5 d-17}{9 (d-5)} t
                - \frac{16}{9}\,,\\
        \kappa'_6(d) \varepsilon_6^4(d;t)
            &= -\frac{(d-4) (3 d-14)}{24} t^2
                + \frac{(3 d-14) (3 d-13)}{12} t
                - \frac{(3 d-14) (d^2-7 d+14)}{3 (3 d-10)}\,,\\
        \kappa'_6(d) \varepsilon_6^5(d;t)
            &= \frac{(d^2-10 d+24)}{6} t^2
                - \frac{2(d^2-11 d+30)}{3} t
                + \frac{4(2 d^2-19 d+45)}{3} \,,\\
        \kappa'_6(d) \varepsilon_6^6(d;t)
            &= \frac{d-4}{3} t^3
                - \frac{2(3 d-13)}{3} t^2
                + \frac{8 (d-5)}{3} t\,,
    \end{align}%
\end{subequations}
with $\kappa_6(d;t)\coloneq (d-5)(d-4)(3d-14)(3d-10)(t-4)^2$ and $\kappa'_6(d)\coloneq (d-4) (d-3)^2 (3 d-14)$.

Note that, unlike $\varepsilon_0^j$ through $\varepsilon_4^j$, the coefficients $\varepsilon_5^j(d;t)$ and $\varepsilon_6^j(d;t)$ diverge at $d=4$.
This is why the $E_i$ appear in the divergent parts of $E_5(4-2\epsilon;t)$ and $E_6(4-2\epsilon;t)$, the implications of which are discussed in \cref{sec:schouten}.
Note also that $\tau_5^{ab}$ and $\tau_6^{ab}$ diverge at the 2-pion threshold, $t=4$.
The Schouten relations also remedy this potentially problematic behavior.

\section{Expressions for the master integrals in  $d=4-2\epsilon$ dimensions} \label{app:expr-4d}

In this section, we give the expressions for the master integrals in $d=4-2\epsilon$ dimensions that are needed for the evaluation of the amplitude.

\subsection{The one-loop bubble in $d=4-2\epsilon$}\label{app:bubble}      
Restating \cref{eq:bubble,eq:Jbub}, we express the bubble as
\begin{equation}
    \Mpi^{-4}I_\bub(4-2\epsilon; t) = \frac{i\exp\Big[\sum_{k=2}^\infty (-\epsilon)^k \frac{\zeta(k)}{k}\Big]}{\epsilon} \left[1 + \sum_{n=1}^\infty (-\epsilon)^n \frac{\Jbub{n}(t)}{n!}\right]\,,
\end{equation}
where $\zeta(k)$ is the Riemann zeta function and
\begin{equation}
    \Jbub{n}(t) \coloneq \int_0^1 \d x\, \log^n\big[1 - x(1-x)t\big]\,.
  \end{equation}
Setting $t=4/(1-\beta^2)$, in the region where $t<0$ we have  $\beta^2>1$ and choose the branch $\beta>1$.  In terms of this, the differential equation~\cref{eq:diffeq-Jbub} is
\begin{equation}\label{eq:diffeq-Jbub-beta}
    (1-\beta^2)\bigg( \beta\frac{\p}{\p\beta}-1\bigg)\Jbub{n}(\beta)=2n\Jbub{n-1}(\beta), \qquad \Jbub{0}(\beta)=1\,.
\end{equation}
This can be solved in terms of the polylogarithms~\cite{NIST:DLMF,NIST:2512}
\begin{equation}
  \Li_{1}(z)\coloneq -\log(1-z)\,, \qquad 
  \Li_{r+1}(z)\coloneq\int_0^z \frac{\d t}{t}\,\Li_{r}(t)\,,
\end{equation}
and the $\epsilon$ orders needed for the evaluations in
this work read 
\begin{subequations}
    \begin{align}
        \label{eq:Jbub-1}
        \Jbub{1}(\beta) 
            &= \beta \Big[\log(\beta_+)-\log(\beta_-)\Big]-2\,,\\
        \label{eq:Jbub-2}
        \Jbub{2}(\beta) 
            &= 8 - 2\beta\big[f_2(\beta_+)-f_2(\beta_-)\big]\,,
            \eqnbreak\qquad
            f_2(z)\coloneq \Li_2(z)+\tfrac12\log(z)^2+2\log(z)\,,\\
        \label{eq:Jbub-3}
        \Jbub{3}(\beta)
            &= -48 + 12\beta\big[f_3(\beta_+)-f_3(\beta_-)\big]
            +3\beta \log(\beta_+)\log(\beta_-) \big[\log(\beta_+)-\log(\beta_-)\big]\,,
            \eqnbreak\qquad
            f_3(z)\coloneq\Li_3(z)+\tfrac1{12}\log(z)^3-\tfrac{\pi^2}{12}\log(z) + f_2(z)\,,
    \end{align}
\end{subequations}
where we have introduced the notation
\begin{equation}
    \beta_+\coloneq \frac{\beta+1}{2\beta}\in\big[1,\tfrac12\big]; \quad \beta_-\coloneq \frac{\beta-1}{2\beta}\in \big[0,\tfrac12\big], \qquad \beta\geq1.
\end{equation}
Although derived for $t<0$, the expressions given here hold in the entire complex plane except for a branch cut along $\beta\in[-1,1]$, i.e., $t\in[4,\infty)$, as is expected from the cut of the amplitude.

Around $\beta=\infty$ (i.e., $t=0$), $\Jbub{n}(\beta)$ has the series expansions
\begin{subequations}\label{eq:Jbub-inf}
    \begin{alignat}{2}
        \Jbub1(\beta) 
            &= \sum_{n=1}^\infty \frac{2}{(2n+1)\beta^{2n}}
            &&= \sum_{n=1}^\infty \frac{-t^n\, n!(n-1)!}{(2n+1)!}\,,\\
        \Jbub2(\beta) 
            &= \frac{8}{15\beta^4} + \frac{64}{105\beta^6} + \frac{568}{945\beta^8} + \O\big(\beta^{-10}\big)
            &&= \frac{t^2}{30} + \frac{t^3}{140} + \frac{11 t^4}{7560} + \O(t^5)\,,\\
        \Jbub3(\beta) 
            &= \frac{16}{35\beta^6} + \frac{16}{21\beta^8} + \O\big(\beta^{-10}\big)
            &&= -\frac{t^3}{140} - \frac{t^4}{420} + \O(t^5)\,.
    \end{alignat}
\end{subequations}

\subsection{The tadpole-bubble products in $d=4-2\epsilon$ dimensions} \label{app:tadbub-4d}

In this section, we state the master integrals defined in \cref{eq:tadbub-2loop,eq:tadbub-3loop}, which following \cref{eq:tadbub} can be expressed in terms of the one-loop tadpole [\cref{eq:tadpole}] and bubble [\cref{eq:bubble,eq:Jbub}] integrals.
We state their $\epsilon$ expansion to the order needed in this work, which is driven by the presence of $1/\epsilon$ counterterms in the two-loop diagrams, and the presence of $T_{30}$ in the divergences of the $E_i$.
The two-loop masters are
\begin{subequations}\label{eq:tadbub-2loop-4d}
    \begin{align}
        \Mpi^{-4}T_{20}(4-2\epsilon)
            &= \frac{1}{\epsilon^2}
            + \frac{2}{\epsilon}
            + \Big[3 + \tfrac{\pi^2}{6}\Big]
            + \Big[4 + \tfrac{\pi^2}{3} - \tfrac23\zeta(3)\Big]\epsilon
            + \O(\epsilon^2)\,,\\
        \Mpi^{-4}T_{11}(4-2\epsilon;t)
            &= \frac{1}{\epsilon^2}
            + \frac{1 - \Jbub1(t)}{\epsilon}
            + \Big[1 + \tfrac{\pi^2}{6} - \Jbub1(t) + \tfrac12 \Jbub2(t)\Big]\eqnbreak
            + \Big[1 + \tfrac{\pi^2}{6} - \tfrac23\zeta(3) - \big(1 + \tfrac{\pi^2}{6}\big)\Jbub1(t) + \tfrac12\Jbub2(t) - \tfrac16\Jbub3(t)\Big]\epsilon
            \eqnbreak+ \O(\epsilon^2)\,,\\
        \Mpi^{-4}T_{02}(4-2\epsilon;t)
            &= \frac{1}{\epsilon^2}
            - \frac{2\Jbub1(t)}{\epsilon}
            + \Big[\tfrac{\pi^2}{6} + \big[\Jbub1(t)\big]^2 + \Jbub2(t)\Big]\eqnbreak
            - \Big[\tfrac23\zeta(3) + \big[\tfrac{\pi^2}{3} - \Jbub2(t)\big]\Jbub2(t) + \tfrac13\Jbub3(t)\Big]\epsilon
            + \O(\epsilon^2)\,,
    \end{align}%
\end{subequations}
and the three-loop ones are
\begin{subequations}\label{eq:tadbub-3loop-4d}
    \begin{align}
        \Mpi^{-6}T_{30}(4-2\epsilon) 
            &= \frac{1}{\epsilon^3}
            + \frac{3}{\epsilon^2}
            + \frac{24+\pi^2}{4 \epsilon}
            + \Big[10 + \tfrac{3\pi^2}{4} - \zeta(3)\Big]
            + \Big[15 + \tfrac{3\pi^2}{2} + \tfrac{19\pi^4}{480} - 3\zeta(3)\Big]\epsilon\eqnbreak
            + \Big[21 + \tfrac{5\pi^2}{2} + \tfrac{19\pi^4}{160} - \big(6 + \tfrac{\pi^2}{4}\big)\zeta(3) - \tfrac{3}{5}\zeta(5)\Big]\epsilon^2
            + \O(\epsilon^3)\,,\\
        \Mpi^{-6}T_{21}(4-2\epsilon;t)
            &= \frac{1}{\epsilon^3}
            + \frac{2-\Jbub1(t)}{\epsilon^2}
            + \frac{12 +\pi^2 - 8\Jbub{1}(t) +2 \Jbub{2}(t)}{4\epsilon}\eqnbreak
            + \Big[4 + \tfrac{\pi^2}{2} - \zeta(3) - \big(3 + \tfrac{\pi^2}{4}\big)\Jbub1(t) + \Jbub2(t) - \tfrac16\Jbub3(t)\Big]
            + \O(\epsilon)\,,\\
        \Mpi^{-6}T_{12}(4-2\epsilon;t)
            &= \frac{1}{\epsilon^3}
            + \frac{1 - 2\Jbub1(t)}{\epsilon^2}
            + \frac{4+ \pi^2 - 8\Jbub1(t)+4 \big[\Jbub1(t)\big]^2 +4 \Jbub2(t)}{4\epsilon}\eqnbreak
            + \Big[1 + \tfrac{\pi^2}{4} - \zeta(3) - \big(2 + \tfrac{\pi^2}{2}\big)\Jbub1(t) + \big[\Jbub1(t)\big]^2 + \big[1 - \Jbub1(t)\big]\Jbub2(t)\eqnbreak\qquad - \tfrac13 \Jbub3(t)\Big]
            + \O(\epsilon)\,,\\
        \Mpi^{-6}T_{03}(4-2\epsilon;t)
            &= \frac{1}{\epsilon^3}
            - \frac{3\Jbub1(t)}{\epsilon^2}
            + \frac{\pi^2 + 12\big[\Jbub1(t)\big]^2 + 6\Jbub2(t)}{4\epsilon}\eqnbreak
            - \Big[\zeta(3) + \big[\Jbub1(t)\big]^3 + 3\big[\tfrac{\pi^2}{4} - \Jbub2(t)\big]\Jbub1(t) + \tfrac12\Jbub3(t)\Big]
            + \O(\epsilon)\,.
   \end{align}%
 \end{subequations}

\subsection{The master integral $E_0$ in $d=4-2\epsilon$} \label{app:E0-4d}
The expansion of $E_0(4-2\epsilon)$ is easily worked out from the parametric representation of the integral and using \texttt{HyperInt}~\cite{Panzer:2014caa},
or by evaluating the expression in~\rcite[eq.~(5.48)]{Cacciatori:2023tzp} using that $\omega_d=2\pi^{\frac{d}{2}}/\Gamma(\frac{d}{2})$, 
or by setting $t=0$ in the expressions for $E_1(4-2\epsilon;t)$ below,
or directly from the dimensional shift in \cref{eq:dimshift-E0}, requiring only knowledge of $T_{30}(4-2\epsilon)$.
In any case,
\begin{multline}\label{eq:E0-4d}
    \Mpi^{-6}E_0(4-2\epsilon)
        = \frac{2}{\epsilon^3}
        + \frac{23}{3\epsilon^2}
        + \frac{\pi^2 + 35}{2\epsilon}
        + \frac{275 + 23\pi^2 - 24\zeta(3)}{12}\\
        + \left(\frac{89 \zeta (3)}{3}+\frac{19 \pi ^4}{240}+\frac{35 \pi ^2}{8}-\frac{189}{8}\right)\epsilon
        + \O(\epsilon^2)\,.
\end{multline}

\subsection{The master integrals $E_1$, $E_2$ and $E_3$ in $d=4-2\epsilon$} \label{app:E1E2E3-4d}
We expand these master integrals by directly applying the dimension shifting relations in \cref{eq:dimshift-E1,eq:dimshift-E2,eq:dimshift-E3}, utilizing the fact that $E_i(2;t)$ are all finite.
For $E_1$, we get
\begin{equation}\label{eq:E1-4d}
    \Mpi^{-6}E_1(4-2\epsilon;t)
        = \frac{2}{\epsilon^3}
        + \frac{23-t}{3\epsilon^2}
        + \frac{t^2-54 t+18 \pi ^2+630}{36 \epsilon} 
        + \bar E_1(4;t) 
        + \O(\epsilon)\,,
\end{equation}
where the finite piece at order $\epsilon^0$ is
\begin{multline}\label{eq:E1-finite}
    \bar E_1(4;t)
        = \frac{t^3+24 t^2-600 t+896}{288} E_1(2;t)
        - \frac{t^4+12 t^3-792 t^2+3968 t+1536}{288} E_2(2;t) \\
        - \frac{(t-16) (t-4) (t^2+40 t+64)}{144} E_3(2;t)
        + \frac{71 t^2+18t+6102}{216} 
        + \frac{(23-t)\pi^2}{12} - 2\zeta(3)\,.
\end{multline}
Likewise for $E_2$,
\begin{equation}\label{eq:E2-4d}
    \Mpi^{-6}E_2(4-2\epsilon;t)
        = \frac{1}{\epsilon^3}
        + \frac{28-t}{12 \epsilon^2}
        + \frac{24-t+2\pi^2}{8\epsilon}
        + \bar E_2(4;t)
        + \O(\epsilon)\,,
\end{equation}
with the finite piece
\begin{multline}\label{eq:E2-finite}
    \bar E_2(4;t)
        = \frac{5 t^2-80 t+96 }{96} E_1(2;t)
        - \frac{5 t^3-116 t^2+376 t+896}{96} E_2(2;t)\\
        - \frac{(t-16) (t-4) (5 t+16)}{48} E_3(2;t) 
        + \frac{(95 t+112) + (28-t)\pi^2}{48} - \zeta(3)\,,
\end{multline}
and lastly for $E_3$,
\begin{equation}\label{eq:E3-4d}
    \Mpi^{-6}E_3(4-2\epsilon;t)
        = -\frac{5}{6 \epsilon^2}
        + \frac{t-20}{8 \epsilon}
        + \bar E_3(4;t)
        + \O(\epsilon)\,,
\end{equation}
with the finite piece
\begin{multline}\label{e:E3eps}
    \bar E_3(4;t)
        = \frac{t^2-12 t-8}{96} E_1(2;t)
        -\frac{t^3-24 t^2+88 t+160 }{96} E_2(2;t) \\
        -\frac{(t-16)(t-4)(t+4)}{48}E_3(2;t)
        +\frac{39 t-232 - 10\pi^2}{48}\,.
\end{multline}
The expressions for the two-dimensional elliptic master integrals are given in \cref{app:expr-2d}.

\subsection{The master integral $E_4$ in $d=4-2\epsilon$} \label{app:E4-4d}
This follows the same procedure as in the previous section, with the main difference being the appearance of $\Jbub{n}(t)$.
Note the absence of $E_4(2;t)$, as remarked in \cref{app:dimshift-E4}.
\begin{equation}\label{eq:E4-4d}
    \Mpi^{-6}E_4(4-2\epsilon;t)
        = \frac{1}{\epsilon^3}
        + \frac{14-9\Jbub1(t)}{6\epsilon^2}
        + \frac{t + 36 + 3\pi^2 - 51\Jbub1(t) + 9\Jbub2(t)}{12 \epsilon}
        + \bar E_4(4;t)
        + \O(\epsilon)\,,
\end{equation}
with the finite piece 
\begin{multline}\label{eq:E4-finite}
    \bar E_4(4;t)
        = \frac{t^2-20 t+160}{96} E_1(2;t) 
        - \frac{(t^2-28 t+120) (t-16) }{96} E_2(2;t)\\
        - \frac{(t-4)(t-16)^2}{48} E_3(2;t)
        + \frac{25t - 136 + 14\pi^2}{24} - \zeta(3)
        - \frac{59+3\pi^2}{8}\Jbub1(t) + \tfrac{17}{8}\Jbub{2}(t)-\tfrac{1}{4}\Jbub{3}(t)\,. 
\end{multline}

\subsection{The master integral $E_5$ in $d=4-2\epsilon$}\label{app:E5-4d}

The expansion of the elliptic master integral $E_5$ is more subtle, because a direct
application of the dimension shifting relation derived in~\cref{app:dimshift-E5E6} gives $1/\epsilon^2$ and $1/\epsilon$ poles depending on the elliptic master integrals in $d=2$. 
But the presence of the elliptic master integrals would spoil the renormalizability of the amplitude.   
These unwanted contributions are removed using the Schouten relations derived in~\cref{sec:schouten}. 
After using these identities, the $\epsilon$ expansion reads
\begin{equation}\label{eq:E5-4d}
    \Mpi^{-6}E_5(4-2\epsilon;t) 
        =\frac{1}{3\epsilon^3}
        + \frac{1-3\Jbub{1}(t)}{3\epsilon^2}
        + \frac{\big[\Jbub{1}(t)\big]^2 - \Jbub{1}(t) + \tfrac12\Jbub{2}(t) + \tfrac{\pi^2}{12}+\tfrac13}{\epsilon} 
        + \bar E_5(4;t)
        + \O(\epsilon)\,.
\end{equation}
Unfortunately, $\bar E_5(4;t)$ is not simply a combination of $E_n(2;t)$ [as is the case for $\bar E_1(4;t)$, etc.] but also of various $\bar E_n^{(r)}(2;t)$.
This can possibly be simplified using additional Schouten relations, the derivation of which would itself be a daunting task, and $E_5(2;t)$ is also not easy to obtain.
Instead, we use $\bar E_5(4;t)$ as-is and evaluate it by solving its differential equation in terms of $E_1(2;t)$.

Rewritten in terms of $\beta$ such that $\beta^ 2=1-4/t$, the finite piece of the differential equation \cref{eq:diffeq-E5} is%
\footnote{
    Note that in this section, $E_n(d;t)$ and $E_n(d;\beta)$ are two different functions related by $t=4/(1-\beta^2)$.
    In general, when the kinematic argument of $E_n$ (or $\Jbub{n}$, etc.) is a Greek letter, it plays the role of $\beta$, not $t$.}
\begin{equation}\label{eq:diffeq-E5bar}
    \beta^2\bigg[\frac{\p^2}{\p\beta^2} - \frac{2}{\beta^2-1}\bigg]\bar E_5(4;\beta) 
        = \mathscr S_5(\beta)\,.
\end{equation}
Serendipitously, the first-order term vanishes.
The source term $\mathscr S_5$ contains $\bar E_1(4;t)$, $\bar E_4(4;t)$ and their derivatives, but since \cref{eq:E1-finite,eq:E4-finite} expresses these solely in terms of $E_1(2;t)$ and $\Jbub{n}(t)$, we have the decomposition
\begin{equation}\label{eq:S5}
    \mathscr S_5(\beta) = \mathscr S_\rat(\beta) + \mathscr S_J(\beta) + \mathscr S_{E_1}(\beta) 
\end{equation}
with rational piece
\begin{equation}\label{eq:S5-rat}
    \mathscr S_\rat(\beta) \coloneq
        \frac{68 \beta^{4} - 44 \beta^{2} + 24 - \beta^{2} \big(\beta^{2}+5\big) \pi^{2}}{6(\beta^2 -1)^{2}}
        + \frac{2 \beta^{2} \zeta(3)}{3(\beta^2 -1)}\,,
\end{equation}
$\Jbub{n}$-dependent (i.e., polylogarithmic) piece
\begin{multline}\label{eq:S5-J}
    \mathscr S_J(\beta)\coloneq 
        \frac{\beta^{2}}{3(\beta^{2}-1)} \Jbub{3}(\beta) 
        + \frac{2}{\beta^2-1}\Jbub{1}(\beta)\Jbub{2}(\beta)
        - \frac{\beta^{4}+5}{(\beta^2-1)^2}\Jbub{2}(\beta)
        - \frac{12}{(\beta^{2}-1)^2} \Jbub{1}(\beta)^{2} \\
        + \frac{(\pi^2-4)\beta^4+(8-\pi^2)\beta^2-52}{2(\beta^{2}-1)^2}\Jbub{1}(\beta)\,,
\end{multline}
and $E_1$-dependent (i.e., elliptic) piece
\begin{equation}\label{eq:S5-E1}
    \begin{gathered}
    \mathscr S_{E_1}\coloneq\bigg[
          s_2(\beta)\frac{\p^2}{\p\beta^2}
        + s_1(\beta)\frac{\p}{\p\beta}
        + s_0(\beta)\bigg]E_1(2;\beta)\,,\\
    s_0(\beta) = \frac23\,  \frac{\beta^2(2\beta^2-1)}{(\beta^2-1)^2}\,,\quad
    s_1(\beta) = \frac{2\beta}{3}\frac{3\beta^4-7\beta^2+3}{\beta^2-1}\,,\quad
    s_2(\beta) = \frac{\beta^2(4\beta^2-3)}{2}\,.
    \end{gathered}
\end{equation}
We solve this differential equation using the Wrońskian method.
A basis of solutions to the homogeneous equation
[i.e., that obtained by setting $\mathscr S_5(\beta)=0$ in \cref{eq:diffeq-E5bar}] is
\begin{equation}\label{eq:E5-homogeneous}
    g_1(\beta) \coloneq \beta^2-1 
    \qquad\text{and}\qquad
    g_2(\beta) \coloneq \frac{\beta^2 - 1}{4}\log\bigg(\frac{\beta+1}{\beta-1}\bigg)-\frac{\beta}{2}\,,
\end{equation}
and the generic solution is then
\begin{equation}\label{eq:E5-generic}
    \bar E_5(4;\beta) = 
        c_1\, g_1(\beta) + c_2\, g_2(\beta)
        - g_1(\beta)\int_{\xi_1}^{\beta} \mathscr S_5(\xi) g_2(\xi) \frac{\d \xi}{\xi^2}
        + g_2(\beta)\int_{\xi_2}^{\beta} \mathscr S_5(\xi) g_1(\xi) \frac{\d \xi}{\xi^2}\,.
\end{equation}
The limits $\xi_1$ and $\xi_2$ are arbitrary as long as the constants $c_1$ and $c_2$ are adjusted accordingly, which can be done by ensuring that the $\beta=\infty$ (i.e., $t=0$) limit from \cref{app:initial} is reproduced. 
Such adjustment is very subtle in practice, and one particular implementation will be discussed in forthcoming work~\cite{LLSV}.

The elliptic piece would be simpler if it only involved $E_1$, not its derivatives, so we apply integration by parts.
Defining
\begin{equation}\label{eq:E5-h}
    h_n(\xi) \coloneq 
          \frac{\p^2}{\p \xi^2}\frac{s_2(\xi)g_n(\xi)}{\xi^2}
        - \frac{\p}{\p \xi}\frac{s_1(\xi)g_n(\xi)}{\xi^2}
        + \frac{s_0(\xi)g_n(\xi)}{\xi^2}\,,
\end{equation}
or, explicitly,
\begin{align}
    \label{eq:h1-h2}
    h_1(\xi) &= \frac{54 \xi^6-57 \xi^4+11 \xi^2-6}{3\xi^2(\xi^2-1)}\,,\\
    h_2(\xi) &= \frac{ (\xi^2 -1) \big(54 \xi^{6}-57 \xi^{4}+11 \xi^{2}-6\big)\log\Big(\frac{\xi+1}{\xi-1}\Big)
        -2 \xi  \big(54 \xi^{6}-93 \xi^{4}+31 \xi^{2}+6\big)
        }{12(\xi^2-1)^2 \xi^2}\,,\notag
\end{align}
we have
\begin{multline}\label{eq:E5-ibp}
      g_2(\beta)\int_{\xi_2}^{\beta} \mathscr S_{E_1}(\xi) g_1(\xi) \frac{\d \xi}{\xi^2}
    - g_1(\beta)\int_{\xi_1}^{\beta} \mathscr S_{E_1}(\xi) g_2(\xi) \frac{\d \xi}{\xi^2}
    \\= 
      g_2(\beta)\int_{\xi_2}^{\beta} h_1(\xi) E_1(2;\xi)\d\xi
    - g_1(\beta)\int_{\xi_1}^{\beta} h_2(\xi) E_1(2;\xi)\d\xi
    + \frac{s_2(\beta)}{\beta^2}E_1(2;\beta) + \Delta_{E_1}\,,
\end{multline}
where $\Delta_{E_1}$ incorporates all boundary terms from the lower integration limit; we will omit it from now on since it can be absorbed into $c_1g_1(\beta) + c_2g_2(\beta)$.
At the upper integration limit, all boundary terms cancel except for $\frac{s_2(\beta)}{\beta^2}\big[g_1(\beta)g'_2(\beta) - g'_1(\beta)g_2(\beta)\big] E_1(2;\beta)$, where the factor in brackets is equal to $1$.

For brevity in the following, we define
\begin{equation}\label{eq:E5-I}
    \begin{aligned}
        \I[\beta; G_1,G_2] \coloneq
            &-G_1\int_{\xi_1}^{\beta} \bigg[\frac{\mathscr S_J(\xi) + \mathscr S_\rat(\xi)}{\xi^2}g_2(\xi) + E_1(2;\xi)h_2(\xi) \bigg] \d\xi\\
            &+G_2\int_{\xi_2}^{\beta} \bigg[\frac{\mathscr S_J(\xi) + \mathscr S_\rat(\xi)}{\xi^2}g_1(\xi) + E_1(2;\xi)h_1(\xi) \bigg] \d\xi\,.
    \end{aligned}
\end{equation}
The reason for introducing $G_1,G_2$ will become apparent in \cref{eq:dE5dt}.
Thus,
\begin{equation}\label{eq:E5-main}
    \bar E_5(4;\beta) = c_1g_1(\beta) + c_2g_2(\beta) + \frac{s_2(\beta)}{\beta^2}E_1(2;\beta) + \I\big[\beta; g_1(\beta),g_2(\beta)\big]\,.
\end{equation}
While this form of $E_5(4;\beta)$ involving a one-dimensional integral is much more tractable than the original $12$-dimensional Feynman integral,
the evaluation of $\I[\beta; G_1,G_2]$ is a formidable task in and of itself, not least because it depends on the ability to evaluate $E_1(2;\xi)$ across the entire integration range.
This is detailed in forthcoming work~\cite{LLSV}.

\subsection{The master integral $E_6$ at $d=4-2\epsilon$}\label{app:E6-4d}

As for $E_5$, the Schouten relations must be applied to the $\epsilon$ expansion of this integral.
After doing that, it reads
\begin{equation}\label{eq:E6-4d}
    \Mpi^{-6}E_6(4-2\epsilon;t)
        = \frac{2+\Jbub{1}(t)}{2 (t-4)\epsilon^2}
        - \frac{4\big[\Jbub{1}(t)\big]^2 + 10\Jbub{1}(t) + \Jbub{2}(t) - 4}{4(t-4)\epsilon}
        + \bar E_6(4;t) + \O(\epsilon)\,,
\end{equation}
where, using the relation~\cref{eq:E6toE5},
\begin{align}\label{eq:E6-finite}
    \bar E_6(4;t)
        &= \bigg[
            \frac{t-16}{48}\frac{\p^2}{\p t^2}
            - \frac{t+16}{24t}\frac{\p}{\p t}
            + \frac{1}{24t}
            \bigg]\bar E_1(4;t)
        + \frac{\bar E_4(4;t)}{2t}
        - \bigg[
            \frac{\p}{\p t} + \frac{1}{t}
            \bigg] \bar E_5(4;t)\eqnbreak
        + \frac{2\Jbub3(t) - 39\Jbub2(t) + 3(\pi^2+67)\Jbub1(t)}{48t}
        + \frac{8\zeta(3) - 11\pi^2}{32t}
        + \frac{3t^2 - 184t - 5490}{1728t}\,,
\end{align}
or in terms of $E_1(2;t)$  
using the expression for $\bar E_4(4;t)$ in \cref{eq:E4-finite},
\begin{align}\label{eq:E6-finiteE12d}
    \bar E_6(4;t)
          &= 
        \bigg[
            \frac{(t-16)(t-4)}{12t^2}\Big(t\frac{\p}{\p t }\Big)^{\kern-2pt2}
            + \frac{t-10}{12t} \Big(t\frac{\p}{\p t }\Big)
            + \frac{1}{24}
            \bigg] E_1(2;t)
        - \bigg[
            \frac{\p}{\p t} + \frac{1}{t}
            \bigg] \bar E_5(4;t)\eqnbreak
        + \frac{-2\Jbub3(t) + 6\Jbub2(t) + 3(4 - \pi^2)\Jbub1(t)}{24t}
        + \frac{\pi^2 - 4\zeta(3) - 20}{12t}\,.
\end{align}
The derivative of $\bar E_5(4;t)$ can be extracted via $\frac{\p}{\p t}=\frac{(\beta^2-1)^2}{8\beta}\frac{\p}{\p\beta}$, \cref{eq:E5-4d} and \cref{eq:E5-main}:
\begin{equation}\label{eq:dE5dt}
    \frac{\p\bar E_5(4;t)}{\p t}
        = 
         \bigg[\frac{t-16}{2t^2}\Big(t\frac{\p}{\p t }\Big) + \frac{t^2-28t+48}{3t^2(t-4)}\bigg]E_1(2;t)
         + \frac{4}{t^2}\Big( c_1 + \tfrac1{4\beta}\Jbub1(\beta)c_2 + \I\big[\beta;1, \tfrac1{4\beta}\Jbub1(\beta)\big]\Big)\,,
\end{equation}
where the integral function $\I$ is the same as in \cref{eq:E5-I}.
As for $\bar E_5$, the remainder of the calculation is deferred to forthcoming work~\cite{LLSV}.

\section{Expressions for the master integrals in two dimensions}\label{app:expr-2d}

At this point, all master integrals have been rewritten in terms of $\Jbub{n}(t)$ (given in \cref{app:tadbub-4d}) and $E_1(2;t)$, so it remains to describe this ``master of master integrals''.
It has been extensively studied and is known in several different forms, among which is the Bessel integral representation~\cite{Groote:2005ay,Bailey:2008ib,Bloch:2014qca}, which holds for $t<16$:
\begin{equation}\label{eq:E1-bessel}
    E_1(2;t) = -8\int_0^\infty x I_0(x \sqrt{t}) \big[K_0(x)\big]^4 \d x\,,
\end{equation}
where $I_\alpha(z)$ and $K_\alpha(z)$ are the modified Bessel functions of the first and second kind, respectively.
This has as a special case $E_0(2)=E_1(2;0)=7\zeta(3)$.

Combining \cref{eq:E1-bessel} with \cref{eq:E2toE1,eq:E3toE1} gives
\begin{align}
    E_2(2;t) &= 2\int_0^\infty x 
        \Big\{
            I_0(x\sqrt{t}) 
            + \tfrac{x\sqrt{t}}{2} I_1(x\sqrt{t})
            \Big\}
        \big[K_0(x)\big]^4 \d x\,,\label{eq:E2-bessel}\\
    E_3(2;t) &= -\frac12\int_0^\infty x 
        \Big\{
            (2+x^2) I_0(x\sqrt{t}) 
            + \tfrac{x(t+2)}{\sqrt{t}} I_1(x\sqrt{t})
            + x^2 I_2(x\sqrt{t})
            \Big\}
        \big[K_0(x)\big]^4 \d x\,.\label{eq:E3-bessel}
\end{align}
When $t<0$, the integrals are kept real by $\sqrt{t}=i\sqrt{|t|}$ and $I_\alpha(iz) = i^\alpha J_\alpha(z)$, where $J_\alpha$ is the (unmodified) Bessel function of the first kind.

The Bessel integral formulation is straightforward but limited to $t<16$ and numerically slow.
The formulation in terms of elliptic functions that is the main result of \rcite{Bloch:2014qca} is applicable everywhere in the complex plane, but is much more complicated to describe, requiring much higher mathematics than what is used elsewhere in this paper.
It is also given in terms of variables whose relationship to our $t$ is highly nontrivial, and, while known in the literature~\cite{Bloch:2014qca}, requires adjustments before being practical.
To keep the scope of the present paper manageable, we therefore defer its description to forthcoming work~\cite{LLSV}.

\section{Unsubtracted expressions for the HVP}\label{app:unsub-results}

Only the on-shell renormalized result $\bar\Pi_T(q^2) \coloneq \Pi_T(q^2)-\Pi_T(0)$ is presented in the main text, but for completeness, we give the unsubtracted result $\Pi_T(q^2)$ in this appendix.
Arranging the chiral power counting as in \cref{eq:main}, the unsubtracted lower-order expressions [compare \cref{eq:PiT-NLO,eq:PiT-NNLO}] are
\begin{align}
    \label{eq:PiT-NLO-unsub}
    \Pi_T^\NLO(t) &= 
        2\bigg[\frac{4-t}{3t}\Jbub1(t) 
        - \frac{1+3\Lpi}{9}\bigg]
        - 4l_5^q - 8h_2^q\,,\\
    \label{eq:PiT-NNLO-unsub}
    \Pi_T^\NNLO(t) &= 
        t\bigg[\frac{4-t}{3t}\Jbub1(t) 
            - \frac{1+3\Lpi}{9}\bigg]^2
        - 4t\bigg[
            \frac{4-t}{3t}\Jbub1(t)
            - \frac{1+3\Lpi}{9}
            \bigg] l_6^q\eqnbreak
        + 8\Lpi\big(2l_5^q - l_6^q\big)
        - 8t c_{56}^q
        - 32 c_{34}^q\,.
\end{align}
From this, \cref{eq:PiT-NLO,eq:PiT-NNLO} are obtained by taking the small-$t$ limit with \cref{eq:Jbub-inf}:
Since \mbox{$\Jbub1(t)=-\frac{t}{6}+\O(t^2)$}, we have
\begin{equation}
    \Pi_T^\NLO(0)   = -\frac{5+3\Lpi}{9} - 4l_5^q - 8h_2^q\,,\qquad
    \Pi_T^\NNLO(0)  = 8\Lpi\big(2l_5^q - l_6^q\big) - 32 c_{34}^q\,.
\end{equation}
The same procedure [ignoring \mbox{$\Jbub2(t)=\O(t^2)$} and \mbox{$\Jbub3(t)=\O(t^3)$}] shows why the constant $(4-\pi^2)/36$ needs to be subtracted in \cref{eq:PiJ} so that $\bar\Pi_J(t)=0$.
Similarly for \cref{eq:PiE}, the subtraction needed for $\bar\Pi_E(0)=0$ is found using \cref{eq:elliptics-t0}.
For the remaining \NNNLO\ expression, we write
\begin{equation}
    \Pi_T^\NNNLO(t) = \bar\Pi_T^\NNNLO(t) + \Pi_T^\NNNLO(0)\,,\qquad
    \Pi_T^\NNNLO(0) = \Pi_\zeta(0) + \Pi_l(0) + \Pi_c(0)  + r_0^q\,,
\end{equation}
where $r_0^q$ is defined in \cref{eq:diagr-00}, and
\begin{align}
    \Pi_\zeta(0)
        &= - \frac{77\zeta(3)}{4}
            + \frac{23\Lpi^2}{3}
            - \frac{703\Lpi}{54}
            + \frac{4753}{432}\,,\\
    \Pi_l(0)
        &= - 4\Lpi^2\big(2l_1^q - l_2^q + 12l_5^q - 6l_6^q\big)
            + \frac{2 - 4\Lpi}{3}\big(l_1^q + 2l_2^q\big)
            + 16\Lpi l_4^q\big(2l_5^q - l_6^q\big)\,,
    \label{eq:Pil-unsub}\\
    \Pi_c(0)
        &= 6c_{31}^q + 6c_{32}^q - 4c_{33}^q - 16c_{44}^q + 64 (l_3^q - l_4^q) c_{34}^q - \Big(
            192c_6^q + 48c_{29}^q - 16c_{30}^q + 12c_{31}^q \eqnbreak\qquad
            + 12c_{32}^q - 8c_{33}^q
            - 192c_{34}^q + 32c_{35}^q - 32c_{44}^q
            + 64c_{46}^q - 128c_{47}^q + 32c_{50}^q \Big) \Lpi\,.
    \label{eq:Pic-unsub}
\end{align}

\section{Implementation details}\label{app:implementation}

\newcommand{\chptlib}{\texttt{ChPTlib}}
\newcommand{\chptprog}{\texttt{ChPT.py}}

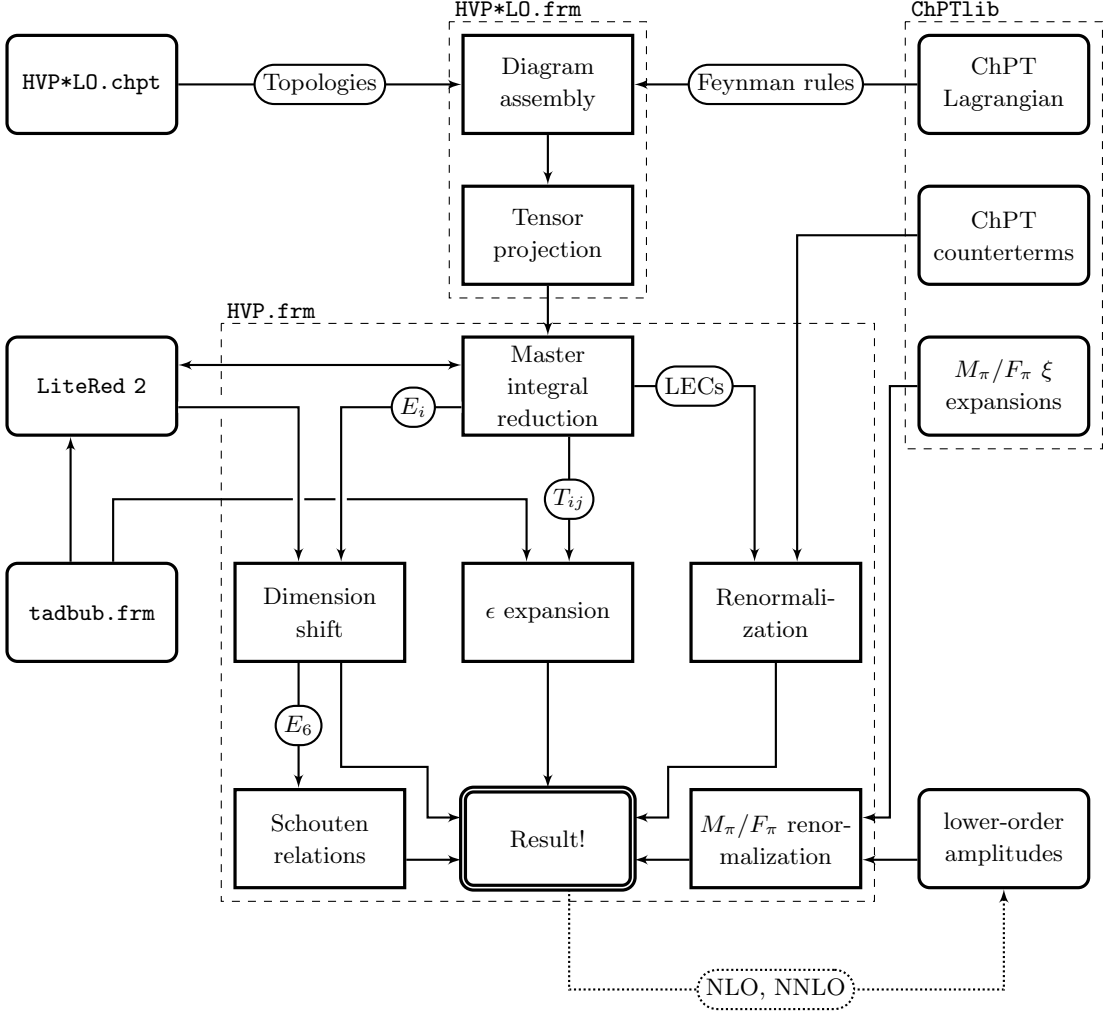
\begin{figure}[tp]
    {\footnotesize\input{implementation.tex}}
    \caption{
        A sketch of the steps involved in the HVP calculations, and the programs performing these steps.
        For NLO and NNLO, only a subset of the steps are of course performed.
        With the exception of third-party programs like \form~\formrefs\ and \texttt{LiteRed 2}~\literedrefs, all necessary code can be found in the linked repositories~\cite{chptlib,hvpfiles}.}
    \label{fig:implementation}
\end{figure}

The calculation of the amplitude was carried out using the \form\ library \chptlib~\cite{chptlib}, which during the preparation of this work evolved from Mattias Sjö's personal \form\ files (themselves inspired by Johan Bijnens' ditto) into something more generally usable (albeit still limited and under development).
We therefore choose to make it available, along with the implementation of the present HVP calculation~\cite{hvpfiles}, so that the reader may reproduce our results.
In this appendix, we briefly describe how to do that.
A graphical outline is given in \cref{fig:implementation}, and a step-by-step summary in the provided \texttt{README} file.%
\footnote{
    Questions regarding the implementation should primarily be directed to Mattias Sjö.}

\chptlib\ is mainly a collection of \form\ procedures, containing an implementation of the bare and renormalized \chpt\ Lagrangian~\cite{Gasser:1983yg,Gasser:1984gg,Bijnens:1999hw,Bijnens:1999sh,Bijnens:2018lez} and utilities for processing \chpt\ amplitudes.
It also contains a Python program, \chptprog\, for automatically generating the necessary \form\ code from a description of the diagram topologies (it is currently not interoperable with automatic Feynman diagram generators).
The diagram description format is intended to be quickly transcribed by hand from a sketch of the diagrams, and to express the momentum routing in a way that directly translates into the $I_{\vec\nu}$ notation of \cref{sec:loops}.
The diagrams in \cref{tab:NLO-NNLO,tab:NNNLO} are found in this format in \texttt{HVP*LO.chpt} (with \texttt{*} standing for \texttt{N}, \texttt{NN} or \texttt{N3}), and running \mbox{\texttt{\chptprog\ --form HVP*LO.chpt}} generates the \form\ code for deriving the Feynman rules, performing all Wick contractions, and symmetrizing.
This is then applied by running the \form\ program \texttt{HVP*LO.frm}, which is generated with \mbox{\texttt{\chptprog\ --form-main HVP*LO.chpt}} and then supplied with, among other things, an implementation of \cref{eq:LT-def}.

With the diagrams generated, all subsequent steps in the calculation are performed by running the \form\ program \texttt{HVP.frm}.
The most time-consuming step is the master integral reduction, which is done by converting the raw amplitude into a Mathematica program that invokes \texttt{LiteRed~2}~\literedrefs, running it, and converting the result back to \form.
After that, the other manipulations described in \cref{sec:more-loops} are performed as outlined in \cref{fig:implementation}.
Of particular note is the standalone file \texttt{tadbub.frm}, which broadcasts the $\epsilon$-expansion (see \cref{app:tadbub-4d}) of the tadpole and bubble integrals, thereby minimizing the risk of clashes between sign and normalization conventions in the various places where this expansion is needed.
The output of \texttt{HVP.frm} is only some \LaTeX\ formatting away from \cref{eq:PiT-NLO,eq:PiT-NNLO,eq:PiT-N3LO,eq:PiE,eq:PiJ,eq:Pil,eq:Pic,eq:Piz}, and most of the crosschecks listed in \cref{sec:results} can be automatically performed by setting various flags listed at the top of the file.

\addcontentsline{toc}{section}{References}
\bibliographystyle{JHEP}
\renewcommand\raggedright{}
\bibliography{references}

\end{document}

%% file: tikz-settings.tex
\newcommand{\tadpoleangle}{130}
\newlength{\tadpolesize}
\newlength{\bubblesize}
\newcommand{\bubbleaspect}{.8}
\newlength{\photonlength}
\newcommand{\diagramscale}{1.25}
\newlength{\vertsize}
\tikzset{
    Ultra thick/.style={line width=2pt},
    miter cap/.style={{Triangle Cap[]}-{Triangle Cap[]}},
    pion/.style = { very thick },
    photon/.style = { thick, decorate, decoration={snake,segment length=.26\photonlength, amplitude=+.05\photonlength} },
    notoph/.style = { thick, decorate, decoration={snake,segment length=.26\photonlength, amplitude=-.05\photonlength} },
    diagrcut/.style={red,decorate,decoration={zigzag,segment length=1mm,amplitude=.25mm}},
    bubble2/.style = {xscale=.8, transform shape},
    elbbub2/.style = {xscale=1.25, transform shape},
    bubble3/.style = {xscale=.64, transform shape},
    threeloop/.style = {scale=1.2, transform shape},
    NLO/.pic = {
        \fill (0,0) circle[radius=\vertsize];
    },
    NNLO/.pic = {
        \fill (-\vertsize,-\vertsize) rectangle (\vertsize, \vertsize);
    },
    NNNLO/.pic = {
        \fill (90:2\vertsize) -- (210:2\vertsize) -- (330:2\vertsize) -- cycle;
    },
    pics/tadpole/.style = {
        code = {
            \draw[pion] (0,0) .. controls (\tadpoleangle:\tadpolesize) and (180-\tadpoleangle:\tadpolesize) .. (0,0)
                foreach \i in {#1} { node[pos=.\i] (node\i) {} }
                ;
        }
    },
    pics/bubble/.style = {
        code = {
            \draw[pion] (0,0) ellipse [x radius=\bubblesize, y radius=\bubbleaspect\bubblesize];
            \foreach \i in {#1} {
                \coordinate (node\i) at
                (canvas polar cs:angle=\i, x radius=\bubblesize, y radius=\bubbleaspect\bubblesize) {};
            }
        }
    },
    pics/bubble/.default = {0}, 
    bubblejoint/.pic = {
        \fill[white] (-\bubblesize,-\bubblesize) rectangle (\bubblesize,\bubblesize);
        \draw[pion] (-\bubblesize,+\bubbleaspect\bubblesize) .. controls (0,+\bubbleaspect\bubblesize) and (0,-\bubbleaspect\bubblesize) .. (+\bubblesize,-\bubbleaspect\bubblesize);
        \draw[pion] (+\bubblesize,+\bubbleaspect\bubblesize) .. controls (0,+\bubbleaspect\bubblesize) and (0,-\bubbleaspect\bubblesize) .. (-\bubblesize,-\bubbleaspect\bubblesize);
    },
}

\newcounter{diagram}
\newenvironment{diagram}[1][]{%
        \begin{tikzpicture}[scale=\diagramscale, transform shape, baseline={(0,0)},#1]%
        \path[use as bounding box] (-\photonlength,-\bubblesize) rectangle (+\photonlength,+\bubblesize);
    }{%
        \end{tikzpicture}%
    }

\definecolor{plotI}{HTML}{0077BB}
\definecolor{plotII}{HTML}{33BBEE}
\definecolor{plotIII}{HTML}{009988}
\definecolor{plotIV}{HTML}{EE7733}
\definecolor{plotV}{HTML}{CC3311}
\definecolor{plotVI}{HTML}{DDAA33}
\definecolor{plotVII}{HTML}{AA3377}

\colorlet{spacelike}{plotI}
\colorlet{subthr}{plotIV}
\colorlet{twopi}{plotIII}
\colorlet{fourpi}{plotVII}
\colorlet{shaded}{gray!30}

%% file: scriptdiagr.tex
\newcommand{\scriptdiagrscale}{0.3}
\newlength{\scriptdiagrleglen}
\newlength{\scriptdiagrleg}
\newcommand{\tad}{%
    {\tikz[scale=\scriptdiagrscale, transform shape]{%
        \tikzset{pion/.style={ thin }}
        \path[use as bounding box] (-1.5\scriptdiagrleglen,0) rectangle (+1.5\scriptdiagrleglen,.6\tadpolesize);
        \draw[pion] (-1.5\scriptdiagrleglen,0) -- (0,0) pic {tadpole} -- (+1.5\scriptdiagrleglen,0);}}}
\newcommand{\bub}{%
    {\tikz[scale=\scriptdiagrscale, transform shape]{%
        \tikzset{pion/.style={ thin }}
        \draw (0,0) pic {bubble};
        \draw[pion] (0,0) + (-\bubblesize,0) -- +(-\scriptdiagrleg,0);
        \draw[pion] (0,0) + (+\bubblesize,0) -- +(+\scriptdiagrleg,0);
        }}}

%% file: diagram-table-lower.tex
\begin{tabular}{r|ccccc}
    \toprule
    &0&1&2&3&4
    \\
    \midrule
    0\arrayfudge
    &
    \begin{diagram}
        \draw[photon] (-\photonlength,0) -- (0,0)
            pic             {NLO}
            ;
        \draw[notoph] (+\photonlength,0) -- (0,0);
    \end{diagram}
    &
    \begin{diagram}
        \draw[photon] (-\photonlength,0) -- (0,0)
            pic             {tadpole}
            ;
        \draw[notoph] (+\photonlength,0) -- (0,0);
    \end{diagram}
    &
    \begin{diagram}
        \draw (0,0) pic {bubble};
        \draw[photon] (-\photonlength,0) -- (-\bubblesize,0);
        \draw[notoph] (+\photonlength,0) -- (+\bubblesize,0);
    \end{diagram}
    &
    &
    \\
    \midrule
    0\arrayfudge
    &
        \begin{diagram}
            \draw[photon] (-\photonlength,0) -- (0,0)
                pic             {NNLO}
                ;
            \draw[notoph] (+\photonlength,0) -- (0,0);
        \end{diagram}
        &
        \begin{diagram}
            \draw[photon] (-\photonlength,0) -- (0,0)
                pic             {tadpole}
                pic             {NLO};
                ;
            \draw[notoph] (+\photonlength,0) -- (0,0);
        \end{diagram}
        &
        \begin{diagram}
            \draw[photon] (-\photonlength,0) -- (0,0)
                pic             {tadpole={50}}
                ;
            \draw (node50) pic {NLO};
            \draw[notoph] (+\photonlength,0) -- (0,0);
        \end{diagram}
        &
        \begin{diagram}
            \draw (0,0) pic {bubble};
            \draw[photon] (-\photonlength,0) -- (-\bubblesize,0)
                pic                                        {NLO}
                ;
            \draw[notoph] (+\photonlength,0) -- (+\bubblesize,0);
        \end{diagram}
        &
        \begin{diagram}
            \draw (0,0) pic {bubble={90}};
            \draw (node90)
                pic                                        {NLO}
                ;
            \draw[photon] (-\photonlength,0) -- (-\bubblesize,0);
            \draw[notoph] (+\photonlength,0) -- (+\bubblesize,0);
        \end{diagram}
        \\
        1\arrayfudge
        &
        \begin{diagram}
            \draw[photon] (-\photonlength,0) -- (0,0)
                pic             {tadpole}
                pic[rotate=180] {tadpole}
                ;
            \draw[notoph] (+\photonlength,0) -- (0,0);
        \end{diagram}
        &
        \begin{diagram}
            \draw[photon] (-\photonlength,0) -- (0,0)
                pic             {tadpole={50}}
                ;
            \draw (node50) pic[scale=.7] {tadpole};
            \draw[notoph] (+\photonlength,0) -- (0,0);
        \end{diagram}
        &
        \begin{diagram}
            \draw (0,0) pic {bubble};
            \draw[photon] (-\photonlength,0) -- (-\bubblesize,0)
                pic[scale=.7, rotate=+\tadpoleangle-90-10] {tadpole}
                ;
            \draw[notoph] (+\photonlength,0) -- (+\bubblesize,0);
        \end{diagram}
        &
        \begin{diagram}
            \draw (0,0) pic {bubble={90}};
            \draw (node90)
                pic[scale=.7]                              {tadpole}
                ;
            \draw[photon] (-\photonlength,0) -- (-\bubblesize,0);
            \draw[notoph] (+\photonlength,0) -- (+\bubblesize,0);
        \end{diagram}
        &
        \begin{diagram}[bubble2]
            \draw (-\bubblesize,0) pic {bubble};
            \draw (+\bubblesize,0) pic {bubble};
            \draw (0,0) pic {bubblejoint};
            \draw[photon] (-\bubblesize,0) + (-\photonlength,0) -- ++ (-\bubblesize,0);
            \draw[notoph] (+\bubblesize,0) + (+\photonlength,0) -- ++ (+\bubblesize,0);
        \end{diagram}
    \\
    \bottomrule
\end{tabular}

%% file: diagram-table.tex
\begin{tabular}{r|ccccc}
    \toprule
    &0&1&2&3&4
    \\
    \midrule
    0\arrayfudge
    &
    \begin{diagram}
        \draw[photon] (-\photonlength,0) -- (0,0)
            pic             {NNNLO}
            ;
        \draw[notoph] (+\photonlength,0) -- (0,0);
    \end{diagram}
    &
    \begin{diagram}
        \draw[photon] (-\photonlength,0) -- (0,0)
            pic             {tadpole}
            pic             {NNLO}
            ;
        \draw[notoph] (+\photonlength,0) -- (0,0);
    \end{diagram}
    &
    \begin{diagram}
        \draw[photon] (-\photonlength,0) -- (0,0)
            pic             {tadpole={50}}
            ;
        \draw (node50) pic {NNLO};
        \draw[notoph] (+\photonlength,0) -- (0,0);
    \end{diagram}
    &
    \begin{diagram}
        \draw (0,0) pic {bubble};
        \draw[photon] (-\photonlength,0) -- (-\bubblesize,0)
            pic                                        {NNLO}
            ;
        \draw[notoph] (+\photonlength,0) -- (+\bubblesize,0);
    \end{diagram}
    &
    \begin{diagram}
        \draw (0,0) pic {bubble={90}};
        \draw (node90)
            pic                       {NNLO}
            ;
        \draw[photon] (-\photonlength,0) -- (-\bubblesize,0);
        \draw[notoph] (+\photonlength,0) -- (+\bubblesize,0);
    \end{diagram}
    \\
    1\arrayfudge
    &
    \begin{diagram}
        \draw[photon] (-\photonlength,0) -- (0,0)
            pic             {tadpole={50}}
            pic             {NLO};
            ;
        \draw (node50) pic {NLO};
        \draw[notoph] (+\photonlength,0) -- (0,0);
    \end{diagram}
    &
    \begin{diagram}
        \draw[photon] (-\photonlength,0) -- (0,0)
            pic             {tadpole={33,67}}
            ;
        \draw (node33) pic {NLO};
        \draw (node67) pic {NLO};
        \draw[notoph] (+\photonlength,0) -- (0,0);
    \end{diagram}
    &
    \begin{diagram}
        \draw (0,0) pic {bubble};
        \draw[photon] (-\photonlength,0) -- (-\bubblesize,0)
            pic                                        {NLO}
            ;
        \draw[notoph] (+\photonlength,0) -- (+\bubblesize,0)
            pic                                        {NLO}
            ;
    \end{diagram}
    &
    \begin{diagram}
        \draw (0,0) pic {bubble={90}};
        \draw (node90)
            pic                                        {NLO}
            ;
        \draw[photon] (-\photonlength,0) -- (-\bubblesize,0)
            pic                                        {NLO}
            ;
        \draw[notoph] (+\photonlength,0) -- (+\bubblesize,0);
    \end{diagram}
    &
    \begin{diagram}
        \draw (0,0) pic {bubble={90,270}};
        \draw (node90)
            pic                       {NLO}
            ;
        \draw (node270)
            pic                       {NLO}
            ;
        \draw[photon] (-\photonlength,0) -- (-\bubblesize,0);
        \draw[notoph] (+\photonlength,0) -- (+\bubblesize,0);
    \end{diagram}
    \\
    2\arrayfudge
    &
    \begin{diagram}
        \draw (0,0) pic {bubble={60,120}};
        \draw (node60)
            pic                       {NLO}
            ;
        \draw (node120)
            pic                       {NLO}
            ;
        \draw[photon] (-\photonlength,0) -- (-\bubblesize,0);
        \draw[notoph] (+\photonlength,0) -- (+\bubblesize,0);
    \end{diagram}
    &
    \begin{diagram}
        \draw[photon] (-\photonlength,0) -- (0,0)
            pic             {tadpole}
            pic[rotate=180] {tadpole}
            pic             {NLO}
            ;
        \draw[notoph] (+\photonlength,0) -- (0,0);
    \end{diagram}
    &
    \begin{diagram}
        \draw[photon] (-\photonlength,0) -- (0,0)
            pic             {tadpole={50}}
            pic[rotate=180] {tadpole}
            ;
        \draw (node50) pic {NLO};
        \draw[notoph] (+\photonlength,0) -- (0,0);
    \end{diagram}
    &
    \begin{diagram}
        \draw[photon] (-\photonlength,0) -- (0,0)
            pic             {tadpole={50}}
            pic             {NLO}
            ;
        \draw (node50) pic[scale=.7] {tadpole};
        \draw[notoph] (+\photonlength,0) -- (0,0);
    \end{diagram}
    &
    \begin{diagram}
        \draw[photon] (-\photonlength,0) -- (0,0)
            pic             {tadpole={50}}
            ;
        \draw (node50)
            pic[scale=.7] {tadpole}
            pic           {NLO}
            ;
        \draw[notoph] (+\photonlength,0) -- (0,0);
    \end{diagram}
    \\
    3\arrayfudge
    &
    \begin{diagram}
        \draw[photon] (-\photonlength,0) -- (0,0)
            pic             {tadpole={50}}
            ;
        \draw (node50) pic[scale=.7] {tadpole={50}};
        \draw (node50) pic           {NLO};
        \draw[notoph] (+\photonlength,0) -- (0,0);
    \end{diagram}
    &
    \begin{diagram}
        \draw[photon] (-\photonlength,0) -- (0,0)
            pic             {tadpole={33,67}}
            ;
        \draw (node33) pic                       {NLO};
        \draw (node67) pic[scale=.6, rotate=-20] {tadpole};
        \draw[notoph] (+\photonlength,0) -- (0,0);
    \end{diagram}
    &
    \begin{diagram}
        \draw (0,0) pic {bubble};
        \draw[photon] (-\photonlength,0) -- (-\bubblesize,0)
            pic[scale=.7, rotate=+\tadpoleangle-90-10] {tadpole}
            pic                                        {NLO}
            ;
        \draw[notoph] (+\photonlength,0) -- (+\bubblesize,0);
    \end{diagram}
    &
    \begin{diagram}
        \draw (0,0) pic {bubble};
        \draw[photon] (-\photonlength,0) -- (-\bubblesize,0)
            pic[scale=.7, rotate=+\tadpoleangle-90-10] {tadpole}
            ;
        \draw[notoph] (+\photonlength,0) -- (+\bubblesize,0)
            pic                                        {NLO}
            ;
    \end{diagram}
    &
    \begin{diagram}
        \draw (0,0) pic {bubble};
        \draw[photon] (-\photonlength,0) -- (-\bubblesize,0)
            pic[scale=.7, rotate=+\tadpoleangle-90-10] {tadpole=50}
            ;
        \draw (node50)
            pic                                        {NLO}
            ;
        \draw[notoph] (+\photonlength,0) -- (+\bubblesize,0);
    \end{diagram}
    \\
    4\arrayfudge
    &
    \begin{diagram}
        \draw (0,0) pic {bubble={90}};
        \draw (node90)
            pic                                        {NLO}
            ;
        \draw[photon] (-\photonlength,0) -- (-\bubblesize,0)
            pic[scale=.7, rotate=+\tadpoleangle-90-10] {tadpole}
            ;
        \draw[notoph] (+\photonlength,0) -- (+\bubblesize,0);
    \end{diagram}
    &
    \begin{diagram}
        \draw (0,0) pic {bubble={90}};
        \draw (node90)
            pic[scale=.7]                              {tadpole}
            ;
        \draw[photon] (-\photonlength,0) -- (-\bubblesize,0)
            pic                                        {NLO}
            ;
        \draw[notoph] (+\photonlength,0) -- (+\bubblesize,0);
    \end{diagram}
    &
    \begin{diagram}
        \draw (0,0) pic {bubble={90}};
        \draw (node90)
            pic[scale=.7]             {tadpole}
            pic                       {NLO}
            ;
        \draw[photon] (-\photonlength,0) -- (-\bubblesize,0);
        \draw[notoph] (+\photonlength,0) -- (+\bubblesize,0);
    \end{diagram}
    &
    \begin{diagram}
        \draw (0,0) pic {bubble={90}};
        \draw (node90)
            pic[scale=.7]             {tadpole={50}}
            ;
        \draw (node50)
            pic                       {NLO}
            ;
        \draw[photon] (-\photonlength,0) -- (-\bubblesize,0);
        \draw[notoph] (+\photonlength,0) -- (+\bubblesize,0);
    \end{diagram}
    &
    \begin{diagram}
        \draw (0,0) pic {bubble={90,270}};
        \draw (node90)
            pic[scale=.7]             {tadpole}
            ;
        \draw (node270)
            pic                       {NLO}
            ;
        \draw[photon] (-\photonlength,0) -- (-\bubblesize,0);
        \draw[notoph] (+\photonlength,0) -- (+\bubblesize,0);
    \end{diagram}
    \\
    5\arrayfudge
    &
    \begin{diagram}
        \draw (0,0) pic {bubble={60,120}};
        \draw (node60)
            pic[scale=.7, rotate=-30] {tadpole}
            ;
        \draw (node120)
            pic                       {NLO}
            ;
        \draw[photon] (-\photonlength,0) -- (-\bubblesize,0);
        \draw[notoph] (+\photonlength,0) -- (+\bubblesize,0);
    \end{diagram}
    &
    \begin{diagram}[bubble2]
        \draw (-\bubblesize,0) pic {bubble};
        \draw (+\bubblesize,0) pic {bubble};
        \draw (0,0) pic {bubblejoint};
        \draw[photon] (-\bubblesize,0) + (-\photonlength,0) -- ++ (-\bubblesize,0)
            pic[elbbub2] {NLO}
            ;
        \draw[notoph] (+\bubblesize,0) + (+\photonlength,0) -- ++ (+\bubblesize,0);
    \end{diagram}
    &
    \begin{diagram}[bubble2]
        \draw (-\bubblesize,0) pic {bubble};
        \draw (+\bubblesize,0) pic {bubble};
        \draw (0,0)
            pic {bubblejoint}
            pic[elbbub2] {NLO}
            ;
        \draw[photon] (-\bubblesize,0) + (-\photonlength,0) -- ++ (-\bubblesize,0);
        \draw[notoph] (+\bubblesize,0) + (+\photonlength,0) -- ++ (+\bubblesize,0);
    \end{diagram}
    &
    \begin{diagram}[bubble2]
        \draw (-\bubblesize,0) pic {bubble={90}};
        \draw (+\bubblesize,0) pic {bubble};
        \draw (0,0) pic {bubblejoint};
        \draw (node90) pic[elbbub2] {NLO};
        \draw[photon] (-\bubblesize,0) + (-\photonlength,0) -- ++ (-\bubblesize,0);
        \draw[notoph] (+\bubblesize,0) + (+\photonlength,0) -- ++ (+\bubblesize,0);
    \end{diagram}
    &
    \begin{diagram}
        \draw[photon] (-\photonlength,0) -- (0,0)
            pic                             {tadpole}
            pic[rotate=+(\tadpoleangle+10)] {tadpole}
            pic[rotate=-(\tadpoleangle+10)] {tadpole}
            ;
        \draw[notoph] (+\photonlength,0) -- (0,0);
    \end{diagram}
    \\
    6\arrayfudge
    &
    \begin{diagram}
        \draw[photon] (-\photonlength,0) -- (0,0)
            pic             {tadpole={50}}
            pic[rotate=180] {tadpole}
            ;
        \draw (node50) pic[scale=.7] {tadpole};
        \draw[notoph] (+\photonlength,0) -- (0,0);
    \end{diagram}
    &
    \begin{diagram}
        \draw[photon] (-\photonlength,0) -- (0,0)
            pic             {tadpole={50}}
            ;
        \draw (node50) pic[scale=.7] {tadpole={50}};
        \draw (node50) pic[scale=.49] {tadpole};
        \draw[notoph] (+\photonlength,0) -- (0,0);
    \end{diagram}
    &
    \begin{diagram}
        \draw[photon] (-\photonlength,0) -- (0,0)
            pic             {tadpole={50}}
            ;
        \draw (node50)
            pic[rotate=+(\tadpoleangle+190), scale=.6] {tadpole}
            pic[rotate=-(\tadpoleangle+190), scale=.6] {tadpole}
            ;
        \draw[notoph] (+\photonlength,0) -- (0,0);
    \end{diagram}
    &
    \begin{diagram}
        \draw[photon] (-\photonlength,0) -- (0,0)
            pic             {tadpole={33,67}}
            ;
        \draw (node33) pic[scale=.6, rotate=+20] {tadpole};
        \draw (node67) pic[scale=.6, rotate=-20] {tadpole};
        \draw[notoph] (+\photonlength,0) -- (0,0);
    \end{diagram}
    &
    \begin{diagram}
        \draw (0,0) pic {bubble};
        \draw[photon] (-\photonlength,0) -- (-\bubblesize,0)
            pic[scale=.7, rotate=+\tadpoleangle-90-10] {tadpole}
            pic[scale=.7, rotate=-\tadpoleangle-90+10] {tadpole}
            ;
        \draw[notoph] (+\photonlength,0) -- (+\bubblesize,0);
    \end{diagram}
    \\
    7\arrayfudge
    &
    \begin{diagram}
        \draw (0,0) pic {bubble};
        \draw[photon] (-\photonlength,0) -- (-\bubblesize,0)
            pic[scale=.7, rotate=+\tadpoleangle-90-10] {tadpole}
            ;
        \draw[notoph] (+\photonlength,0) -- (+\bubblesize,0)
            pic[scale=.7, rotate=-\tadpoleangle+90+10] {tadpole}
            ;
    \end{diagram}
    &
    \begin{diagram}
        \draw (0,0) pic {bubble};
        \draw[photon] (-\photonlength,0) -- (-\bubblesize,0)
            pic[scale=.7, rotate=+\tadpoleangle-90-10] {tadpole={50}}
            ;
        \draw (node50)
            pic[scale=.49]                             {tadpole}
            ;
        \draw[notoph] (+\photonlength,0) -- (+\bubblesize,0);
    \end{diagram}
    &
    \begin{diagram}
        \draw (0,0) pic {bubble={90}};
        \draw (node90)
            pic[scale=.7]                              {tadpole}
            ;
        \draw[photon] (-\photonlength,0) -- (-\bubblesize,0)
            pic[scale=.7, rotate=+\tadpoleangle-90-10] {tadpole}
            ;
        \draw[notoph] (+\photonlength,0) -- (+\bubblesize,0);
    \end{diagram}
    &
    \begin{diagram}
        \draw (0,0) pic {bubble={90}};
        \draw (node90)
            pic[scale=.7]             {tadpole={50}}
            ;
        \draw (node50)
            pic[scale=.49]            {tadpole}
            ;
        \draw[photon] (-\photonlength,0) -- (-\bubblesize,0);
        \draw[notoph] (+\photonlength,0) -- (+\bubblesize,0);
    \end{diagram}
    &
    \begin{diagram}
        \draw (0,0) pic {bubble={90}};
        \draw (node90)
            pic[rotate=+(\tadpoleangle+190), scale=.7] {tadpole}
            pic[rotate=-(\tadpoleangle+190), scale=.7] {tadpole}
            ;
        \draw[photon] (-\photonlength,0) -- (-\bubblesize,0);
        \draw[notoph] (+\photonlength,0) -- (+\bubblesize,0);
    \end{diagram}
    \\
    8\arrayfudge
    &
    \begin{diagram}
        \draw (0,0) pic {bubble={60,120}};
        \draw (node60)
            pic[scale=.7, rotate=-30] {tadpole}
            ;
        \draw (node120)
            pic[scale=.7, rotate=+30] {tadpole}
            ;
        \draw[photon] (-\photonlength,0) -- (-\bubblesize,0);
        \draw[notoph] (+\photonlength,0) -- (+\bubblesize,0);
    \end{diagram}
    &
    \begin{diagram}
        \draw (0,0) pic {bubble={90,270}};
        \draw (node90)
            pic[scale=.7]             {tadpole}
            ;
        \draw (node270)
            pic[scale=.7, rotate=180] {tadpole}
            ;
        \draw[photon] (-\photonlength,0) -- (-\bubblesize,0);
        \draw[notoph] (+\photonlength,0) -- (+\bubblesize,0);
    \end{diagram}
    &
    \begin{diagram}[bubble2]
        \draw (-\bubblesize,0) pic {bubble};
        \draw (+\bubblesize,0) pic {bubble};
        \draw (0,0) pic {bubblejoint};
        \draw[photon] (-\bubblesize,0) + (-\photonlength,0) -- ++ (-\bubblesize,0)
            pic[scale=.7, rotate=+\tadpoleangle-90-10] {tadpole}
            ;
        \draw[notoph] (+\bubblesize,0) + (+\photonlength,0) -- ++ (+\bubblesize,0);
    \end{diagram}
    &
    \begin{diagram}[bubble2]
        \draw (-\bubblesize,0) pic {bubble={90}};
        \draw (+\bubblesize,0) pic {bubble};
        \draw (0,0) pic {bubblejoint};
        \draw (node90) pic[scale=.7] {tadpole};
        \draw[photon] (-\bubblesize,0) + (-\photonlength,0) -- ++ (-\bubblesize,0);
        \draw[notoph] (+\bubblesize,0) + (+\photonlength,0) -- ++ (+\bubblesize,0);
    \end{diagram}
    &
    \begin{diagram}[bubble2]
        \draw (-\bubblesize,0) pic {bubble};
        \draw (+\bubblesize,0) pic {bubble};
        \draw (0,0)
            pic {bubblejoint}
            pic[xscale=.49,yscale=.7] {tadpole}
            ;
        \draw[photon] (-\bubblesize,0) + (-\photonlength,0) -- ++ (-\bubblesize,0);
        \draw[notoph] (+\bubblesize,0) + (+\photonlength,0) -- ++ (+\bubblesize,0);
    \end{diagram}
    \\
    9\arrayfudge
    &
    \begin{diagram}[bubble3]
        \draw (0,0)             pic {bubble};
        \draw (-2\bubblesize,0) pic {bubble};
        \draw (+2\bubblesize,0) pic {bubble};
        \draw (-\bubblesize,0) pic {bubblejoint};
        \draw (+\bubblesize,0) pic {bubblejoint};
        \draw[photon] (-2\bubblesize,0) + (-\photonlength,0) -- ++ (-\bubblesize,0);
        \draw[notoph] (+2\bubblesize,0) + (+\photonlength,0) -- ++ (+\bubblesize,0);
    \end{diagram}
    &
    \begin{diagram}[threeloop]
        \draw[photon] (-\photonlength,0) -- (0,0)
            pic[xscale=1.25] {tadpole}
            pic[xscale=.5]   {tadpole}
            ;
        \draw[notoph] (+\photonlength,0) -- (0,0);
    \end{diagram}
    &
    \begin{diagram}[threeloop]
        \draw[photon] (-\photonlength,0) -- (0,0)
            pic[xscale=1.25] {tadpole={50}}
            ;
        \renewcommand{\bubbleaspect}{1}
        \draw (node50) + (0,-.05\bubblesize)
            pic[scale=.49] {bubble}
            ;
        \draw[notoph] (+\photonlength,0) -- (0,0);
    \end{diagram}
    &
    \begin{diagram}[threeloop]
        \renewcommand{\bubbleaspect}{1}
        \draw (0,0)
            pic            {bubble}
            pic[yscale=.4] {bubble}
            ;
        \draw[photon] (-\photonlength,0) -- (-\bubblesize,0);
        \draw[notoph] (+\photonlength,0) -- (+\bubblesize,0);
    \end{diagram}
    &
    \begin{diagram}[threeloop]
        \renewcommand{\bubbleaspect}{1}
        \draw (0,0)
            pic {bubble=150}
            ;
        \draw (node150)
            pic[rotate=-150, scale=.49] {bubble}
            ;
        \draw[photon] (-\photonlength,0) -- (-\bubblesize,0);
        \draw[notoph] (+\photonlength,0) -- (+\bubblesize,0);
    \end{diagram}
    \\
    10\arrayfudge
    &
    \begin{diagram}[threeloop]
        \renewcommand{\bubbleaspect}{1}
        \draw (0,0)
            pic {bubble=90}
            ;
        \draw (node90)
            pic[scale=.49] {bubble}
            ;
        \draw[photon] (-\photonlength,0) -- (-\bubblesize,0);
        \draw[notoph] (+\photonlength,0) -- (+\bubblesize,0);
    \end{diagram}
    &
    \begin{diagram}[threeloop]
        \renewcommand{\bubbleaspect}{1}
        \draw (0,0)
            pic            {bubble}
            pic[xscale=.4] {bubble}
            ;
        \draw[photon] (-\photonlength,0) -- (-\bubblesize,0);
        \draw[notoph] (+\photonlength,0) -- (+\bubblesize,0);
    \end{diagram}
    &&&\\
    \bottomrule
\end{tabular}

%% file: implementation.tex
\usetikzlibrary{arrows}
\tikzset{
    flowchart box/.style={draw, very thick, text width=6.2em, text centered, minimum height=4em},
    calcstep/.style={flowchart box},
    resource/.style={flowchart box, rounded corners},
    terminal/.style={resource, double},
    dummy/.style={flowchart box, draw=none},
    overpass/.style={preaction={draw, white, line width=.15cm}},
    oneway/.style={draw, thick, -latex'},
    twoway/.style={draw, thick, latex'-latex'},
    what/.style={pos=#1, draw, fill=white, rounded corners=.8em, minimum height=1.6em}, what/.default={.5},
    }

\newlength{\shift}
\newcommand{\N}[1]{($(#1)+(0,+\shift)$)}
\renewcommand{\S}[1]{($(#1)+(0,-\shift)$)}
\renewcommand{\E}[1]{($(#1)+(+\shift,0)$)}
\newcommand{\W}[1]{($(#1)+(-\shift,0)$)}
\newcommand{\NE}[1]{($(#1)+(+\shift,+\shift)$)}
\newcommand{\NW}[1]{($(#1)+(-\shift,+\shift)$)}
\newcommand{\SE}[1]{($(#1)+(+\shift,-\shift)$)}
\newcommand{\SW}[1]{($(#1)+(-\shift,-\shift)$)}

\newcommand{\filebox}[3]{%
    \draw[dashed] \NW{#1.north west} rectangle \SE{#2.south east};
    \node[anchor=south west, inner sep=2pt] at \NW{#1.north west} {\texttt{#3}};
    }

\begin{tikzpicture}[node distance=3cm]
    \node[calcstep               ]  (diag) at (0,0) {Diagram assembly};
    \node[dummy,     left of=diag]  (dum1)          {};
    \node[resource,  left of=dum1]  (chpt)          {\texttt{HVP*LO.chpt}};
    \node[dummy,    right of=diag]  (dum2)          {};
    \node[resource, right of=dum2]  (lagr)          {\chpt\ Lagrangian};
    \begin{scope}[node distance=2cm]
        \node[calcstep, below of=diag]  (proj)      {Tensor projection};
        \node[resource, below of=lagr]  (rnrm)      {\chpt\ counterterms};
        \node[resource, below of=rnrm]  (mfxi)      {$\Mpi$/$\Fpi$ $\xi$ expansions};
        \node[calcstep, below of=proj]  (mred)      {Master integral reduction};
    \end{scope}
    \node[dummy,     left of=mred]  (dum3)          {};
    \node[dummy,    right of=mred]  (dum4)          {};
    \node[resource,  left of=dum3]  (lred)          {\texttt{LiteRed 2}};
    \node[resource, below of=lred]  (tbub)          {\texttt{tadbub.frm}};
    \node[calcstep, below of=mred]  (epsx)          {$\epsilon$ expansion};
    \node[calcstep,  left of=epsx]  (dims)          {Dimension shift};
    \node[calcstep, right of=epsx]  (lecs)          {Renormali\-zation};
    \node[calcstep, below of=dims]  (schn)          {Schouten relations};
    \node[terminal, below of=epsx]  (res!)          {Result!};
    \node[calcstep, right of=res!]  (mass)          {$\Mpi$/$\Fpi$ renormalization};
    \node[resource, right of=mass]  (lowr)          {lower-order amplitudes};
    \begin{scope}[node distance=2cm]
        \node[dummy,    below of=mass]  (dum5)      {};
    \end{scope}

    \setlength{\shift}{.16cm}
    \filebox{diag}{proj}{HVP*LO.frm}
    \filebox{lagr}{mfxi}{ChPTlib}
    \filebox{dum3}{mass}{HVP.frm}

    \setlength{\shift}{.28cm}

    \draw[oneway] (chpt) -- (diag) node[what] {Topologies};
    \draw[oneway] (lagr) -- (diag) node[what] {Feynman rules};
    \draw[oneway] (diag) -- (proj);
    \draw[oneway] (proj) -- (mred);
    \draw[oneway] (epsx) -- (res!);
    \draw[oneway] \E{tbub.north} |- ($(dims.north)!.5!(dum3.south)$) -| \W{epsx.north};
    \draw[oneway] \W{tbub.north} -- \W{lred.south};
    \draw[oneway] \E{mred.south} -- \E{epsx.north} node[what] {$T_{ij}$};
    \draw[oneway, overpass] \S{mred.west} -| \E{dims.north} node[what=.2] {$E_i$};
    \draw[oneway] \W{dims.south} -- \W{schn.north} node[what] {$E_{5,6}$};
    \draw[oneway] \S{schn.east} -- \S{res!.west};
    \draw[oneway] \S{mass.west} -- \S{res!.east};
    \draw[oneway] \E{dims.south} -- \NE{schn.north} -| \NE{schn.east} -- \N{res!.west};
    \draw[oneway] (lecs) -- \N{mass.north} -| \NW{mass.west} -- \N{res!.east};
    \draw[oneway] (mred) -| \W{lecs.north} node[what=.25] {LECs};
    \draw[oneway] (rnrm) -| \E{lecs.north};
    \draw[oneway] (mfxi) -- ($(mfxi.west)!.5!(dum4.east)$) |- \N{mass.east};
    \draw[oneway] \S{lowr.west} -- \S{mass.east};
    \draw[twoway] \N{mred.west} -- \N{lred.east};
    \draw[oneway, overpass] \S{lred.east} -| \W{dims.north};
    \draw[oneway, densely dotted] \E{res!.south} |- (dum5.center) -| (lowr) node[what=0] {NLO, NNLO};

\end{tikzpicture}